\documentclass[aps,prb,twocolumn,amsmath,amssymb,nofootinbib,superscriptaddress,floatfix]{revtex4}
\usepackage{amsmath}
\usepackage{amssymb}
\usepackage{amsthm}
\usepackage[pdftex,colorlinks,citecolor=blue,linkcolor=blue,urlcolor=blue]{hyperref} 
\usepackage{graphicx}
\usepackage{dcolumn} 
\usepackage{bm} 
\usepackage{longtable}
\usepackage{ulem}   
\usepackage{comment} 
\usepackage{datetime}

\usepackage{tabularx}
\usepackage{float}
\restylefloat{table}

\usepackage{dcolumn,booktabs}
\newcolumntype{d}[1]{D{,}{,}{#1}} 
\newcolumntype{.}[1]{D{.}{.}{#1}} 

\newcommand{\be}{\begin{equation}}
\newcommand{\ee}{\end{equation}}
\newcommand{\bea}{\begin{eqnarray}}
\newcommand{\eea}{\end{eqnarray}}
\newcommand{\barr}{\begin{array}}
\newcommand{\earr}{\end{array}}

\long\def\begincomment#1\endcomment{}

\newcommand{\cH}{\mathcal{H}}

\newcommand{\tr}{\mathop{\mathrm{tr}}}

\newcommand{\Z}{\mathbb{Z}}

\newcommand{\tU}{\text{U}}

\newcommand{\cblue}[1]{\textcolor{black}{#1}}

\newcommand{\ti}{\text{i}}

\newcommand{\frm}[1]{
\fbox{\parbox{3.1in}{
\parindent=0pt #1
}
}
}

\newcommand{\middlewave}[1]{\raisebox{0.5em}{\uwave{\hspace{#1}}}}

\usepackage{tikz}
\usetikzlibrary{arrows,matrix,calc,scopes,decorations.markings}

\newcommand{\CocycleTriangleTWO}[5]{
\begin{tikzpicture}[scale=0.9,baseline]
\coordinate (c) at (0,0);
\coordinate (d) at (0,1.1);
\coordinate (a) at (210:1.1);
\coordinate (b) at (330:1.1);

\draw (a) node[left] {\scalebox{0.85}{$#1$}} -- (b) node[right] {\scalebox{0.85}{$#2$}} -- (c) node[below] {\scalebox{0.85}{$#3$}} -- (d) node[above] {\scalebox{0.85}{$#4$}} -- (a) -- (c);
\draw (b) -- (d);
\draw[->,>=latex,line width=0.65pt] (a) -- ($ (b) ! 0.5 ! (a) $); 
\draw[-<,>=latex,line width=0.65pt] (a) -- ($ (a) ! 0.5 ! (d) $);
\draw[-<,>=latex,line width=0.65pt] (b) -- ($ (b) ! 0.5 ! (d) $);

\ifnum #5=1 {
\draw[-<,>=latex,line width=0.65pt] (b) -- ($ (b) ! 0.5 ! (c) $);
\draw[->,>=latex,line width=0.65pt] (c) -- ($ (c) ! 0.5 ! (d) $);
\draw[-<,>=latex,line width=0.65pt] (a) -- ($ (a) ! 0.5 ! (c) $);}
\fi
\ifnum #5=2 {
\draw[-<,>=latex,line width=0.65pt] (b) -- ($ (b) ! 0.5 ! (c) $);
\draw[-<,>=latex,line width=0.65pt] (c) -- ($ (c) ! 0.5 ! (d) $);
\draw[-<,>=latex,line width=0.65pt] (a) -- ($ (a) ! 0.5 ! (c) $);}
\fi
\ifnum #5=3 {
\draw[->,>=latex,line width=0.65pt] (b) -- ($ (b) ! 0.5 ! (c) $);
\draw[-<,>=latex,line width=0.65pt] (c) -- ($ (c) ! 0.5 ! (d) $);
\draw[-<,>=latex,line width=0.65pt] (a) -- ($ (a) ! 0.5 ! (c) $);}
\fi
\ifnum #5=4 {
\draw[->,>=latex,line width=0.65pt] (b) -- ($ (b) ! 0.5 ! (c) $);
\draw[-<,>=latex,line width=0.65pt] (c) -- ($ (c) ! 0.5 ! (d) $);
\draw[->,>=latex,line width=0.65pt] (a) -- ($ (a) ! 0.5 ! (c) $);}
\fi
\end{tikzpicture}
}

\newcommand{\CocycleTriangle}[5]{
\begin{tikzpicture}[scale=0.9,baseline]
\coordinate (c) at (0,0);
\coordinate (d) at (0,1.1);
\coordinate (a) at (210:1.1);
\coordinate (b) at (330:1.1);

\draw (a) node[left] {\scalebox{0.85}{$#1$}} -- (b) node[right] {\scalebox{0.85}{$#2$}} -- (c) node[below] {\scalebox{0.85}{$#3$}} -- (d) node[above] {\scalebox{0.85}{$#4$}} -- (a) -- (c);
\draw (b) -- (d);
\draw[-<,>=latex,line width=0.65pt] (a) -- ($ (a) ! 0.5 ! (b) $);
\draw[-<,>=latex,line width=0.65pt] (a) -- ($ (a) ! 0.5 ! (d) $);
\draw[-<,>=latex,line width=0.65pt] (b) -- ($ (b) ! 0.5 ! (d) $);

\ifnum #5=1 {
\draw[-<,>=latex,line width=0.65pt] (b) -- ($ (b) ! 0.5 ! (c) $);
\draw[->,>=latex,line width=0.65pt] (c) -- ($ (c) ! 0.5 ! (d) $);
\draw[-<,>=latex,line width=0.65pt] (a) -- ($ (a) ! 0.5 ! (c) $);}
\fi
\ifnum #5=2 {
\draw[-<,>=latex,line width=0.65pt] (b) -- ($ (b) ! 0.5 ! (c) $);
\draw[-<,>=latex,line width=0.65pt] (c) -- ($ (c) ! 0.5 ! (d) $);
\draw[-<,>=latex,line width=0.65pt] (a) -- ($ (a) ! 0.5 ! (c) $);}
\fi
\ifnum #5=3 {
\draw[->,>=latex,line width=0.65pt] (b) -- ($ (b) ! 0.5 ! (c) $);
\draw[-<,>=latex,line width=0.65pt] (c) -- ($ (c) ! 0.5 ! (d) $);
\draw[-<,>=latex,line width=0.65pt] (a) -- ($ (a) ! 0.5 ! (c) $);}
\fi
\ifnum #5=4 {
\draw[->,>=latex,line width=0.65pt] (b) -- ($ (b) ! 0.5 ! (c) $);
\draw[-<,>=latex,line width=0.65pt] (c) -- ($ (c) ! 0.5 ! (d) $);
\draw[->,>=latex,line width=0.65pt] (a) -- ($ (a) ! 0.5 ! (c) $);}
\fi
\end{tikzpicture}
}

\newcommand{\tetrahedraTWO}[4]{
\begin{tikzpicture}[scale=1,baseline]
\coordinate (a) at (0,-0.1);
\coordinate (b) at (0.82,-0.79);
\coordinate (c) at (2.5,0);
\coordinate (d) at (1.25,1.5);
\coordinate (e) at ($ (a) ! 0.32 ! (c) $);
\coordinate (f) at ($ (a) ! 0.525 ! (c) $);
\draw (a) -- (e);
\draw (b) -- (d);
\draw (f) -- (c);
\draw (a) -- (b) -- (c) -- (d) -- cycle;
\draw[->,>=latex, line width=0.75pt] (a) -- ($ (a) ! 0.5 ! (b) $);
\draw[-<,>=latex, line width=0.75pt] (b) -- ($ (b) ! 0.5 ! (c) $);
\draw[-<,>=latex, line width=0.75pt] (c) -- ($ (c) ! 0.5 ! (d) $);
\draw[<-,>=latex, line width=0.75pt] (f) -- (c);
\draw[-<,>=latex, line width=0.75pt] (b) -- ($ (b) ! 0.5 ! (d) $);
\draw[-<,>=latex, line width=0.75pt] (a) -- ($ (a) ! 0.5 ! (d) $);

\node[left] at(a) {\scalebox{0.7}{$#1$}};
\node[below] at(b) {\scalebox{0.7}{$#2$}};
\node[right] at(c) {\scalebox{0.7}{$#3$}};
\node[above] at(d) {\scalebox{0.7}{$#4$}};

\node[below] at ($ (a) ! 0.4 ! (b) $) {\scalebox{0.7}{$#2{#1}^{-1}$}};
\node[below] at ($ (b) ! 0.65 ! (c) $) {\scalebox{0.7}{$#2{#3}^{-1}$}};
\node[right] at ($ (c) ! 0.5 ! (d) $) {\scalebox{0.7}{$#3{#4}^{-1}$}};
\node[below] at (f) {\scalebox{0.7}{$#1{#3}^{-1}$}};
\node[right] at ($ (b) ! 0.49 ! (d) $) {\scalebox{0.7}{$#2{#4}^{-1}$}};
\node[left] at ($ (a) ! 0.51 ! (d) $) {\scalebox{0.7}{$#1{#4}^{-1}$}};
\end{tikzpicture}
}

\newcommand{\tetrahedra}[4]{
\begin{tikzpicture}[scale=1,baseline]
\coordinate (a) at (0,-0.1);
\coordinate (b) at (0.82,-0.79);
\coordinate (c) at (2.5,0);
\coordinate (d) at (1.25,1.5);
\coordinate (e) at ($ (a) ! 0.32 ! (c) $);
\coordinate (f) at ($ (a) ! 0.525 ! (c) $);
\draw (a) -- (e);
\draw (b) -- (d);
\draw (f) -- (c);
\draw (a) -- (b) -- (c) -- (d) -- cycle;
\draw[-<,>=latex, line width=0.75pt] (a) -- ($ (a) ! 0.5 ! (b) $);
\draw[-<,>=latex, line width=0.75pt] (b) -- ($ (b) ! 0.5 ! (c) $);
\draw[-<,>=latex, line width=0.75pt] (c) -- ($ (c) ! 0.5 ! (d) $);
\draw[<-,>=latex, line width=0.75pt] (f) -- (c);
\draw[-<,>=latex, line width=0.75pt] (b) -- ($ (b) ! 0.5 ! (d) $);
\draw[-<,>=latex, line width=0.75pt] (a) -- ($ (a) ! 0.5 ! (d) $);

\node[left] at(a) {\scalebox{0.7}{$#1$}};
\node[below] at(b) {\scalebox{0.7}{$#2$}};
\node[right] at(c) {\scalebox{0.7}{$#3$}};
\node[above] at(d) {\scalebox{0.7}{$#4$}};

\node[below] at ($ (a) ! 0.4 ! (b) $) {\scalebox{0.7}{$#1{#2}^{-1}$}};
\node[below] at ($ (b) ! 0.65 ! (c) $) {\scalebox{0.7}{$#2{#3}^{-1}$}};
\node[right] at ($ (c) ! 0.5 ! (d) $) {\scalebox{0.7}{$#3{#4}^{-1}$}};
\node[below] at (f) {\scalebox{0.7}{$#1{#3}^{-1}$}};
\node[right] at ($ (b) ! 0.49 ! (d) $) {\scalebox{0.7}{$#2{#4}^{-1}$}};
\node[left] at ($ (a) ! 0.51 ! (d) $) {\scalebox{0.7}{$#1{#4}^{-1}$}};
\end{tikzpicture}
}

\begin{document}


\title{Bosonic Anomalies, Induced Fractional Quantum Numbers and Degenerate Zero 
Modes: 
the anomalous edge physics of Symmetry-Protected Topological States
}


\author{Juven C. Wang} \email{juven@mit.edu}  
\affiliation{Department of Physics, Massachusetts Institute of Technology, Cambridge, MA 02139, USA}
\affiliation{Perimeter Institute for Theoretical Physics, Waterloo, ON, N2L 2Y5, Canada}

\author{Luiz H. Santos}  \email{lsantos@perimeterinstitute.ca}
\affiliation{Perimeter Institute for Theoretical Physics, Waterloo, ON, N2L 2Y5, Canada}

\author{Xiao-Gang Wen} \email{xwen@perimeterinstitute.ca}
\affiliation{Perimeter Institute for Theoretical Physics, Waterloo, ON, N2L 2Y5, Canada}
\affiliation{Department of Physics, Massachusetts Institute of Technology, Cambridge, MA 02139, USA}
\affiliation{Institute for Advanced Study, Tsinghua University, Beijing,
100084, P. R. China}


\begin{abstract}

The boundary of symmetry-protected topological states (SPTs) can harbor new quantum anomaly phenomena.
In this work, 
we characterize the bosonic anomalies introduced by the 1+1D non-onsite-symmetric gapless edge modes 
of 2+1D bulk bosonic SPTs with a generic finite Abelian group symmetry (isomorphic to 
$G=\prod_i Z_{N_i}=Z_{N_1} \times Z_{N_2} \times Z_{N_3} \times \dots$). 
We demonstrate that some classes of SPTs (termed ``Type II") trap fractional quantum numbers (such as fractional $Z_N$ charges) at the 0D kink of the symmetry-breaking domain walls;
while 
some classes of SPTs (termed ``Type III") have degenerate zero energy modes (carrying the projective representation protected by the unbroken part of the symmetry),
either near the 0D kink of a symmetry-breaking domain wall, or on a symmetry-preserving 1D system dimensionally reduced from a thin 2D tube 
with a monodromy defect 1D line embedded. 
More generally, the energy spectrum and conformal dimensions of gapless edge modes under an external gauge flux insertion (or twisted by a branch cut, i.e., a monodromy defect line) through the 1D ring
can distinguish many SPT 
classes. 
We provide a manifest correspondence from the physical phenomena, the induced fractional quantum number and the zero energy mode degeneracy, 
to the mathematical concept of cocycles that appears in the group cohomology classification of SPTs, thus achieving a concrete physical materialization of the cocycles.
The aforementioned edge properties are formulated 
in terms of a long wavelength
continuum field theory involving scalar chiral bosons, as well as in terms of 
Matrix Product Operators and discrete quantum lattice models. 
Our lattice approach yields a regularization with anomalous non-onsite symmetry for the field theory description.
We also formulate some bosonic anomalies in terms of the Goldstone-Wilczek formula. 
\end{abstract}

\maketitle
\tableofcontents

\section{Introduction}

Symmetry dictates the conservation law and the corresponding conserved current on classical actions in classical physics, such as by Noether's theorem.\cite{Noether}
However, as it is now well-known,  there is a potential obstruction of some classical symmetry to be promoted to a consistent symmetry in the quantum level.
This is the paradigm of ``quantum anomalies.''\cite{'tHooft:1979bh}

Quantum anomalies occur in our real-world physics, such as pion decaying to two photons via Adler-Bell-Jackiw chiral anomaly.\cite{{Adler:1969gk},{Bell:1969ts},{Supplementary Material}}
Anomalies also constrain beautifully on the Standard Model of particle physics, in particular to the Glashow-Weinberg-Salam theory, via anomaly-cancellations 
of gauge and gravitational couplings. 
The above two familiar examples of anomalies concern chiral fermions and continuous symmetry (e.g. U(1), SU(2), SU(3)).
Out of curiosity, we ask 
``Are there concrete examples of quantum anomalies for bosons instead? And anomalies for discrete symmetries? 
Are they potentially testable experimentally in the lab in the near future?'' 

In this work, we address the question affirmatively and demonstrate that ``bosonic anomalies for discrete symmetries'' 
can be expected on the boundary of some interacting bosonic symmetry-protected topological states(SPTs)
in condensed matter systems.\cite{{Chen:2011pg},2013arXiv1301.0861C}
(Such interacting bosonic SPTs may be realized in the future by applying the ultracold bosonic gas controlled by optical lattice,\cite{Bloch:2008zzb}
see a recent proposal and reference therein.\cite{1404.2818})
Our work thus will address some of the interplays between ``symmetry,'' ``quantum anomaly,'' and ``topology.''

There has been rapid progress 
on exploring the entangled quantum states with gapless edge modes protected by some global symmetry.
The classic 
example is the one dimensional(1D, one dimensional space and one dimensional time, or 1+1D)
 Haldane spin-1 chain with SO(3) spin rotational symmetry.\cite{{H8364},{AKL8877}}
Another renown example are topological insulators, which are protected by fermion number conservation U(1) symmetry and time reversal symmetry $Z_2^T$.\cite{{KM0501},
{BZ0602},
{KM0502},
{MB0706},
{FKM0703},
{QHZ0824}}
Topological insulator may be realized in a non-interacting free fermion system,  
while there are 
so-called the bosonic SPTs, 
which can only happen in an 
interacting bosonic system.
%


{In attempting to understand various phases
of interacting bosonic systems, it is important to try to characterize them in terms of
unique physical properties. The goal of this paper is to address this question for bosonic
SPTs in $2$D.}
{Let us motivate our question
in the simplest scenario of the $1$D SPTs given by the spin-$1$ Haldane chain. 
The Hamiltonian conserves spin rotation and time-reversal symmetries and the ground state is formed
by singlets in the bulk. Bulk excitations are formed by breaking singlets, a process that requires an energy gap. 
Its non-trivial property resides on the edges, both of which contain an effective spin-$1/2$ transforming projectively 
under rotation or time-reversal symmetry. Since the edge spin is effectively ``{free}'', 
it renders 
a $2$-fold degeneracy (per edge) in the spectrum.
 Hence, here the mathematical concept of projective representations is directly connected to spectral zero energy mode degeneracy.}

\begin{figure}[h!]
\includegraphics[width=0.5\textwidth]{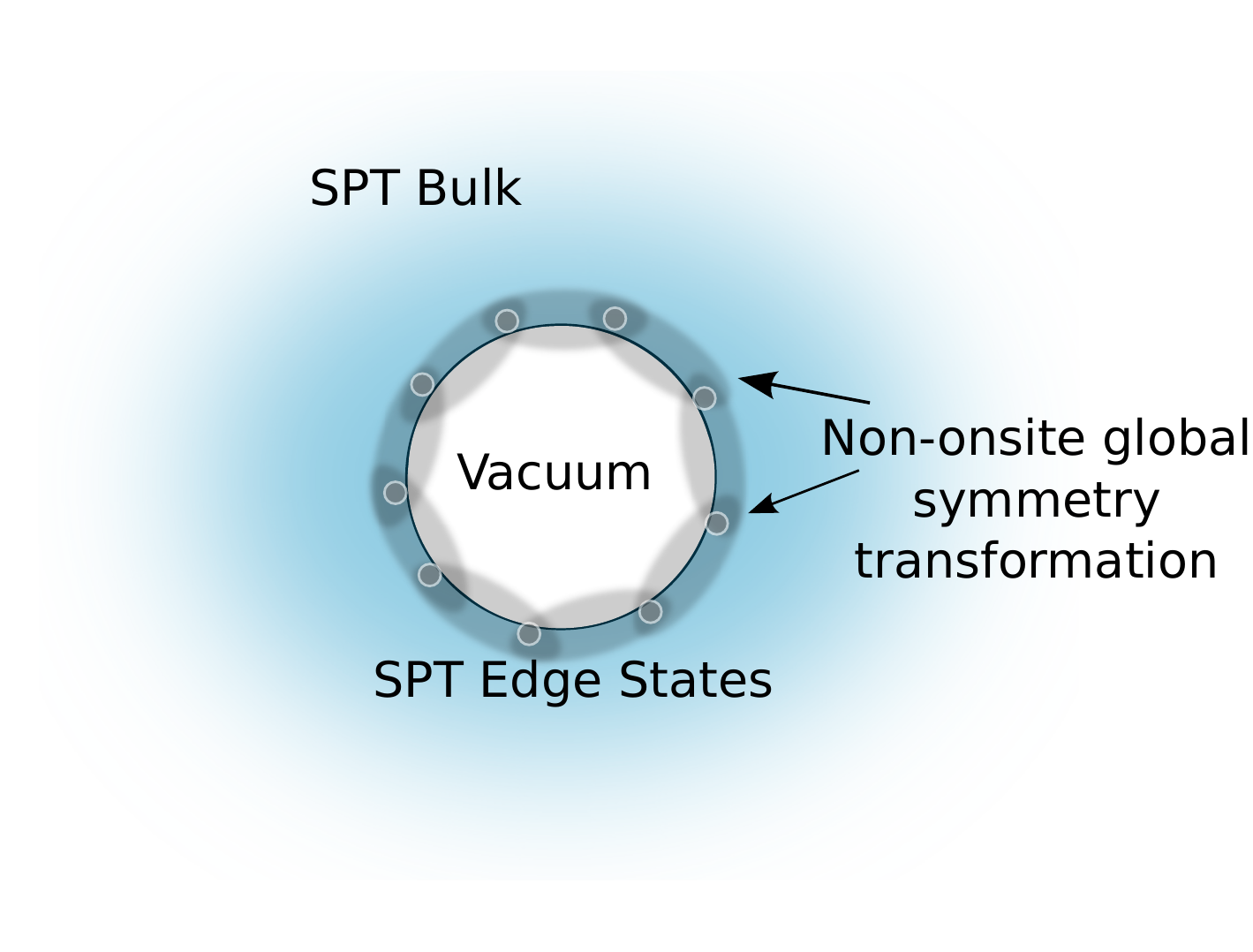}
\caption{The boundary of 2D SPT state harbors 1D gapless edge modes if the global symmetry is preserved (not broken spontaneously or explicitly).
The global symmetry transformation $S$ of 1D edge mode acts in a non-onsite manner, where $S$
cannot be written as a tensor product form on each site. (i.e. The symmetry operator $S$ acts on more than a single site for its tensor operators, where
we show schematically $S$ acts on two neighbor sites.)
}
\label{fig:1}
\end{figure}

{In this work, we will show that edge modes of bosonic SPTs in $2$D can also provide
physical signatures of the bulk state.  
We will study the 2D bosonic SPTs with 1D edge modes on the boundary(see Fig.\ref{fig:1}), protected by a global symmetry $G$ of 
a generic cyclic group $G=\prod_i Z_{N_i}= Z_{N_1} \times Z_{N_2} \times Z_{N_3} \times \dots$ (to which any finite Abelian group is isomorphic).
Our basic result is that point defects on the 1D edge are associated to induced $Z_{N}$ charge (referred as Type II bosonic anomaly in Sec.\ref{sec: Type II frac})
or protected degeneracies(referred as Type III bosonic anomaly in Sec.\ref{sec: Type III zero}) for some classes of SPTs.
}

{The edge modes of our focus 
have the 
property
that they can only be gapped out if the symmetry is broken. In a description around a gapless 
1+1D Luttinger liquid-like fixed point, this means that putative interacting energy-gap-opening terms (sine-Gordon cosine terms) violate
the symmetry and are therefore forbidden (which does not rule out the possibility that a gap may
open by symmetry breaking). 
The suppression of all these gap opening terms is a manifestation 
that counter-propagating modes carry different global charges, which,
consequentially implies that back-scattering processes violate the symmetry.
Thus an important step in capturing the edge properties of SPTs 
is to construct the symmetry transformation that endow counter propagating modes with this
anomalous property. We will study this \emph{anomalous non-onsite symmetry} explicitly. 
}

Recently, several theoretical approaches have been developed to understand bosonic SPTs, such as using group cohomology,\cite{Chen:2011pg,{2013arXiv1301.0861C},{Mesaros:2012yd}} 
lattice models,\cite{{LevinGu},{2011PhRvB..84w5141C},{Chen:2012hc},{Santos:2013uda},{Chen:201301}} 
matrix product states,\cite{{2011PhRvB..84w5141C},{Santos:2013uda}} 
field theory techniques,\cite{{Lu:2012dt},{Vishwanath:2012tq},{Ye:2013upa},{Bi:2013oza},{Santos:2013uda},{Chen:201301}} 
or projective 
construction.\cite{{Essin:2013rca},{Ye:2013wma},{1210.0907},{1210.0909},{1212.0863}}
One of the goals of this paper is to address the connections among miscellaneous approaches by working out a few specific examples.
To this end, we 
specifically highlight 
three learned aspects about 
SPTs-\\%
\underline{$[1]$. \emph{Non-onsite symmetry on the edge}}: 
An important feature of SPT is that the \emph{global symmetry}  
\cblue{acting on a local Hamiltonian of edge modes}  
is realized 
\emph{non-onsite}.\cite{{2011PhRvB..84w5141C},{Chen:2012hc},{Santos:2013uda}}
For a given symmetry group $G$, the non-onsite symmetry means that its symmetry transformation \emph{cannot} be written as a tensor product form on each site,\cite{{Chen:2011pg},{2011PhRvB..84w5141C}} 
\be
U(g)_{\text{non-onsite}} \neq \otimes _i U_i(g),
\ee 
for $g \in G$  of the symmetry group. On the other hand, the onsite symmetry transformation $U(g)$ can be written in a tensor product form acting on each site $i$,\cite{Chen:2011pg,2011PhRvB..84w5141C} i.e.
$U(g)_{\text{onsite}}= \otimes _i U_i(g)$, for $g \in G$. 
(The symmetry transformation acts as an operator $U(g)$ with $g \in G$, transforming the state $| v \rangle$ globally by  $U(g) | v \rangle$.)
Therefore, to study the SPT edge mode, one should realize how the non-onsite symmetry acts on the 
boundary as in Fig.\ref{fig:1}.
\\
\underline{$[2]$. \emph{Group cohomology construction}}: 
It has been proposed that $d+1$ dimensional($d+1$D) SPTs of symmetry-group-$G$
interacting boson system can be constructed by the 
number of distinct cocycles in the $d+1$-th cohomology group, $\cH^{d+1}(G,\tU(1))$, with $\tU(1)$ coefficient.\cite{Chen:2011pg,{Dijkgraaf:1989pz}} 
(See also the first use of cocycle in the high energy context by Jackiw in Ref.\onlinecite{{Jackiw:1984rd},{Treiman:1986ep}})
While another general framework of cobordism theory is subsequently proposed\cite{Kapustin:2014tfa}
to account for subtleties when symmetry $G$ involves time-reversal,\cite{Vishwanath:2012tq} in our work we will focus on a 
finite Abelian symmetry group $G=\prod_i Z_{N_i}$,
where the group cohomology is a complete classification.\\
\underline{$[3]$. \emph{Surface anomalies}}:
It has been proposed that the edge modes of SPTs are the source of gauge anomalies, while that of intrinsic topological orders are the source of gravitational anomalies.\cite{Wen:2013oza}
SPT boundary states are known to show at least one of three properties:\\
$\bullet$(1) symmetry-preserving gapless edge modes,\\
$\bullet$(2) symmetry-breaking gapped edge modes, \\
$\bullet$(3)\;symmetry-preserving gapped edge modes with surface topological order.\cite{Vishwanath:2012tq,{Burnell:2013bka},{MetlitskiFisher},{ChongSenthil},{Bonderson:2013pla}}\\

\noindent
{\bf 
Bosonic Anomalies realized on the SPT edge}

The three aspects $\bullet$(1),$\bullet$(2),$\bullet$(3) above had hinted at the bosonic anomalies harbored on the boundary of interacting bosonic SPTs.
In this work, we focus on \emph{characterizing the bosonic anomalies as precisely as possible}, and attempt 
to connect our \emph{bosonic anomalies to the notion defined in the high energy physics context.}
In short, we aim to \emph{make connections between the meanings of boundary bosonic anomalies studied in both high energy physics and condensed matter theory}.

We will examine a generic finite Abelian 
$G=\prod_i Z_{N_i}$ bosonic SPTs,
and study {\it what is truly anomalous} about the edge under the case of $\bullet$(1) and $\bullet$(2) above. 
(Since it is forbidden to have any intrinsic topological order in a 1D edge, we do not have scenario  $\bullet$(3).)
We focus on addressing the properties of its 1+1D edge modes, their anomalous non-onsite symmetry and bosonic anomalies 
from three different perspectives,
(i) quantum lattice models, (ii) matrix product states, and (iii) quantum field theory; while connecting them to cocycles of group cohomology.

We shall now define the meaning of quantum anomaly in a language appreciable by both high energy physics and condensed matter communities -

\frm{The {quantum anomaly} is \emph{an obstruction of a symmetry of a theory to be fully-regularized for a full quantum theory as an onsite symmetry on the UV-cutoff lattice 
in the same spacetime dimension}.} 

According to this definition, 
to characterize our bosonic anomalies, we will find several possible obstructions to regulate the symmetry at the quantum level:

\noindent
$\star$ \underline{Obstruction of onsite symmetries}:
Consistently we will find throughout our examples to fully-regularize our SPTs 1D edge theory on the 1D lattice Hamiltonian requires the
\emph{non-onsite symmetry}, namely, \emph{realizing the symmetry anomalously}. 
\cblue{The {non-onsite symmetry} on the edge cannot be ``dynamically gauged'' on its own spacetime dimension,\cite{{LevinGu},{2011PhRvB..84w5141C},{Chen:2012hc},{Santos:2013uda},{Wen:2013oza}}
thus this also implies the following obstruction.}

\begin{figure}[h!]
 \includegraphics[width=0.45\textwidth]{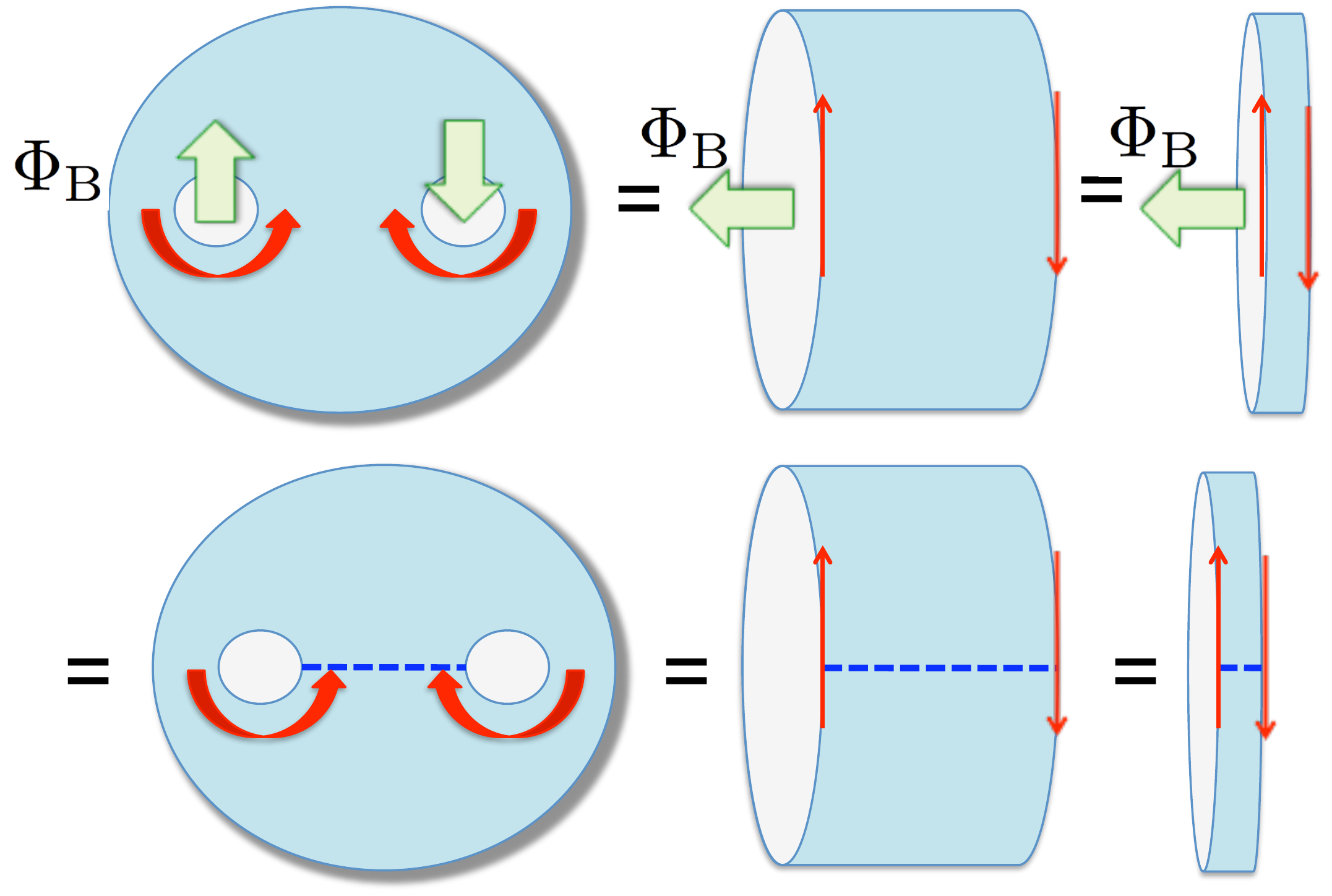}
\caption{ The intuitive way to view the bulk-boundary correspondence for edge modes of SPTs (or intrinsic topological order) under
the flux insertion, or equivalently the monodromy defect / branch cut (blue dashed line) modifying the bulk and the edge Hamiltonians.
SPTs locate on a large sphere with two holes with flux-in and flux-out, is analogous to, a Laughlin type flux insertion through a cylinder,
inducing 
anomalous edge modes(red arrows) moving along the opposite directions on two edges.
}
\label{fig:flux_cut_analogy}
\end{figure}

\noindent
$\star$ \underline{Obstruction of the same spacetime dimension}:
\cblue{We will show 
that the physical observables 
for gapless edge modes (the case $\bullet$(1)) 
are their energy spectral shifts\cite{Santos:2013uda}  
under symmetry-preserving external flux insertion through 
a compact 1D ring.  
The energy spectral shift is caused by the Laughlin-type flux insertion of Fig.\ref{fig:flux_cut_analogy}.
The \emph{flux insertion} can be equivalently regarded as an effective \emph{branch cut} 
modifying the Hamiltonian (blue dashed line in Fig.\ref{fig:flux_cut_analogy}) connecting from the edge to an extra dimensional bulk.
Thus the spectral shifts also indicate the transportation of quantum numbers from one edge to the other edge.
This can be regarded as the anomaly requiring an \emph{extra dimensional bulk}. 
}

\noindent
$\star$ \underline{Non-perturbative effects}:
We know that the familiar 
Adler-Bell-Jackiw anomaly of \emph{chiral fermions},\cite{{Adler:1969gk},{Bell:1969ts}} observed 
in the pion decay in particle-physics can be captured
by the perturbative 1-loop Feynman diagram. 
However, importantly, the result is non-perturbative, being exact from low energy IR to high energy UV.
This effect can be further confirmed via Fujikawa's path integral method\cite{Fujikawa:1979ay} non-perturbatively.
Instead of the well-known \emph{chiral fermionic anomalies}, do we have \emph{bosonic anomalies} with these non-perturbative effects?

Indeed, yes, we will show two other kinds of bosonic anomalies with non-perturbative effects 
with symmetry-breaking gapped edges (\cblue{the case $\bullet$(2)}): One kind of consequent anomalies for Type II SPTs
under $Z_{N_1}$ symmetry-breaking domain walls 
is 
the {\bf induced fractional $Z_{N_2}$ charge} 
trapped near 0D kink of gapped domain walls.
Amazingly, through a fermionization/bosonization procedure, 
we can apply the field-theoretic Goldstone-Wilczek method to capture the 1-loop Feynman diagram effect \emph{non-perturbatively}, 
as this fractional charge is known to be robust without higher-loop diagrammatic corrections.\cite{Goldstone:1981kk}
We will term this a Type II bosonic anomaly.

The second kind of anomalies for symmetry-breaking gapped edge (\cblue{the case $\bullet$(2)})
is that the edge is gapped under $Z_{N_1}$ symmetry-breaking domain walls, with a consequent 
{\bf degenerate zero energy ground states} due to the projective representation of other symmetries $Z_{N_2} \times Z_{N_3}$.
The zero mode degeneracy is found to be  {$\gcd(N_1,N_2,N_3)$}-fold.
We will term this a Type III bosonic anomaly. 


The paper is organized as follows. 
In Sec.\ref{sec: GC}, we start with some basic results in group cohomology and its $n$-cocycles. The readers who are not familiar with group cohomology 
may either take the chance to learn the basics, or skip it 
and proceed to Sec.\ref{sec:Type I}. 
We set up Type I,\cite{Santos:2013uda}  II, III SPT lattice construction in Sec.\ref{sec:Type I}, its matrix product operators and its low energy field theory.
Remarkably, the Type III non-onsite symmetry transformation is distinct from the Type I,  Type II;  
it introduces a new quantum number, a different charge vector coupling $Q$ for the conserved current term.
\emph{Although the Type III symmetry $G$ is Abelian, its symmetry transformation operator has a non-commutative non-Abelian feature
thus  \cblue{yielding} degenerate zero energy modes.}
In Sec.\ref{sec: Type II frac} and \ref{sec: Type III zero}, we study the physical observables for bosonic anomalies of these SPT: induced fractional quantum numbers and degenerate zero energy modes. 
In Sec.\ref{sec:flux}, we work on the {\it twisted sector} : the effect of gauge 
flux insertion through a 1D ring effectively captured by using a branch cut or so-called monodromy defect\cite{Wen:2013ue}  modifying the original Hamiltonian.\cite{Santos:2013uda}
The {\it twisted non-onsite symmetry transformation} and  {\it twisted lattice Hamiltonians} are studied,
which spectral shift response under flux insertion provides physical observables to distinguish different SPTs,\cite{{Santos:2013uda},{Zaletel}} 
\cblue{applicable} for all Type I, Type II, Type III SPTs.
Our main results are summarized in Table \ref{table3}, \ref{table1}, \ref{table2}.\\

\begin{widetext} 
\begin{center}
\begin{table} [H]
\begin{tabular}{|c||c|c|c|c|}
\hline
Group Cohomology&  \multicolumn{4}{c}{Bosonic Anomalies and Physical Observables} \vline \\[0mm]\hline
 3-cocycle & $p$ in $\cH^3(G,\tU(1))$  & induced fractional charge  &  degenerate zero energy modes & $\tilde{\Delta}(\tilde{\mathcal{P}})$ under flux/monodromy    \\[0mm]  \hline \hline
Type I $p_1$: Eq.(\ref{type1}) & $\color{blue}{ \Z_{N_1}}$ & No & No & Yes  \\[0mm]  \hline
Type II $p_{12}$: Eq.(\ref{type2}) &   $\color{blue}{\Z_{N_{12}}}$ & Yes. {${\frac{p_{12}}{N_{12}}}$ of $Z_{N_2}$ charge}.  &  No &  Yes \\[1mm] \hline 
Type III $p_{123}$: Eq.(\ref{type3})&  $\color{blue}{\Z_{N_{123}}}$ & No  & Yes. $N_{123}$ degeneracy. &  Yes   \\[1mm] \hline
\end{tabular}
\caption{A summary of bosonic anomalies as 1D edge physical observables to detect the 2+1D SPT of $G=Z_{N_1} \times Z_{N_2} \times Z_{N_3}$ symmetry, here we use $p_i$, $p_{ij}$, $p_{ijk}$ to label the
SPT class index in the third cohomology group $\cH^3(G,\tU(1))$.
For Type II class $p_{12} \in \color{blue}{\Z_{N_{12}}}$, we can use a unit of $Z_{N_1}$-symmetry-breaking domain wall to induce a ${\frac{p_{12}}{N_{12}}}$ fractional $Z_{N_2}$ charge, see Sec.\ref{sec: Type II frac}. 
For Type III class $p_{123} \in \color{blue}{\Z_{N_{123}}}$, we can either use $Z_{N_1}$-symmetry-breaking domain wall
or use  $Z_{N_1}$-symmetry-preserving flux insertion\cite{finite_size} (effectively a monodromy defect) through 1D ring to trap $N_{123}$ multiple degenerate zero energy modes, see Sec.\ref{sec: Type III zero}.
For Type I class $p_{1} \in \color{blue}{ \Z_{N_1}}$, our proposed physical observable is the energy spectrum (or conformal dimension $\tilde{\Delta}(\tilde{\mathcal{P}})$
as a function of momentum ${\mathcal{P}}$, see Ref.\onlinecite{Santos:2013uda}) shift under the flux insertion. This energy spectral shift also works for all other (Type II, Type III) classes, see Sec.\ref{sec:flux}.
This table serves as topological invariants for Type I, II, III bosonic SPT in the context of Ref.\onlinecite{Wen:2013ue}.} 
\label{table3}
\end{table}
\end{center}
\end{widetext}


\noindent(NOTE: Our notation for finite cyclic group is either $Z_N$ or $\mathbb{Z}_N$, though mathematically they are the same.
We denote $Z_N$ for the symmetry group $G$, the discrete gauge $Z_N$ flux, or the $Z_N$ variables.   
We denote $\mathbb{Z}_N$ only for the classes of SPT classification. In addition, we denote $n+1$D as $n$ dimensional space and $1$ dimensional time, and denote $n$D as $n$ dimensional space.
We also denote ${\gcd(N_i,N_j)}\equiv N_{ij}$ and $\gcd(N_i,N_j,N_l) \equiv N_{ijl}$ 
with gcd standing for the greatest common divisor.)

\section{Group Cohomology and Cocycles \label{sec: GC}}

In this section, we will gather the information known and predicted by the group cohomology approach.\cite{Chen:2011pg} 
First, it has been predicted that the $d+1$-D bosonic SPTs can be 
\cblue{constructed} by a mathematical object: the $(d+1)$-th Borel cohomology group $\cH^{d+1}(G,\tU(1))$ of $G$ over $G$-module U(1).\cite{Chen:2011pg,{Dijkgraaf:1989pz}} (It is almost complete classification for bosons, if without considering time-reversal symmetry.)
The SPT classification itself as $\cH^{d+1}(G,\tU(1))$ also forms a group structure.
Throughout the paper, we 
study a generic cyclic group $G=\prod_{i=1}^m Z_{N_i}= Z_{N_1} \times Z_{N_2} \times Z_{N_3} \times \dots$. 
It is generic enough 
in the sense that any finite Abelian group is isomorphic to such a finite cyclic group $G$.
We can thus compute its third cohomology group(see also Ref.\onlinecite{{Chen:2011pg},{deWildPropitius:1996gt}}), 
\be \label{eq:group-cohomology}
\cH^3(G,\tU(1)) = \prod_{1 \leq i < j < l \leq m}  \Z_{N_i} \times \Z_{\gcd(N_i,N_j)} \times  \Z_{\gcd(N_i,N_j,N_l)}. 
\ee
Here $\gcd(N_i,N_j, \dots)$ stands for the greatest common divisor among the numbers $(N_i,N_j, \dots)$.
For simplicity, we denote ${\gcd(N_i,N_j)}\equiv N_{ij}$ and ${\gcd(N_i,N_j,N_l)} \equiv N_{ijl}$.
This cohomology group predicts that there are $\Z_{N_i} \times \Z_{ N_{ij}} \times  \Z_{ N_{ijl}}$ distinct classes for SPTs.
One can find explicit 3-cocycles, such that each distinct 3-cocycles labels the distinct classes in SPTs.
(More generally, $(d+1)$-cocycles for $(d+1)$-th cohomology group $\cH^{d+1}(G,\tU(1))$.)
The $n$-cochain is a mapping $\omega_{}^{}(A_1,A_2,\dots,A_n)$:  $G^n \to \tU(1)$ (which inputs
$A_i \in G$, $i=1,\dots, n$, and outputs a $\tU(1)$ phase).
The $n$-cochains satisfy the group multiplication rules:
\be
(\omega_{1}\cdot\omega_{2})(A_1,\dots,A_n)= \omega_{1}^{}(A_1,\dots,A_n)\cdot \omega_{2}^{}(A_1,\dots,A_n), 
\ee
thus form an Abelian group.
The $n$-cocycles is a $n$-cochain 
additionally
satisfying the $n$-cocycle-conditions $\delta \omega=1$. 
The $3$-cocycle-condition (a pentagon relation) is
\be \label{eq:cocycle-conditions}
\delta \omega(A,B,C,D)=\frac{ \omega(B,C,D) \omega(A,BC,D) \omega(A,B,C) }{\omega(AB,C,D)\omega(A,B,CD) }=1\;\;\;\;\;
\ee
with $A,B,C,D \in G$. One should 
check that the distinct $3$-cocycles are not equivalent by $3$-coboundaries, i.e. 
any $\omega_1(A,B,C)$ is equivalent to $\omega_2(A,B,C)$ if they are identified by a $3$-coboundaries $\delta \Omega(A,B,C)$.
\be \label{eq:coboundary-conditions}
\frac{\omega_1(A,B,C)}{\omega_2(A,B,C)}=\delta \Omega(A,B,C)=\frac{ \Omega(B,C) \Omega(A,BC) }{\Omega(AB,C)\Omega(A,B) }
\ee
with some 2-cochain $\Omega(B,C)$. 
The 3-cochain forms a group $C^3$,
the 3-cochain satisfies the 3-cocycle conditions Eq.(\ref{eq:cocycle-conditions}) further forms a subgroup $Z^3$,
and the 3-coboundaries satisfies Eq.(\ref{eq:coboundary-conditions}) further forms a subgroup $B^3$ (since $ \delta^2 \Omega(A,B,C)=1$). Overall
\be
B^3 \subset Z^3 \subset C^3
\ee
The third cohomology group is exactly a kernel $Z^3$ (the group of 3-cocycles) mod out image $B^3$ (the group of 3-coboundary) relation:
\be
\cH^3(G,\tU(1))= Z^3 /B^3.
\ee
For any 
finite Abelian group $G$, we can derive the distinct 3-cocycles satisfying Eq.(\ref{eq:cocycle-conditions}) (but 
not identified as 3-coboundary by Eq.(\ref{eq:coboundary-conditions})):
\bea
\omega_{\text{I}}^{(i)}(A,B,C)    &=& 
\exp \Big( \frac{2 \pi \ti p_{i}  }{N_{i}^{2}} \; 
a_{i}(b_{i} +c_{i} -[b_{i}+c_{i}]) \Big)  \label{type1} \;\;\;\;\;\\
\omega_{\text{II}}^{(ij)}(A,B,C) &=&             
\exp \Big( 
\frac{2 \pi \ti p_{ij} }{N_{i}N_{j}}  \;
a_{i}(b_{j} +c_{j} - [b_{j}+c_{j}]) \Big)  \label{type2} \;\;\;\;\;\\
\omega_{\text{III}}^{(ijl)} (A,B,C) &=& \exp \Big( \frac{2 \pi \ti
p_{ijl}  }{{\gcd}(N_{i}, N_{j},N_{l})} \;
a_{i}b_{j}c_{l} \Big),            \label{type3}
\eea
so-called Type I, Type II, Type III 3-cocycles\cite{deWildPropitius:1996gt} respectively.
Here $A,B,C \in G$. We denote that $A=(a_1,a_2,a_3,\dots)$, where $a_i \in Z_{N_i}$, and similarly for $B,C$.
And $[b_{i}+c_{i}]$ are defined as the $(b_{i}+c_{i}) \text{mod} N_i$, the module elements in $Z_{N_i}$.
In Table \ref{table1}, we summarize some data of group cohomology and their corresponding realization as SPT by using
(i) quantum lattice model, (ii) matrix product states, and (iii) quantum field theory approach. 
In Sec.\ref{sec:Type I}, 
we will demonstrate their explicit construction 
for Type I, Type II, Type III 3-cocycles and their corresponding Type I, Type II, Type III  SPTs. 
%
%
%
%
%

\begin{widetext}
\begin{center}
\begin{table} [H]
\begin{tabular}{|c||c|c|c|c|c|}
\hline
 3-cocycle & \text{min. symm. group} $G$ &$\cH^3(G,\tU(1))$& lattice model's $S$; $H$ & MPO's $S$ & field theory's $S$  \\[0mm]  \hline \hline
Type I $p_1$: Eq.(\ref{type1}) & $Z_{N}$ & \color{blue}{$\Z_{N}$}&  Eq.(\ref{eq:Type I symmetry explicit}); Eq.(\ref{eq:Type II Hamiltonian lattice})& Eq.(\ref{eq:MPOType II S 12}) & Eq.(\ref{eq:globalS_Type II_1}) \\[0mm]  \hline
Type II $p_{12}$: Eq.(\ref{type2}) & $Z_{N_1} \times Z_{N_2}$ & $\Z_{N_1} \times \Z_{N_2} \times \color{blue}{\Z_{N_{12}}}$  &  Eq.(\ref{S1symp12}); 
Eq.(\ref{eq:Type II Hamiltonian lattice})  &
Eq.(\ref{eq:MPOType II S 12})& Eq.(\ref{eq:globalS_Type II_1}) \\[1mm] \hline 
Type III $p_{123}$: Eq.(\ref{type3})& $Z_{N_1} \times Z_{N_2} \times Z_{N_3}$ &  
 $\prod\limits_{1 \leq i < j \leq 3} \Z_{N_i} \times \Z_{N_{ij}} \times  \color{blue}{\Z_{N_{123}}}$
 &  Eq.(\ref{eq:Type III_S123}); Eq.(\ref{eq:Type II Hamiltonian lattice})& Eq.(\ref{eq:Type III_S})  & Eq.(\ref{eq:globalS_Type III})\\[3.8mm] \hline
\end{tabular}
\caption{  Given a generic finite Abelian global symmetry group (isomorphic to a cyclic group $G=\prod_{i=1}^m Z_{N_i}$ $= Z_{N_1} \times Z_{N_2} \times Z_{N_3} \times \dots$), here we provide the data of group cohomology and their corresponding realization as symmetry protected topological (SPT) states by using
(i) quantum lattice models, (ii) matrix product operators(MPO), and (iii) quantum field theory approach. 
The classification labels $p_1, p_{12}, p_{ijk}$ belong to the Type I {\textcolor{blue}{{$\Z_{N}$}}} class, Type II {\textcolor{blue}{${\Z_{N_{12}}}$}} ($\equiv\Z_{\gcd(N_1,N_2)}$) class, 
Type III {\textcolor{blue} {${\Z_{N_{123}}}$}} ($\equiv\Z_{\gcd(N_1,N_2,N_3)}$) class (all labeled in blue color in the table) respectively. 
}
\label{table1}
\end{table}
\end{center}
\end{widetext}

\section{SPTs with ${Z}_{N_1} \times Z_{N_2} \times Z_{N_3}$ symmetry\label{sec:Type I}} 

\begin{figure}[h!]
\includegraphics[width=0.2\textwidth]{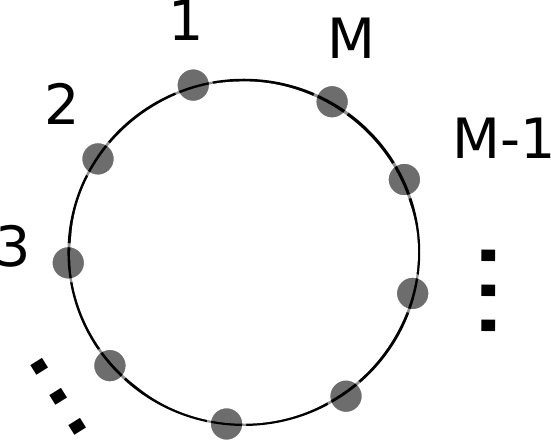}
\caption{ The illustration of 1D lattice model with $M$-sites on a compact ring.
}
\label{fig:2}
\end{figure}

We will now go further to consider the edge modes of lattice Hamiltonian with $G={Z}_{N_1} \times {Z}_{N_2}  \times {Z}_{N_3}$ symmetry
on a compact ring with $M$ sites (Fig.\ref{fig:2}).
Since there are at most three finite Abelian subgroup indices shown in Eq.(\ref{type1}),(\ref{type2}),(\ref{type3}),
such a finite group with three Abelian discrete subgroups is the minimal example containing necessary and sufficient information to explore finite Abelian SPTs.
Such a symmetry-group $G$ may have nontrivial SPT class of Type I, Type II and Type III SPTs.
Apparently 
the Type I SPTs studied in our previous work happen,\cite{Santos:2013uda} which are the
class of $p_u \in \mathbb{Z}_{N_u}$ 
in $\cH^3({Z}_{N_1} \times {Z}_{N_2} \times {Z}_{N_3},\tU(1))$ of Eq.(\ref{eq:group-cohomology}).
Here and below we denote $u, v, w \in \{1,2,3 \}$ and $u,v,w$ are distinct. 
We will also introduce is 
the new class where ${Z}_{N_u}$ and ${Z}_{N_v}$ rotor models ``{\it talk to each other}.''
This will be the mixed Type II class $p_{uv} \in \mathbb{Z}_{N_{uv}}$, 
where symmetry transformation of ${Z}_{N_1}$ global symmetry will affect the ${Z}_{N_2}$ rotor models, while 
similarly ${Z}_{N_2}$ global symmetry will affect the ${Z}_{N_1}$ rotor models. 
There is a new class where three ${Z}_{N_1}$, ${Z}_{N_2}$, ${Z}_{N_3}$ rotor models {\it directly talk to each other}.
This will be the exotic Type III class 
$p_{123} \in \mathbb{Z}_{ N_{123}}$,
where the symmetry transformation of ${Z}_{N_u}$ global symmetry will affect the mixed ${Z}_{N_v},{Z}_{N_w}$ rotor models {\it in 
a mutual way}. 

\cblue{
To verify that our model construction corresponding to the Type I, Type II, Type III 3-cocycle in Eq.(\ref{type1}),(\ref{type2}),(\ref{type3}),
we will implement a technique called ``Matrix Product Operators'' in Sec.\ref{sec:MPO-Type I}. 
We would like to realize a discrete lattice model in Sec.\ref{sec: lattice Type I II}  and a continuum field theory in Sec.\ref{sec:Type II field}, 
to capture the essence of these classes of SPTs.
}



\subsection{Matrix Product Operators and Cocycles \label{sec:MPO-Type I}}

\cblue{
There are various advantages to put a quantum system on a discretize lattice, 
better than viewing it as a continuum field theory. For example, one advantage is that
 the symmetry transformation can be regularized so to understand its property such as onsite or non-onsite.
Another advantage is that we can simulate our model by considering a discretized finite system with a finite dimensional Hilbert space.
For our purpose, to regularize a quantum system on a discrete lattice,
we will firstly use the matrix product operators (MPO) formalism (see Ref.\onlinecite{{MPO},{Chen:2012hc}} and Reference therein) to
formulate our symmetry transformations 
corresponding to non-trivial 3-cocycles in the third cohomology group in $\cH^3({Z}_{N_1} \times {Z}_{N_2},\tU(1))=\mathbb{Z}_{N_1} \times \mathbb{Z}_{N_2} \times \mathbb{Z}_{N_{12}}$.} 

First we formulate the unitary operator $S$ 
as the 
MPO: 
\be \label{eq:MPO_general}
S =\sum_{ \{j,j'\}} \tr[T^{j_1 j_1'}_{\alpha_1 \alpha_2}  T^{j_2 j_2'}_{\alpha_2 \alpha_3} \dots T^{j_M j_M'}_{\alpha_M \alpha_1}] |   j'_1, \dots, j'_M\rangle \langle j_1, \dots, j_M |.
\ee
with the its coefficient taking the trace (tr) of a series of onsite tensor $T(g)$ on a lattice, and input a state
$| j_1, \dots, j_M \rangle$ and output another state $|   j'_1, \dots, j'_M\rangle$.
\cblue{
$T=T(g)$ is a tensor with multi-indices and with dependency on a group element $g \in G$ for a symmetry group.
This is the operator formalism of matrix product states (MPS).
Here {\it physical indices} $j_1,j_2, \dots, j_M$ and $j'_1,j'_2, \dots, j'_M$ are labeled by input/output physical eigenvalues (here ${Z}_N$ rotor angle), the subindices $1,2,\dots, M$ are the physical site indices.
%
There are also {\it virtual indices} $\alpha_1, \alpha_2, \dots, \alpha_M$ which are traced in the end.
Summing over all the operation from $\{j,j'\}$ indices, we shall reproduce the symmetry transformation operator $S$. 
}%

\cblue{
What MPO really helps us 
is that}
{\emph{by contracting MPO tensors $T(g)$ of $G$-symmetry transformation ${S}$ (here $g\in G$) of three neighbored-sites in different sequence on the 
effective 1D lattice of SPT edge modes, 
it can reveal the {\bf nontrivial projective phase} corresponds to {\bf the nontrivial 3-cocycles
of the cohomology group}}.}\\

To find out the projective representation $e^{i \theta(g_a,g_b,g_c)}$. 
Below we use the facts of tensors $T(g_a), T(g_b), T(g_c)$ acting on three neighbored sites $a$, $b$, $c$ with group elements ${g_a,g_b,g_c}$. 
We know a generic projective relation:
\be \label{eq:projective}
T(g_a\cdot g_b)=P_{g_a,g_b}^\dagger T(g_a) T(g_b)P_{g_a,g_b}.
\ee
Here $P_{g_a,g_b}$ is the projection operator.
We contract three neighbored-site tensors in two different orders,
\be \label{eq:projective-3cocycle}
{(P_{g_a,g_b} }{ \otimes I_3) P_{g_a g_b, g_c}} \simeq e^{\ti \theta(g_a,g_b,g_c)} {( I_1 \otimes P_{g_b,g_c} ) P_{g_a,g_bg_c}}.
\ee
The left-hand-side contracts the sites $a,b$ first then with the site $c$, while the right-hand-side contracts the sites $b,c$ first then with the site $a$.
Here $\simeq$ means the equivalence is up to a projection out of un-parallel states. 
We can derive $P_{g_a,g_b}$ by observing 
that $P_{g_a,g_b}$ inputs one state and outputs two states.\cite{Supplementary Material}

For Type I SPT class, this MPO formalism has been done quite carefully in Ref.\onlinecite{Chen:2012hc},\onlinecite{Santos:2013uda}. 
Here we generalize it to other SPTs, below we input a group element with $g=(k_1,k_2,k_3)$ and $k_1 \in Z_{N_1}, k_2 \in Z_{N_2}, k_3 \in Z_{N_3}$. 
Without losing generality, we focus on the symmetry
Type I index $p_1 \in \Z_{N_1}$, Type II index $p_{12} \in \Z_{N_{12}}$, Type III index $p_{123} \in  \Z_{N_{123}}$.
By index relabeling, we can fulfill all SPT symmetries 
within the classification in Eq.(\ref{eq:group-cohomology}).


\begin{widetext}
We propose our $T(g)$ tensor for Type I,\cite{Chen:2012hc,{Santos:2013uda}} II symmetry with $p_{12} \in \Z_{N_{12}}$, $p_{12} \in \Z_{N_{12}}$ as
\bea \label{eq:Type II T 12}
&&(T^{\phi^{(1)}_{in}, \phi^{(1)}_{out},\phi^{(2)}_{in}, \phi^{(2)}_{out}})^{(p_1,p_{12})}_{\varphi_\alpha^{(1)},\varphi_\beta^{(1)},\varphi_\alpha^{(2)},\varphi_\beta^{(2)},N_1}(\frac{2\pi k_1}{N_1} )=\delta(\phi^{(1)}_{out}-\phi^{(1)}_{in}-\frac{2\pi k_1 }{N_1})\delta(\phi^{(2)}_{out}-\phi^{(2)}_{in}) \\
&&
\cdot \int d\varphi_\alpha^{(1)} d\varphi_\beta^{(1)} | \varphi_\beta^{(1)} \rangle \langle \varphi_\alpha^{(1)} | \delta(\varphi_\beta^{(1)}-\phi_{in}^{(1)}) e^{\ti p_1 k_1(\varphi^{(1)}_\alpha-\phi^{(1)}_{in})_r
/N_1} 
\cdot \int d\tilde{\varphi}_\alpha^{(2)} d\tilde{\varphi}_\beta^{(2)} | \tilde{\varphi}_\beta^{(2)} \rangle \langle \tilde{\varphi}_\alpha^{(2)} | \delta(\tilde{\varphi}_\beta^{(2)}-\tilde{\phi}_{in}^{(2)}) e^{\ti p_{12} k_{1}(\tilde{\varphi}^{(2)}_\alpha-\tilde{\phi}^{(2)}_{in})_r/N_1}.  \nonumber
\eea
We propose the Type III $T(g)$ tensor with $p_{123} \in  \Z_{N_{123}}$ as
\bea  \label{eq:Type III T}
&&(T^{\phi^{(1)}_{in}, \phi^{(1)}_{out},\phi^{(2)}_{in}, \phi^{(2)}_{out},\phi^{(3)}_{in}, \phi^{(3)}_{out}})^{(p_{123})}_{\varphi_\alpha^{(1)},\varphi_\beta^{(1)},\varphi_\alpha^{(2)},\varphi_\beta^{(2)},
\varphi_\alpha^{(3)},\varphi_\beta^{(3)},N_1,N_2,N_3}(\frac{2\pi k_1}{N_1},\frac{2\pi k_2}{N_2},\frac{2\pi k_3}{N_3}) 
\nonumber\\
&&=\prod_{u,v,w \in \{1,2,3\}} \int d\varphi_\alpha^{(u)}  | \phi_{in}^{(u)} \rangle \langle \varphi_\alpha^{(u)} |  
\exp[{\ti \,p_{123}  \epsilon^{uvw} k_u\frac{(\varphi^{(v)}_\alpha \phi^{(w)}_{in} )_r}{N_u}  {\frac{N_1N_2N_3}{2\pi\gcd({N_1,N_2,N_3})}}}] 
 \cdot |   \phi^{(u)}_{out} \rangle \langle \phi^{(u)}_{in}|.  \label{eq:Type III T_2}
\eea
\end{widetext}

Here we consider a lattice with both $\phi^{(u)}$, $\varphi^{(u)}$ as  $Z_{N_u}$ rotor angles.
The tilde notation $\tilde{\phi}^{(u)}$, $\tilde{\varphi}^{(u)}$, for example on $\tilde{\phi}^{(2)}$, means that the variables are in units of $\frac{2\pi}{N_{12}}$, but not in $\frac{2\pi}{N_2}$ unit
(The reason will become explicit later when we regularize the Hamiltonian on a lattice in Sec.\ref{sec: lattice Type I II}).

Take Eq.(\ref{eq:Type II T 12}), by computing the projection operator $P_{g_a,g_b}$ via Eq.(\ref{eq:projective}), we derive the projective phase:
\be \label{eq:proj Type I}
e^{\ti \theta(g_a,g_b,g_c)} =e^{\ti p_1\frac{2\pi}{N} m_c\frac{m_a+m_b-[m_a+m_b]_N}{N}} =
\omega_{\text{I}}^{(i)}(m_c,m_a,m_b) 
\ee
which the complex phase in the projective representation 
indeed induces the Type I 3-cocycle $\omega_{\text{I}}^{(i)}(m_c,m_a,m_b)$ of Eq.(\ref{type1}) 
in the third cohomology group $\cH^3({Z}_N,\tU(1))=\mathbb{Z}_N$. 
(Up to the index redefinition $p_{1} \to -p_{1}$.)
We further derive the projective phase as Type II 3-cocycle of 
Eq.(\ref{type2}),
\bea 
&&e^{\ti \theta(g_a,g_b, g_c)}=e^{\ti p_{12} ({ \frac{2\pi {m^{(1)}_c} }{N_1}  })\big( (m^{(2)}_a+{m^{(2)}_b}) - [m^{(2)}_a+{m^{(2)}_b}]_{N_2} \big) /N_2} \nonumber \\
&&=\omega_{\text{II}}^{(ij)}(m_3,m_1,m_2) \label{eq:proj Type II}
\eea
up to the index redefinition $p_{12} \to -p_{12}$. 
Here $[m_a+m_b]_N$ with subindex $N$ means taking the value module $N$.

Take Eq.(\ref{eq:Type III T}), we can also derive the projective phase
$e^{\ti \theta(g_a, g_b, g_c)}$ 
of Type III $T(g)$ tensor as 
\bea \label{}
&&e^{\ti \theta(g_a,g_b, g_c)}=e^{ \ti 2\pi p_{123} \epsilon^{u  v w}\big( \frac{{m^{(u)}_c} }{N_{u}} \frac{m^{(v)}_a}{N_v} \frac{m^{(w)}_b}{N_w} \big)\frac{N_1N_2N_3}{N_{123} }} \nonumber \\
&&\simeq \omega_{\text{III}}^{(uvw)}(m_c,m_a,m_b).
\eea
Adjust $p_{123}$ index (i.e. setting only the $p_{123}$ index in ${m^{(1)}_c}{m^{(2)}_a}{m^{(3)}_b}$ to be nonzero, while others $p_{213}=p_{312}=0$), 
and compute Eq.(\ref{eq:projective-3cocycle}) with only $p_{123}$ index,
we can recover the projective phase reveals Type III 3-cocycle in Eq.(\ref{type3}).

By Eq.(\ref{eq:MPO_general}), we verify that $T(g)$ of Type I, II in Eq.(\ref{eq:Type II T 12}) renders the symmetry transformation operator 
${S}^{(p_1,p_{12})}_{N_1}$: 
\bea  \label{eq:MPOType II S 12} 
{S}^{(p_1,p_{12})}_{N_1}=&&\prod_{j=1}^M  e^{\ti 2\pi L^{(1)}_j/N_1} \cdot \exp[ \ti \frac{p_1}{N_1}  ( \phi^{(1)}_{j+1}-\phi^{(1)}_{j} )_r]  \nonumber\\
&&\cdot \exp[ \ti \frac{p_{12}}{N_1}  ( \tilde{\phi}^{(2)}_{j+1}-\tilde{\phi}^{(2)}_{j} )_r].  
\eea
here $j$ are the site indices, from $1$ to $M$ shown in Fig.\ref{fig:2}.


By Eq.(\ref{eq:MPO_general}), we verify that $T(g)$ of Type III in Eq.(\ref{eq:Type III T}) renders the symmetry transformation operator 
${S}^{(p_{123})}_{N_1,N_2,N_3}$:
\bea  \label{eq:Type III_S}
S^{(p_{123})}_{N_1,N_2,N_3}
={\prod_{j=1}^M} ( \prod^M_{\substack{  {u,v,w \in \{1,2,3\}}   } }  e^{\ti 2\pi L^{(u)}_j/N_u} \cdot W^{\text{III}}_{j,j+1} ). 
\eea
with
\be
W^{\text{III}}_{j,j+1} \equiv \prod_{u,v,w \in \{1,2,3\}} e^{\Big( { \ti {\frac{N_1N_2N_3}{2\pi N_{123}}} \epsilon^{u v w} \frac{p_{123}}{N_{u}}  \big(   \phi^{(v)}_{j+1} \phi^{(w)}_{j} \big) } \Big)}. 
\ee
For both Eq.(\ref{eq:MPOType II S 12}) and Eq.(\ref{eq:Type III_S}),
there is an onsite piece $\langle \phi_j^{(u)}  | e^{i 2\pi L^{(u)}_j/N_u} | \phi_j^{(u)} \rangle$ 
and 
also extra non-onsite symmetry transformation parts: namely, 
$\exp[ \ti \frac{p_1}{N_1}  ( \phi^{(1)}_{j+1}-\phi^{(1)}_{j} )_r]$,
$\exp[ \ti \frac{p_{12}}{N_1}  ( \tilde{\phi}^{(2)}_{j+1}-\tilde{\phi}^{(2)}_{j} )_r]$, and $W^{\text{III}}_{j,j+1}$.
We introduce an angular momentum operator $L^{(u)}_j$ conjugate to $\phi^{(u)}_j$,
such that the $e^{i 2\pi L^{(u)}_j/N_u}$ shifts the rotor angle by $\frac{2\pi}{N_{u}}$ unit, from $|\phi^{(u)}_{j} \rangle$ to $|  \phi^{(u)}_{j} +\frac{2\pi}{N_{u}} \rangle$. 
The subindex $r$ means that we 
further {\it regularize the variable to a discrete compact rotor angle}. 

Meanwhile $p_{1}=p_{1} \text{ mod } N_{1}$, $p_{12}=p_{12} \text{ mod } N_{12}$ and $p_{123}=p_{123} \text{ mod } N_{123}$, 
these demonstrate that our MPO construction fulfills all classes in Eq.(\ref{eq:group-cohomology}) as we desire.
So far we have achieved the SPT symmetry transformation operators Eq.(\ref{eq:MPOType II S 12}),(\ref{eq:Type III_S}) via MPO.
Other technical 
derivations on MPO formalism are 
preserved in Supplemental Materials. 

\subsection{Lattice model \label{sec: lattice Type I II}}

To construct a lattice model, we require the minimal ingredients: (i) $Z_{N_u}$ operators (with $Z_{N_u}$ variables).
(ii) Hilbert space (the state-space where $Z_{N_u}$ operators act on) consists with $Z_{N_u}$ variables-state.
Again we denote $u=1,2,3$ for $Z_{N_1}$,$Z_{N_2}$,$Z_{N_3}$ symmetry.
We can naturally choose the $Z_{N_u}$ variable $\omega_u \equiv e^{\ti\,2\pi/N_u}$, such that 
$\omega_u^{N_u}=1$. 
Here and below we will redefine the quantum state and operators from the MPO basis in Sec.\ref{sec:MPO-Type I} to a lattice basis via:
\be
{\phi}^{(u)}_j \to \phi_{u,j},\;\;\; 
{L}^{(u)}_l \to {L}^{}_{u,l}. \label{eq:MPO-to-lattice}
\ee
The natural physical states on a single site are 
the $Z_{N_u}$ rotor angle state
$| \phi_u=0 \rangle, | \phi_u=2\pi/N_u \rangle, \dots, | \phi_u=2\pi(N_u-1)/N_u \rangle$.

One can find a dual state of rotor angle state $|\phi_u \rangle$, the angular momentum $|L_u \rangle$, such that the basis from $|\phi_u \rangle$ can transform to 
$|L_u \rangle$
via the Fourier transformation, $| \phi_u \rangle=\sum^{N_u-1}_{L_u=0} \frac{1}{\sqrt{N_u}} e^{\ti L_u \phi_u} | L_u \rangle$. 
One can find two proper operators $\sigma^{(u)}_{},\tau^{(u)}_{}$ which make $| \phi_u \rangle$ and $| L_u \rangle$ their own eigenstates respectively. 
With a site index $j$ ($j = 1,...,M$),
we can project $\sigma^{(u)}_{j},\tau^{(u)}_{j}$ operators into the rotor angle 
$| \phi_{u,j} \rangle$ basis, so we can derive $\sigma^{(u)}_{j},\tau^{(u)}_{j}$ operators as $N_u \times N_u$ matrices.
Their forms are :
\be \label{eq: sigma operator}
\sigma^{(u)}_{j}   =  {\begin{pmatrix} 
1 & 0 & 0 & 0 \\
0 & \omega_u & 0 & 0\\  
0 & 0 & \ddots  & 0\\  
0 & 0 & 0 & \omega_u^{N_u-1}
\end{pmatrix}}_j = \langle \phi_{u,j} | e^{\ti \hat{\phi}^{(u)}_{j}} | \phi_{u,j} \rangle
\ee
\be
\tau^{(u)}_{j}   =  {\begin{pmatrix} 
0 & 0 & 0 & \dots &0& 1  \\
1 & 0 & 0 & \dots &0& 0 \\
0 & 1 & 0 & \dots &0& 0 \\
0 & 0 & 1 & \dots &0& 0 \\
\vdots &0 & 0 & \dots &1 & 0 
\end{pmatrix}}_j =\langle \phi_{u,j} | e^{\ti2 \pi \hat{L}^{(u)}_j/N} | \phi_{u,j} \rangle ,  
\ee
Operators and variables satisfy the analogue property mentioned in Ref.\onlinecite{Santos:2013uda}, 
such as $(\tau^{(u)})^{N_u}_{j} = (\sigma^{(u)})^{N_u}_{j} =\openone$, 
$\tau^{(u)\dagger}_{j}\,\sigma^{(u)}_{j}\,\tau^{(u)}_{j} = \omega_u\,\sigma^{(u)}_{j}$. 
It also enforces the canonical conjugation relation on $\hat{\phi}^{(u)}$ and $\hat{L}^{(u)}$ operators, i.e.
$[\hat{\phi}_j^{(u)},\hat{L}_l^{(v)} ]= \ti \,\delta_{(j,l)}\delta_{(u,v)}$ with the symmetry group index $u,v$
and the site indices $j,l$.
Here $| \phi \rangle$ and $| L \rangle$ are eigenstates of $\hat{\phi}$ and $\hat{L}$ operators respectively.

The linear combination of all $| \phi_1 \rangle$ $| \phi_2 \rangle$ $| \phi_3 \rangle$ states form a complete  $N_1 \times N_2 \times N_3$-dimensional Hilbert space on a single site.

\subsubsection{symmetry transformations \label{Sec: type I II sym transf}}

\noindent
{\bf Type I, II $Z_{N_1} \times Z_{N_2}$ symmetry transformations}\\

Firstly we 
warm up with a generic ${Z}_N$ lattice model realizing the SPT edge modes on a 1D ring with $M$ sites (Fig.\ref{fig:2}).
It has been emphasized in Ref.\onlinecite{Chen:2011pg,Chen:2012hc} that the SPT edge modes have a special non-onsite symmetry transformation, which means that
its symmetry transformation cannot be written as a tensor product form on each site, thus
$
U(g)_{\text{non-onsite}} \neq \otimes _i U_i(g).
$ 
In general, the symmetry transformation contain a onsite part and another non-onsite part.
The trivial class of SPT (trivial bulk insulator) with unprotected gapped edge modes can be achieved by a simple 
Hamiltonian
as
$-\lambda \, \sum^{M}_{j=1}  ( \tau_{j} + \tau^{\dagger}_{j} )$. (Notice that for the simplest $Z_2$ symmetry, the $\tau_j$ operator  reduces to a spin operator $(\sigma_z)_j$.)
The simple way to find an onsite operator which this Hamiltonian respects and which acts at each site is the $\prod^{M}_{j=1} \tau_j$, a series of $\tau_j$.
On the other hand, to capture the \emph{non-onsite} symmetry transformation, we can use a \emph{domain wall} variable 
in Ref.\onlinecite{Santos:2013uda}, where the symmetry transformation
contains information stored non-locally between different sites (here we will use the minimum construction: 
symmetry stored non-locally between {\it two nearest neighbored sites}). 
\cblue{Based on the understanding of previous work,\cite{LevinGu,Chen:2012hc,{Santos:2013uda}}
we propose this {non-onsite} symmetry transformation $U_{j,j+1}$ with a domain wall $(N_{\text{dw}})_{j,j+1}$ operator acting non-locally on site $j$ and $j+1$ as},
\be \label{eq:domain-wall}
U_{j,j+1} 
\equiv \exp \big( \ti \frac{p}{N}  \frac{2\pi}{N}(\delta N_{\text{dw}})_{j,j+1} \big) \equiv\exp[ \ti \frac{p}{N}  (\phi_{1,j+1} - \phi_{1,j})_r],   
\ee
\cblue{The justification of non-onsite symmetry operator Eq.(\ref{eq:domain-wall}) realizing SPT edge symmetry is based on
MPO formalism already done in Sec.\ref{sec:MPO-Type I}.  
}
The domain wall operator $(\delta N_{\text{dw}})_{j,j+1}$ counts the number of units of ${Z}_N$ angle between sites $j$ and $j+1$, so
indeed $(2\pi/N)(\delta N_{\text{dw}})_{j,j+1}$$=(\phi_{1,j+1} - \phi_{1,j})_r$. The subindex $r$ means that we need to further {\it regularize the variable to a discrete ${Z}_N$ angle}.
Here we insert a $p$ index, which is just an available free index with $p=p \text{ mod } N$. 
From Sec.\ref{sec:MPO-Type I}, $p$ is indeed the classification index for the $p$-th of $\mathbb{Z}_N$ class in the third cohomology group $\cH^3(Z_N,\tU(1))=\mathbb{Z}_N$.

Now the question is 
how should we fully regularize this $U_{j,j+1}$ operator into terms of $Z_N$ operators $\sigma^{\dagger}_{j}$ and $\sigma_{j+1}$.
We see the fact that the $N$-th power of $U_{j,j+1}$ renders a constraint
\bea  \label{eq:constraint Type I}
U_{j,j+1}^N 
&=&(\exp[ \ti  \phi_{1,j}]^\dagger \exp[ \ti  \phi_{1,j+1}] )^p=(\sigma^{\dagger}_{j}\sigma_{j+1})^p. \;\;\;
\eea
(Since $\exp[\, \ti\,  \phi_{1,j}]_{ab}= \langle \phi_a | e^{\, \ti \, \phi_j} | \phi_b \rangle = \sigma_{ab,j}$.)
More explicitly, we can write it as a polynomial ansatz 
$ 
U_{j,j+1}=
\exp[\frac{\ti}{N} \sum^{N-1}_{a=0}\, q_{a}\,(\sigma^{\dagger}_{j}\sigma_{j+1})^{a}]
$.
The non-onsite symmetry operator $U_{j,j+1}$ reduces to a problem of solving polynomial coefficients $q_a$ by the constraint Eq.(\ref{eq:constraint Type I}).
Indeed we can solve the constraint explicitly, thus the non-onsite symmetry transformation operator 
acting on a $M$-site ring from $j=1, \dots, M$ is derived: 
\bea \label{eq:Type I symmetry explicit}
U_{j,j+1}=e^
{
-\ti \frac{2\pi}{N^2}p\, 
\Big\{
\left(
\frac{N-1}{2}\,
\right)
\openone
+
\sum^{N-1}_{a=1}\,
\frac
{
(\sigma^{\dagger}_{j}\sigma_{j+1})^a
}
{
(\omega^a - 1)
}
\Big\}}.
\eea
%


For a lattice SPTs model with $G=Z_{N_1} \times Z_{N_2}$, 
we can convert MPO's symmetry transformation Eq.(\ref{eq:MPOType II S 12})
to a lattice variable via Eq.(\ref{eq:Type I symmetry explicit}).
We obtain the $Z_{N_u}$ symmetry transformation (here and below $u,v \in \{ 1,2\}, u\neq v$):
\bea \label{S1symp12}
&&\bullet \;\; S^{(p_u,p_{uv})}_{N_u}
\equiv \prod^{M}_{j=1}  e^{\ti 2\pi L^{}_{u,j}/N_u} \cdot \exp[ \ti \frac{p_u}{N_u}  ( \phi_{u,j+1}-\phi_{u,j} )_r]  \nonumber \\
&& \cdot \exp[ \ti \frac{p_{uv}}{N_u}  ( \tilde{\phi}_{v,j+2}-\tilde{\phi}_{v,j} )_r]   \label{eq:symmetry form lattice Type II S1} \nonumber \\
&&= \prod^{M}_{j=1}\tau^{(u)}_{j}\,
\cdot U^{(N_u,p_u)}_{j,j+1}
\cdot U^{(N_u,p_{uv})}_{j,j+2} \nonumber \\
&&=\prod^{M}_{j=1}\tau^{(u)}_{j}\, \cdot e^{(-\ti \frac{2\pi}{N_u^2}p_u\, 
\Big\{ \left( \frac{N_u-1}{2}\,\right) \openone + \sum^{N_u-1}_{a=1}\,\frac {(\sigma^{(u)\dagger}_{j}\sigma^{(u)}_{j+1})^a} {((\omega_u)^a - 1)}
\Big\})} \;\;\;\; \;\;\; \nonumber \\
&& \cdot e^{ ( -\ti\frac{2\pi }{N_{uv} N_{u}} p_{uv} \, \Big\{ (\frac{N_{uv}-1}{2}) \openone+ \sum^{N_{uv}-1}_{a=1}\, \frac { \left( \tilde{\sigma}^{(v) \dagger}_{j}\tilde{\sigma}^{(v)}_{j+2} \right)^a}
{\omega_{uv}^a -1}\Big\} ) }.
\eea
The operator is unitary, i.e. $S^{(p_u,p_{uv})}_{N_u} S^{(p_u,p_{uv})\dagger}_{N_u} =1$. Here $\sigma_{M+j}\equiv\sigma_{j}$.
The intervals of rotor angles are
\bea 
&&\phi_{1,j} \in \{ n \frac{2\pi}{N_1} | n \in \mathbb{Z}\},\;\;\; \phi_{2,j} \in \{ n \frac{2\pi}{N_2} | n \in \mathbb{Z}\}, \nonumber \\
&&\tilde{\phi}_{1,j}, \tilde{\phi}_{2,j} \in \{ n \frac{2\pi}{ N_{12} } | n \in \mathbb{Z}\}. \label{eq:field regularize}
\eea
where $\phi_{1,j}$ is $Z_{N_1}$ angle, $\phi_{2,j}$ is $Z_{N_2}$ angle, $\tilde{\phi}_{1,j}$ and $\tilde{\phi}_{2,j}$ are $Z_{N_{12}}$ angles (recall ${\gcd{(N_1,N_2)}} \equiv N_{12}$).
There are 
some  remarks on our above formalism:\\
(i) First,  the $Z_{N_1},Z_{N_2}$ symmetry transformation Eq.(\ref{S1symp12}) 
including both the Type I indices $p_1$, $p_2$ and also Type II indices $p_{12}$ and $p_{21}$. Though $p_1$, $p_2$ are distinct indices, but $p_{12}$ and $p_{21}$ indices are the same index, $p_{12}+p_{21} \to p_{12}$. The invariance $p_{12}+p_{21}$ describes the same SPT symmetry class.\\
(ii) The second remark, 
for Type I non-onsite symmetry transformation (with $p_1$ and $p_2$) are chosen to act on the nearest-neighbor sites (NN: site-$j$ and site-$j+1$); but 
the Type II non-onsite symmetry transformation (with $p_{12}$ and $p_{21}$) are  chosen to be the next nearest-neighbor sites (NNN: site-$j$ and site-$j+2$). The reason is that
we have to avoid the nontrivial Type I and Type II symmetry transformations cancel or interfere with each other. Though in the Sec.\ref{sec:Type II field}, 
we will reveal that the low energy field theory description of non-onsite symmetry transformations for both NN and NNN having the same form in the continuum limit.
In the absence of Type I index, we can have Type II non-onsite symmetry transformation act on nearest-neighbor sites.\\ 
(iii) The third remark, the domain wall picture mentioned in Eq.(\ref{eq:domain-wall}) of Sec.\ref{sec:Type I} for Type II $p_{12}$ class still hold. But here the lattice regularization is different for terms with $p_{12},p_{21}$ indices.
In order to have distinct $Z_{\gcd{({N}_1,{N}_2)}}$ class with the identification $p_{12} =p_{12}$ mod $N_{12}$.
We will expect that, performing the ${N_u}$ times $Z_{N_u}$ symmetry transformation on the Type II $p_{uv}$ 
non-onsite piece, 
renders a constraint 
\be \label{eq: Type II constraint}
(U^{(N_u,p_{uv})}_{j,j+2})^{N_u}=(\tilde{\sigma}^{(v)\dagger}_{j}\tilde{\sigma}^{(v)}_{j+2})^{p_{uv}},  
\ee

To impose the identification $p_{12} =p_{12}$ mod ${N_{12}}$ and $p_{21} =p_{21}$ mod ${N_{12}}$
so that we have distinct $Z_{\gcd{({N}_1,{N}_2)}}$ classes for the Type II symmetry class
(which leads to impose the constraint $(\tilde{\sigma}^{(1)}_{j})^{N_{12}}=(\tilde{\sigma}^{(2)}_{j})^{N_{12}} = \openone$),
we can regularize the $\tilde{\sigma}^{(1)}_{j}$, $\tilde{\sigma}^{(2)}_{j}$  operators in terms of $Z_{\gcd{(N_1,N_2)}}$ variables. 
With ${\omega}_{12} \equiv {\omega}_{21}  \equiv e^{i \frac{2 \pi}{N_{12}} }$, we have  
$\omega_{12}^{N_{12}} =1$.
The  $\tilde{\sigma}^{(u)}_{j}$ matrix has $N_u \times N_u$ components, for $u=1,2$. It is block diagonalizable
with $\frac{N_u}{N_{12}} $ subblocks, and each subblock with ${N_{12}} \times {N_{12}}$ components.
Our regularization provides the nice property:
$\tau^{(1)\dagger}_{j}\,\tilde{\sigma}^{(1)}_{j}\,\tau^{(1)}_{j} = \omega_{12}\,\sigma^{(1)}_{j}$
and $\tau^{(2)\dagger}_{j}\,\tilde{\sigma}^{(2)}_{j}\,\tau^{(2)}_{j} = \omega_{12}\,\sigma^{(2)}_{j}$.
Use the above procedure to regularize Eq.(\ref{eq:MPOType II S 12}) 
on a discretized lattice and solve the constraint Eq.(\ref{eq: Type II constraint}), we obtain
an explicit form of lattice-regularized symmetry transformations Eq.(\ref{S1symp12}). 
For more details on our lattice regularization, see Supplemental Materials. 

\noindent
{\bf Type III symmetry transformations}\\

To construct a Type III SPT with a Type III 3-cocycle Eq.(\ref{type3}),
the key observation is that the 3-cocycle inputs, for example, $a_1 \in Z_{N_1}$, $b_2 \in Z_{N_2}$, $c_3 \in Z_{N_3}$ and outputs a U(1) phase.
This implies that the ${Z}_{N_1}$ symmetry transformation will affect the mixed ${Z}_{N_2},{Z}_{N_3}$ rotor models, etc.
This observation guides us to write down the tensor $T(g)$ in Eq.(\ref{eq:Type III T}) 
and we obtain the symmetry transformation $S^{(p)}_N=S^{(p_{123})}_{N_1,N_2,N_3}$ as Eq.(\ref{eq:Type III_S}):
\bea  \label{eq:Type III_S123}
\bullet &&S^{(p_{123})}_{N_1,N_2,N_3}
={\prod_{j=1}^M} ( \prod^M_{\substack{  {u,v,w \in \{1,2,3\}}   } }  \tau_j^{(u)} \cdot W^{\text{III}}_{j,j+1} ). 
\eea
There is an onsite piece $\tau_j \equiv \langle \phi_j | e^{i 2\pi L^{(u)}_j/N} | \phi_j \rangle$ and also an extra non-onsite symmetry transformation part $W^{\text{III}}_{j,j+1}$.
This non-onsite symmetry transformation $W^{\text{III}}_{j,j+1}$, acting on the site $j$ and $j+1$, is defined by the following, and can be further regularized on the lattice:
\be \label{eq:Type III_W}
\bullet \;\;W^{\text{III}}_{j,j+1}
={ \prod_{u,v,w \in \{1,2,3\}}  \Big( \sigma_{j}^{(v)\dagger} \sigma_{j+1}^{(v)}  \Big)^{ \epsilon^{u v w}   p_{123} {  \frac{ {\log(\sigma_{j}^{(w)})} N_v N_w}{2\pi N_{123}}} }}.
\;\;\;\;\; 
\ee
here we separate $Z_{N_1}$,$Z_{N_2}$,$Z_{N_3}$ non-onsite symmetry transformation to $W^{\text{III}}_{j,j+1; N_1}$,$W^{\text{III}}_{j,j+1; N_2}$,$W^{\text{III}}_{j,j+1; N_3}$ respectively.
Eq.(\ref{eq:Type III_S123}),(\ref{eq:Type III_W}) are fully regularized in terms of $Z_N$ variables on a lattice, 
although they contain \emph{anomalous non-onsite symmetry} operators.\cite{Supplementary Material}

\subsubsection{lattice Hamiltonians \label{sec:lattice Hamiltonians: type I II}}

We had mentioned the trivial class of SPT Hamiltonian (the class of $p=0$) 
for 1D gapped edge:
\be
H^{(0)}_{N}=-\lambda \, \sum^{M}_{j=1}  ( \tau_{j} + \tau^{\dagger}_{j} )
\ee
Apparently, the 
Hamiltonian is symmetry preserving respect to $S^{(0)}_{N} \equiv\prod^{M}_{j=1} \tau_j$, 
i.e. $S^{(0)}_{N} H^{(0)}_{N} (S^{(0)}_{N})^{-1} =H^{(0)}_{N}$. 
In addition, this Hamiltonian has a symmetry-preserving gapped ground state.

To extend our lattice Hamiltonian construction to $p \neq 0$ class, intuitively we can view the nontrivial SPT Hamiltonians as close relatives of the trivial 
Hamiltonian (which preserves the onsite part of the symmetry transformation with $p=0$), which 
satisfies the symmetry-preserving constraint, i.e.
\be
S^{(p)}_{N} H^{(p)}_{N} (S^{(p)}_{N})^{-1} =H^{(p)}_{N},
\ee
More explicitly, to construct a SPT Hamiltonian of $Z_{N_1} \times Z_{N_2}\times Z_{N_3}$ symmetry
obeying translation 
and symmetry transformation 
invariant (here and below $u, v, w \in \{1,2,3 \}$ and $u,v,w$ are distinct): 
\bea \label{principle type II}
&&\bullet\;\; [H^{(p_u,p_{uv},p_{uvw})}_{N_1,N_2,N_3}  ,T]=0, \;\; \\ 
&&\bullet\;\;  [H^{(p_u,p_{uv},p_{uvw})}_{N_1,N_2,N_3}  ,
S^{(p)}_{N} ]
=0
\eea
Here $T$ is a translation operator by one lattice site,
satisfying
$
T^{\dagger}\,X_{j}\,T = X_{j+1},
\quad
j = 1,...,M,
$
for any operator $X_{j}$ on the ring such that
$
X_{M+1}
\equiv
X_{1}
$.
Also $T$ satisfies
$
T^{M}
=
\openone
$.
We can immediately derive the following SPT Hamiltonian satisfying the rules, 
\be \label{eq:Type II Hamiltonian lattice}
\bullet \;\;
H^{(p_u,p_{uv},p_{uvw})}_{N_1,N_2,N_3} 
\equiv
- \lambda
\!
\sum^{M}_{j=1} 
\!\!
\sum^{N-1}_{\ell = 0}
\!\!
\left(
S^{(p)}_{N} 
\right)^{\!\!-\ell}
\!\!\!
( \tau_{j} + \tau^{\dagger}_{j} )  
\!\!
\left(
S^{(p)}_{N} 
\right)^{\ell} +\dots,
\ee
where we define our notations: $S^{(p)}_{N} \equiv \prod_{ u, v, w \in \{1,2,3 \} }S^{(p_u,p_{uv},p_{uvw})}_{N_u} $  
and $\tau_{j} \equiv \tau_{j}^{(1)} \otimes \openone_{N_2 \times N_2}  \otimes \openone_{N_3 \times N_3} + \openone_{N_1 \times N_1} \otimes \tau_{j}^{(2)}  \otimes \openone_{N_3 \times N_3} 
+\openone_{N_1 \times N_1}   \otimes \openone_{N_2 \times N_2}  \otimes \tau_{j}^{(3)} $.
Here $\tau_{j}$ is a matrix of ${(N_1 \times N_2 \times N_3 )\times (N_1 \times N_2 \times N_3)}$-components.
The tower series of sum over power of $(S^{(p)}_{N})$ over $(\tau_{j} + \tau^{\dagger}_{j} )$ will be shifted upon $S^{(p)}_{N} H^{(p)}_{N} (S^{(p)}_{N})^{-1}$, but the overall sum of this Hamiltonian 
is a symmetry-preserving invariant.

\subsection{Field Theory \label{sec:Type II field}}

From a full-refualrized lattice model 
in the previous section, 
we attempt to take the low energy 
limit to realize its corresponding field theory, 
by identifying the commutation relation
$[\hat{\phi}^{(u)}_j,\hat{L}^{(v)}_l ]= \ti \,\delta_{(j,l)} \delta_{(u,v)}$ (here $j, l$ are the site indices, $u, v \in \{ 1,2, 3\}$ 
are the $Z_{N_1},Z_{N_2},Z_{N_3}$ rotor model indices) in the continuum as 
\bea 
&&[\phi_u(x_1), \frac{1}{2\pi}\partial_x \phi_{v}'(x_2)]=  \, \ti \,  \delta(x_1-x_2) \delta_{(u,v)} \label{eq:commutation} 
\eea
which means the $Z_{N_1},Z_{N_2},Z_{N_3}$ lattice operators $\hat{\phi}^{(1)}_j, \hat{L}^{(1)}_l$, $\hat{\phi}^{(2)}_j, \hat{L}^{(2)}_l$, $\hat{\phi}^{(3)}_j, \hat{L}^{(3)}_l$ and field operators $\phi_1,\phi_1'$, $\phi_2,\phi_2'$, $\phi_3,\phi_3'$ are identified by 
\be
\hat{\phi}^{(u)}_j \to \phi_u (x_j),\;\;\; 
\hat{L}^{(u)}_l \to \frac{1}{2\pi}\partial_x \phi_{u}'(x_l). \label{eq:discrete-to-cont}
\ee
We view $\phi_u$ and $\phi_u'$ as the dual rotor angles as before, the relation follows as Sec.\ref{sec:Type II field}. 
We have no difficulty to formulate a K matrix multiplet chiral boson field theory 
(non-chiral `doubled' version of Ref.\onlinecite{Floreanini:1987as}'s action) as 
\be \label{eq:Kmatrix_edge}
\text{{\bf S}}_{\text{SPT},\partial \mathcal{M}^2}= 
\frac{1}{4\pi} \int dt\; dx \; \big( K_{IJ} \partial_t \phi_{I} \partial_x \phi_{J} -V_{IJ}\partial_x \phi_{I}   \partial_x \phi_{J} \big) + \dots.
\ee
requiring a rank-6 symmetric $K$-matrix, 
\be
K_{\text{SPT}} =\bigl( {\begin{smallmatrix} 
0 &1 \\
1 & 0
\end{smallmatrix}}  \bigl) \oplus \bigl( {\begin{smallmatrix} 
0 &1 \\
1 & 0 
\end{smallmatrix}}  \bigl)  
\oplus \bigl( {\begin{smallmatrix} 
0 &1 \\
1 & 0 
\end{smallmatrix}}  \bigl).  
\ee
with a chiral boson multiplet $\phi_I(x)=(\phi_1 (x),$ $\phi_1 ' (x),$ $\phi_2 (x),$ $\phi_2' (x),$ $\phi_3 (x),$ $\phi_3' (x) )$.
The commutation relation Eq.(\ref{eq:commutation}) becomes: $ [\phi_I(x_1), K_{I'J} \partial_x \phi_{J}(x_2)]= {2\pi} \ti  \delta_{I I'} \delta(x_1-x_2)$.
The continuum limit of Eq.(\ref{eq:symmetry form lattice Type II S1}) 
becomes\cite{field tranf regularized}
\bea \label{eq:globalS_Type II_1}
\bullet \;\;  S^{(p_u,p_{uv})}_{N_u}
=
\exp[
\frac{\ti}{N_u}\,
(
\int^{L}_{0}\,dx\,\partial_{x}\phi_{u}'
+
p_u\,\int^{L}_{0}\,dx\,\partial_{x}\phi_{u} \nonumber \\
+
0\int^{L}_{0}\,dx\,\partial_{x}\phi_{v}'
+
p_{uv}\,\int^{L}_{0}\,dx\,\partial_{x} \tilde{\phi}_{v}
)
] \;\;\;
\eea
Notice that we carefully input a tilde on some $\tilde{\phi}_v$ fields. We stress the lattice regularization of $\tilde{\phi}_v$
is different from ${\phi}_v$, see Eq.(\ref{eq:field regularize}), which is analogous to $\tilde{\sigma}^{(1)}$, $\tilde{\sigma}^{(2)}$ and $\sigma^{(1)}$, $\sigma^{(2)}$ in Sec.\ref{Sec: type I II sym transf}. 
We should mention two remarks: 
First, there are higher order terms beyond $\text{{\bf S}}_{\text{SPT},\partial \mathcal{M}^2}$'s quadratic terms when taking continuum limit of lattice. 
At the low energy limit, it shall be reasonable to drop higher order terms. 
Second, in the nontrivial SPT class (some topological terms $p_i \neq 0$, $p_{ij} \neq 0$), the $\det(V)\neq 0$ and all eigenvalues are non-zeros, so the edge modes are gapless.
In the trivial insulating class (all topological terms $p = 0$), the $\det(V)= 0$, so the edge modes may be gapped (consistent with Sec.\ref{sec:lattice Hamiltonians: type I II}).
Use Eq.(\ref{eq:commutation}), we derive the 1D edge global symmetry transformation
$S^{(p_u,p_{uv})}_{N_u}$, 
as an example, we consider $S^{(p_1,p_{12})}_{N_1}$,\cite{field tranf regularized} 
\begin{equation} \label{eq:S1}
S^{(p_1,p_{12})}_{N_1}
 {\begin{pmatrix} 
\phi_1 (x)   \\
\phi_1 ' (x)  \\
\tilde{\phi}_2 (x)   \\
\tilde{\phi}_2' (x)
\end{pmatrix}} (S^{(p_1,p_{12})}_{N_1})^{-1}
=
 {\begin{pmatrix} 
\phi_1 (x)   \\
\phi_1 ' (x)  \\
\tilde{\phi}_2 (x)   \\
\tilde{\phi}_2' (x)
\end{pmatrix}} 
+\frac{2\pi}{N_1}{\begin{pmatrix} 
1   \\
p_1 \\
0\\
p_{12}  
\end{pmatrix}}.  
\end{equation}
We can see how $p_{12}$, $p_{21}$ identify the same index
by doing a $M$ matrix with $M \in  \text{SL}(4, \mathbb{Z})$ transformation on the K matrix Chern-Simons theory,
which redefines the $\phi$ field, but still describe the same theory. That means: $K \to K'= M^T K M$ and $\phi \to \phi'=M^{-1} \phi$,
and so the symmetry charge vector $q \to q' =M^{-1} q$. By choosing
\bea
&&M=
\bigg( \begin{smallmatrix}
 1 & 0 & 0 & 0 \\
 {p_1} & 1 &  {p_{21}} & 0 \\
 0 & 0 & 1 & 0 \\
  {p_{12}} & 0 & {p_2} & 1
\end{smallmatrix} \bigg),
 \text{   then the basis is changed to} \nonumber\\
&&
K'=
\bigg( \begin{smallmatrix}
 2 {p_1} & 1 &  {p_{12}}+ {p_{21}} & 0 \\
 1 & 0 & 0 & 0 \\
  {p_{12}}+ {p_{21}} & 0 & 2 {p_2} & 1 \\
 0 & 0 & 1 & 0
 \end{smallmatrix} \bigg),
q_1'=
\bigg( \begin{smallmatrix}
1\\
0\\
0\\
0
\end{smallmatrix}
\bigg),\;
q_2'=
\bigg( \begin{smallmatrix}
0\\
0\\
1\\
0
\end{smallmatrix}
\bigg). \nonumber
\eea
The theory labeled by $K_{\text{SPT}},q_1,q_2$ is equivalent to the one  labeled by $K',q_1',q_2'$.
Thus we show that $p_{12}+p_{21} \to p_{12}$ identifies the same index.
There are other ways using the gauged or probed-field version of topological gauge theory (either on the edge or in the bulk) 
to identify the gauge theory's symmetry transformation,\cite{Ye:2013upa}
or the bulk braiding statistics\cite{Chenggu} to determine this Type II classification
$
p_{12}\; \text{mod}(\gcd(N_1,N_2))
$.

The nontrivial fact that when $p_{12}=N_{12}$ is a trivial class, the symmetry transformation in Eq.(\ref{eq:S1}) 
may not go back to 
the trivial symmetry under the condition $\int^{L}_{0}\,dx\,\partial_{x} \tilde{\phi}_{1}$ $=\int^{L}_{0}\,dx\,\partial_{x} \tilde{\phi}_{2}$ $=2\pi$, implying a soliton
can induce fractional charge (for details see Sec.\ref{sec: Type II frac}).

Our next goal is deriving Type III symmetry transformation Eq.(\ref{eq:Type III_S}). 
By taking the continuum limit of
\bea
&& \epsilon^{(u=1,2,3)(v)(w)}  \phi^{{j+1},(v)}_{in} \phi^{j,(w)}_{in} \\ 
&&=\big((\phi^{{j+1},(v)}_{in}-\phi^{{j},(v)}_{in}) \phi^{j,(w)}_{in}-(\phi^{{j+1},(w)}_{in}-\phi^{{j},(w)}_{in}) \phi^{j,(v)}_{in} \big) \nonumber \\
&&\to 
\big(\partial_x \phi^{(v)}_{in}(x) \phi^{(w)}_{in}(x)-\partial_x \phi^{(w)}_{in}(x) \phi^{(v)}_{in}(x) \big)
\eea
we can massage the continuum limit of Type III symmetry transformation Eq.(\ref{eq:Type III_S}) to ($\gcd({N_1,N_2,N_3}) \equiv N_{123}$)
\begin{widetext}
\bea \label{eq:globalS_Type III}
{\bullet \;S^{(p_{123})}_{N_1,N_2,N_3}=\prod_{u,v,w \in \{1,2,3\}} \exp\big[ \frac{\ti}{N_u}\,
(
\int^{L}_{0}\,dx\,\partial_{x}\phi_{u}' )\big] \cdot
 \exp\big[ \ti {\frac{N_1N_2N_3}{2\pi N_{123}}} \frac{p_{123}}{N_{u}} \int^{L}_{0}\,dx\, \epsilon^{u v w} \partial_x \phi^{}_{v}(x) \phi^{}_{w}(x) \big]\;
}\;\;\;\;\eea
\end{widetext}
Here $u,v,w \in \{ 1,2,3\}$ are the label of the symmetry group $Z_{N_1}$, $Z_{N_2}$, $Z_{N_3}$'s indices.
Though this Type III class is already known in the group cohomology sense, 
{ this Type III field theory symmetry transformation result is entirely new and not yet been well-explored in the literature,}  
especially not yet studied in the field theory in the SPT context. 
Our result is an extension along the work of Ref.\onlinecite{{Lu:2012dt}},\onlinecite{{Ye:2013upa}}.

The commutation relation leads to 
\bea
&&[\phi_I(x_i), K_{I'J}  \phi_{J}(x_j)]= 
- {2\pi} \ti\, \delta_{I I'}  \tilde{h}(x_i-x_j). \;\;\;\;\;\;
\eea
Here $\tilde{h}(x_i-x_j) \equiv h(x_i-x_j) -1/2$, where
$h(x)$ is the Heaviside step function, with $h(x)=1$ for $x \geq 0$ and $h(x)=0$ for $x < 0$. 
And $\tilde{h}(x)$ is ${h}(x)$ shifted by 1/2, i.e. $\tilde{h}(x)=1/2$ for $x \geq 0$ and $h(x)=-1/2$ for $x < 0$.
The shifted 1/2 value is for consistency condition for the integration-by-part and the commutation relation.
Use these relations, 
we derive the global symmetry transformation $S^{(p_{123})}_{N_1,N_2,N_3}$ acting on the rotor fields $\phi_u(x),\phi_u'(x)$ (here $u \in \{ 1,2,3\}$) on the 1D edge by 

\begin{widetext}
\bea \label{Type III S phi}
&& 
(S^{(p_{123})}_{N_1,N_2,N_3}) \phi_u(x) {(S^{(p_{123})}_{N_1,N_2,N_3})}^{-1}=\phi_u(x)+\frac{2\pi}{N_{u}} \\
\label{Type III S phi'}
&& 
{(S^{(p_{123})}_{N_1,N_2,N_3}) \phi_u'(x) {(S^{(p_{123})}_{N_1,N_2,N_3})}^{-1}=\phi_u'(x) - \epsilon^{uvw} Q \frac{2\pi}{N_{v}} ( 2 \phi^{}_{w}(x)-\frac{(\phi^{}_{w}(L)+\phi^{}_{w}(0))}{2}})
\eea
\end{widetext}
where one can define a Type III symmetry charge $Q \equiv {p_{123}  \frac{N_1N_2N_3}{2\pi N_{123}}}$.
Here the 1D edge is on a compact circle with the length $L$, here $\phi^{}_{w}(L)$ are $\phi^{}_{w}(0)$ taking value at the position $x=0$ (also $x=L$).
(In the case of infinite 1D line, we shall replace $\phi^{}_{w}(L)$ by $\phi^{}_{w}(\infty)$ and replace $\phi^{}_{w}(0)$ by $\phi^{}_{w}(-\infty)$. ) 
But $\phi^{}_{w}(L)$ may differ from $\phi^{}_{w}(0)$ by $2\pi n$ with some 
number $n$ if there is a nontrivial winding, i.e.
\be
\phi^{}_{w}(L)=\phi^{}_{w}(0)+ 2\pi n=2\pi \frac{n_w}{N_w}+ 2\pi n,
\ee
where we apply the fact that $\phi^{}_{w}(0)$ is a $Z_{N_w}$ rotor angle.
So Eq.(\ref{Type III S phi'}) effectively results in a shift
$
+ \epsilon^{uvw} p_{123} {\frac{N_u }{ N_{123}}}    {(2\pi {n_w} + \pi {N_w}{n} )}  
$
and a rotation $\epsilon^{uvw} Q \frac{2\pi}{N_{v}} ( 2 \phi^{}_{w}(x))$.
Since ${\frac{N_u }{ N_{123}}}$ is necessarily an integer, the symmetry transformation $(S^{(p_{123})}_{N_1,N_2,N_3}) \phi_u'(x) {(S^{(p_{123})}_{N_1,N_2,N_3})}^{-1}$ will
shift by a $2 \pi$ multiple if $p_{123} \, {\frac{N_u }{ N_{123} }}  \,{N_w}{n}$ is an even integer.\\

By realizing the field theory symmetry transformation, we have obtained all classes of SPT edge field theory
within the group cohomology $\cH^3(Z_{N_1} \times Z_{N_2} \times Z_{N_3},\tU(1))$ 
with $p_u \in \mathbb{Z}_{N_u}$, $p_{uv} \in \mathbb{Z}_{N_{uv}}$, $p_{123} \in \mathbb{Z}_{ N_{123}}$.

\color{black}




\section{Type II Bosonic Anomaly: Fractional Quantum Numbers trapped at the Domain Walls
\label{sec: Type II frac}}

We now apply the tools we develop in Sec.\ref{sec:Type I} 
to capture physical observables for these SPTs.
We propose the experimental/numerical signatures for certain SPT with Type II class $p_{12} \neq 0$ with (at least) two symmetry group $Z_{N_1} \times Z_{N_2}$,
also as a way to study the physical measurements for Type II bosonic anomaly.
We show that when the $Z_{N_1}$ symmetry is broken by $Z_{N_1}$ domain wall 
created on a ring, there will be some fractionalized $Z_{N_2}$ charges induced near the kink.
We will demonstrate our field theory method can easily capture this effect.

\subsection{Field theory approach: fractional $Z_{N}$ charge trapped at the kink of $Z_N$ symmetry-breaking Domain Walls
\label{sec:Type II bnom field}
}

Consider the $Z_{N_1}$ domain wall is created on a ring (the $Z_{N_1}$ symmetry is broken), then the $Z_{N_1}$ domain wall can be captured by 
$\phi_{1}(x)$ for $x \in [0,L)$ takes some constant value $\phi_0$ while $\phi_{1}(L)$ shifted by $2\pi \frac{n_1}{N_1}$ away from $\phi_0$.
This means that $\phi_{1}(x)$ has the fractional winding number:
\be
\int^{L}_{0}\,dx\,\partial_{x} \phi_{1} =\phi_{1}(L) -\phi_{1}(0) = 2\pi \frac{n_1}{N_1},
\ee

\begin{figure}[h!]
\includegraphics[width=0.32\textwidth]{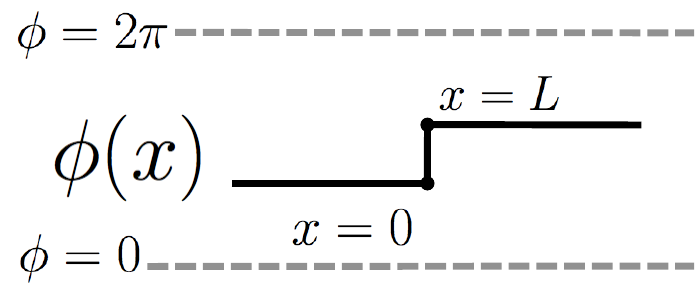} 
\caption{ We expect some fractional charge trapped near a single kink around $x=0$ (i.e. $x=0+\epsilon$) and $x=L$ (i.e. $x=0-\epsilon$) in the domain walls.
For $Z_{N_1}$-symmetry breaking domain wall with a kink jump $\Delta \phi_1=2\pi \frac{n_{12}}{N_{12}}$, we predict that the fractionalized $(n_{12}/N_{12})p_{21}$ units of $Z_{N_2}$ charge are induced.}
\label{fig:domain_walls_kink}
\end{figure}

Also recall Eq.(\ref{eq:globalS_Type II_1}) that the Type II $p_{21} \neq 0$ (and $p_{1} = 0, p_{2} = 0$) $Z_{N_2}$ symmetry transformation
\bea
 \;\;  S^{(p_2,p_{21})}_{N_2}
=
\exp[
\frac{ \ti}{N_2}\,
(
p_{21}\,\int^{L}_{0}\,dx\,\partial_{x} \tilde{\phi}_{1}  
+
\int^{L}_{0}\,dx\,\partial_{x}\phi_{2}'
)
], \;\;\;
\eea
can measure the induced $Z_{N_2}$ charge on a state $| \Psi_{\text{domain}} \rangle$ with this domain wall feature as
\bea \label{eq:induced charge type II}
&&S^{(p_2,p_{21})}_{N_2} | \Psi_{\text{domain}} \rangle =\exp[
\frac{ \ti \, p_{21} }{N_2}\,
(
 \tilde{\phi}_{1}(L) -\tilde{\phi}_{1}(0)
)
] | \Psi_{\text{domain}} \rangle \nonumber\\
&&=\exp[
(
2\pi \ti \frac{n_{12} \; p_{21} }{N_{12}\, N_2}
)
] | \Psi_{\text{domain}} \rangle.
\eea
We also adopt two facts that:
First, $\int^{L}_{0}\,dx\,\partial_{x} \tilde{\phi}_{1}  = 2\pi \frac{n_{12}}{N_{12}}$ with some integer ${n_{12}}$, where the $\tilde{\phi}_{1}$ is regularized in a unit of $2\pi/{N_{12}}$. 
Second, as $Z_{N_2}$ symmetry is not broken, both $\phi_{2}$ and $\phi_{2}'$ have no domain walls, then the above evaluation takes into account that $\int^{L}_{0}\,dx\,\partial_{x}\phi_{2}'=0$.
This implies that induced charge is fractionalized $(n_{12}/N_{12})p_{21}$ (recall $p_{12}, p_{21}  \in  \mathbb{Z}_{N_{12}}$ ) $Z_{N_2}$ charge. This is the fractional charge trapped at the configuration of a single kink in Fig.\ref{fig:domain_walls_kink}.

\begin{figure}[h!]
\includegraphics[width=0.32\textwidth]{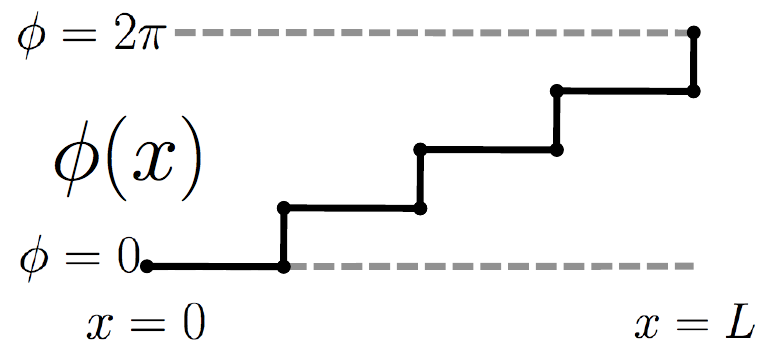} 
\caption{ A nontrivial winding $\int^{L}_{0}\,dx\,\partial_{x} {\phi}(x) = 2\pi$.
This is like a soliton a soliton (or particle) insertion.
For $N_{12}$ number of $Z_{N_1}$-symmetry breaking domain walls, we predict that the integer $p_{21}$ units of total induced $Z_{N_2}$ charge on a 1D ring.
 In average, each kink captures a $p_{21}/N_{12}$ fractional units of $Z_{N_2}$ charge.}
\label{fig:domain_walls_inst}
\end{figure}

On the other hand, one can imagine a series of $N_{12}$ number of 
$Z_{N_1}$-symmetry-breaking domain wall each breaks to different vacuum expectation value(v.e.v.) where the domain wall in the region $[0,x_1)$,$[x_1,x_2)$, $\dots$, $[x_{N_{12}-1},x_{N_{12}}=L)$
with their symmetry-breaking $\phi_1$ value at $0$, 
$2\pi \frac{1}{N_{12}}$, $2\pi \frac{2}{N_{12}}$, $\dots$, $2\pi \frac{N_{12}-1}{N_{12}}$.
This means a nontrivial winding number, like a soliton effect (see Fig.\ref{fig:domain_walls_inst}), 
$
\int^{L}_{0}\,dx\,\partial_{x} \tilde{\phi}_{1} 
= 2\pi 
$ 
and 
$
S^{(p_2,p_{21})}_{N_2} | \Psi_{\text{domain wall}} \rangle$ 
 $=\exp[
(
2\pi \ti \frac{  p_{21} }{\, N_2}
)
] | \Psi_{\text{domain wall}} \rangle
$
capturing
$p_{21}$ integer units of $Z_{N_2}$ charge at $N_{12}$ kinks for totally $N_{12}$ domain walls, in the configuration of Fig.\ref{fig:domain_walls_inst}. In average, each kink captures the $p_{21}/N_{12}$ fractional units of $Z_{N_2}$ charge.


Similarly, we can consider the $Z_{N_2}$ domain wall is created on a ring (the $Z_{N_2}$ symmetry is broken), then the $Z_{N_2}$ domain wall can be captured by 
$\phi_{2}(x)$ soliton profile for $x \in [0,L)$.  
We consider a series of $N_{12}$ number of 
$Z_{N_2}$-symmetry-breaking 
domain walls, each breaks to different v.e.v.
(with an overall profile of 
$
\int^{L}_{0}\,dx\,\partial_{x} \tilde{\phi}_{2} = 2\pi 
$).
By
$
S^{(p_1,p_{12})}_{N_1} | \Psi_{\text{domain wall}} \rangle 
=\exp[
(
2\pi \ti \frac{  p_{12} }{\, N_1}
)
] | \Psi_{\text{domain wall}} \rangle
$, the $N_{12}$ kinks of domain wall captures
$p_{12}$ integer units of $Z_{N_1}$ charge for totally $N_{12}$ domain wall, as in Fig.\ref{fig:domain_walls_inst}. 
In average, each domain wall captures $p_{12}/N_{12}$ fractional units of $Z_{N_1}$ charge.


\subsection{Goldstone-Wilczek formula and Fractional Quantum Number \label{sec:Goldstone-Wilczek}}

It is interesting to view our result above in light of the Goldstone-Wilczek (G-W) approach.\cite{Goldstone:1981kk} 
We warm up by computing  $1/2$-fermion charge found by Jackiw-Rebbi\cite{Jackiw:1975fn} using G-W method 
We will then do a more general case for SPT. 
The construction, valid for 1D systems, works as follows.\\

\noindent
{\bf Jackiw-Rebbi model:}  Consider a Lagrangian describing spinless fermions $\psi(x)$ coupled to a classical background profile $\lambda(x)$
via a term $\lambda\,\psi^{\dagger}\sigma_{3} \psi$. In the high temperature phase, the v.e.v. of $\lambda$ is zero and no mass is generated
for the fermions. In the low temperature phase, the $\lambda$ acquires two degenerate vacuum values $\pm  \langle \lambda \rangle$ 
that are related by a ${Z}_2$ symmetry. Generically we have
\be
\langle \lambda \rangle\,\cos\big( \phi(x) - \theta(x) \big),
\ee
where we use the bosonization dictionary 
$
\psi^{\dagger}\sigma_{3} \psi
\rightarrow
\cos(\phi(x))
$
and a phase change $\Delta\theta = \pi$ captures the existence of a domain wall
separating regions with opposite values of the v.e.v. of $\lambda$.
From the fact that the fermion density 
$
\rho(x)
=
\psi^{\dagger}(x)\psi(x) 
= 
\frac{1}{2\pi} \partial_{x} \phi(x)
$ (and the current $J^\mu=\psi^{\dagger}\gamma^\mu\psi =\frac{1}{2\pi}\epsilon^{\mu \nu} \partial_\nu \phi$),
it follows that the induced charge $Q_{\text{dw}}$ on the kink by a domain wall is
\be
Q_{\text{dw}} 
= 
\int^{x_0 + \varepsilon}_{x_0 - \varepsilon}\,dx\,\rho(x)
=
\int^{x_0 + \varepsilon}_{x_0 - \varepsilon}\,dx\,\frac{1}{2\pi} \partial_{x} \phi(x)
=
\frac{1}{2},
\ee
where $x_0$ denotes the center of the domain wall.\\

\begin{figure}[t]
\includegraphics[width=0.35\textwidth]{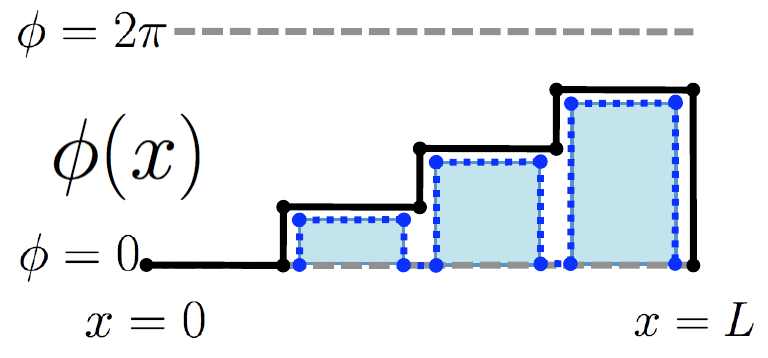} 
\caption{A profile of several domain walls, each with kinks and anti-kinks(in blue color).
For $Z_{N_1}$ symmetry-breaking domain wall, each single kink can trap fractionalized $Z_{N_2}$ charge.
However, overall there is no nontrivial winding, $\int^{L}_{0}\,dx\,\partial_{x} {\phi_1}(x) = 0$ (i.e. no net soliton insertion), so there
is no net induced charge on the whole 1D ring.}
\label{fig:domain_walls_antis}
\end{figure}

\noindent
{\bf Type II Bosonic Anomalies:} 
We now consider the case where the ${Z}_{N_1}$ symmetry is spontaneously broken
into different ``{vacuum}'' regions. This can be captured by an effective term in the Hamiltonian 
of the form
\begin{equation}
H_{sb}
=
-\lambda\,\cos \big(\phi_{1}(x) - \theta(x) \big), \quad \lambda > 0,
\end{equation}
and the ground state
is obtained, in the large $\lambda$ limit, by phase locking $\phi_{1} = \theta$, which opens a gap in the spectrum.

Different domain wall regions are described by different choices of the profile $\theta(x)$, as discussed in Sec.\ref{sec:Type II bnom field}. In particular, if we have $\theta(x) = \theta_{k}(x)$
and
$
\theta_{k}(x) = (k-1)\,2\pi/N_{12}, 
$ for
$
x \in [(k-1)L/N_{12}, k L/N_{12} )$, $k = 1,...,N_{12}$.
then we see that that, a domain wall separating regions $k$ and $k+1$ (where the phase difference is $2\pi/N_{12}$)
induces a ${Z}_{N_2}$ charge given by
\bea \label{eq: G-W Q}
&&\delta\,Q_{k,k+1}
=
\int^{k L/N_{12} + \varepsilon}_{k L/N_{12} - \varepsilon}\,dx\,\delta\rho_{2}(x) \nonumber
\\
&&=
\frac{1}{2\pi} \int^{k L/N_{12} + \varepsilon}_{k L/N_{12} - \varepsilon}\,dx\,
\frac{p_{12}}{N_{2}}\,\partial_{x} \phi^{}_{1}
=
\frac{p_{12}}{N_2 N_{12}}.
\eea
This implies a fractional of ${p_{12}}/{N_{12}}$ induced $Z_{N_2}$ charge on a single kink of $Z_{N_1}$-symmetry breaking domain walls, consistent with Eq.(\ref{eq:induced charge type II}).

\begin{figure}[h!]
\includegraphics[width=0.4\textwidth]{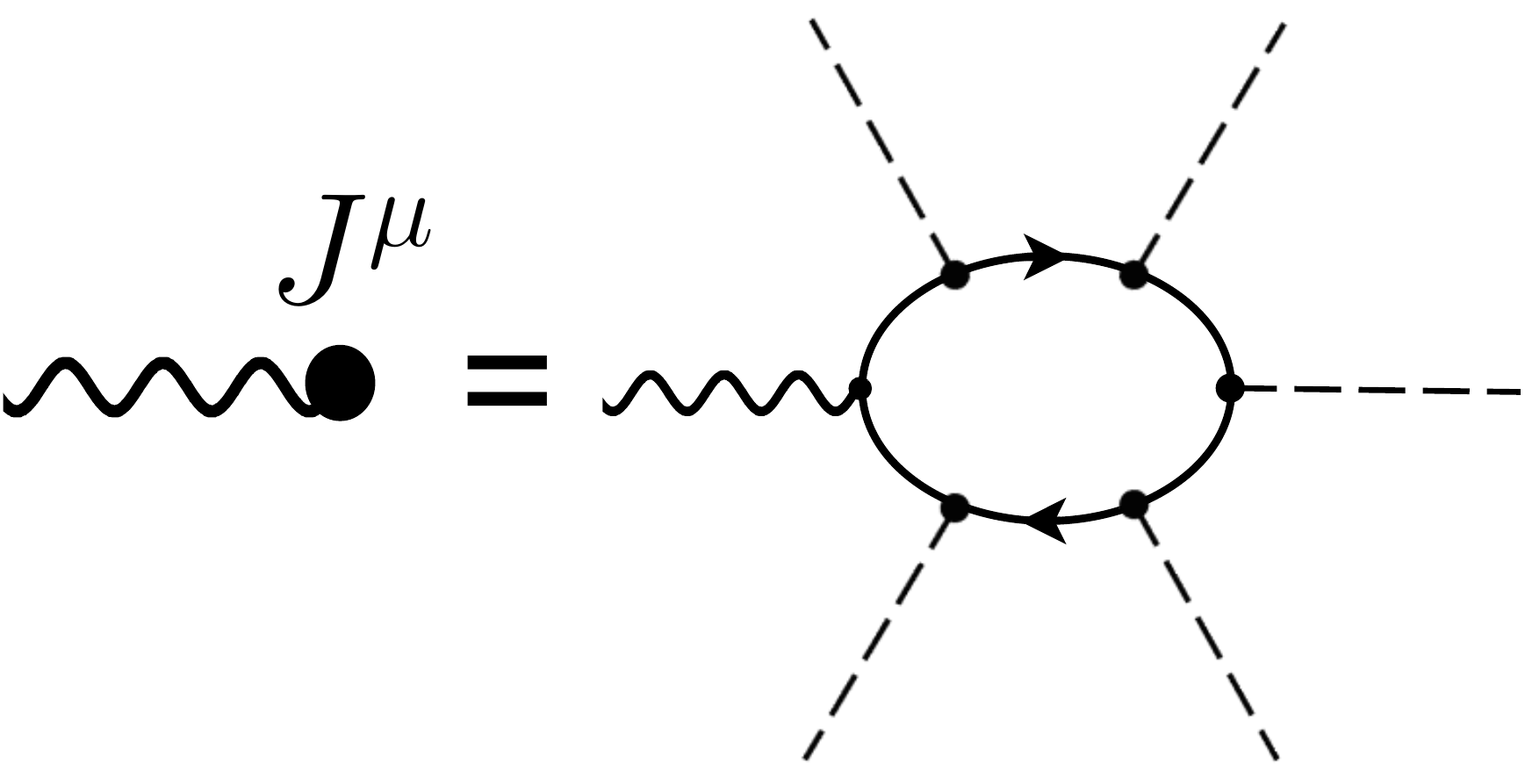} 
\caption{ In the fermionized language, one can capture the anomaly effect on induced (fractional) charge/current under soliton background by the 1-loop diagram.\cite{Goldstone:1981kk}
With the solid line --- represents fermions, 
the wavy line  \protect\middlewave{.42cm}\, 
 represents the external (gauge) field coupling to the induced current $J^\mu$ (or charge $J^0$), and the dashed line - - represents the scalar soliton (domain walls) background.
Here in Sec.\ref{sec:Goldstone-Wilczek}, instead of fermionizing the theory, we directly address in the bosonized language to capture the bosonic anomaly.}
\label{fig:soliton_current}
\end{figure}

Some remarks follow:
If the system is placed on a ring, 
(i) First, with net soliton (or particle) insertions, then the total charge induced is non-zero, see Fig.\ref{fig:domain_walls_inst}.\\
(ii) Second, without net soliton (or particle) insertions, 
then the total charge induced is obviously zero, as domain walls necessarily
come in pairs with opposite charges on the kink and the anti-kink, see Fig.\ref{fig:domain_walls_antis}.\\
(iii) One can also capture this bosonic anomaly in the fermionized language using the 1-loop diagram under soliton background,\cite{Goldstone:1981kk} shown in Fig.\ref{fig:soliton_current}.\\
(iv) A related phenomena has also been examined recently where 
fractionalized boundary excitations cause that 
the symmetry-broken boundary cannot be proliferated to restore the symmetry.\cite{Lu:2013wna}

\subsection{Lattice approach: Projective phase observed at Domain Walls}

Now we would like to formulate a fully regularized lattice approach to derive the induced fractional charge, 
and compare 
to the complementary field theory done in Sec.\ref{sec:Type II bnom field} and Goldstone-Wilczek approach in Sec.\ref{sec:Goldstone-Wilczek}.
Below our notation follows Sec.\ref{sec:Type I}.
Recall that in the case of a system with onsite symmetry, such as $Z_{N}$ rotor 
model on a 1D ring with a simple Hamiltonian of $\sum_j (\sigma_j + \sigma_j^\dagger)$,
there is an on-site symmetry transformation $S=\prod_j \tau_j $ acting on the full ring. 
We can simply take a segment (from the site $r_1$ to $r_2$) of the symmetry transformation defined as a $D$ operator $D(r_1,r_2) \equiv \prod_{j=r_1}^{r_2} \tau_j $.
The $D$ operator does the job to flip the measurement on $\langle \sigma_\ell \rangle$. What we mean is that $\langle \psi | \sigma_{\ell} |\psi \rangle$
and $\langle \psi' | \sigma_{\ell} |\psi' \rangle \equiv\langle \psi | D^\dagger\sigma_{\ell} D |\psi \rangle=e^{i 2\pi/N}\langle \psi | \sigma_{\ell} |\psi \rangle$ are distinct by a phase $e^{i 2\pi/N}$
as long as $\ell \in [r_1, r_2]$. Thus $D$ operator creates domain wall profile.

For our case of SPT edge modes with non-onsite symmetry studied here, we are readily to generalize the above and take a line segment of non-onsite symmetry transformation with symmetry $Z_{N_u}$ (from the site $r_1$ to $r_2$) and define it as 
a $D_{N_u}$ operator, $D_{N_u}(r_1,r_2)\equiv \prod_{j=r_1}^{r_2} \tau^{(u)}_j \prod_{j=r_1}^{r_2} U_{j,j+1} W^{III}_{j,j+1}$ (from the expression of $S_{N_u}$, with the onsite piece  $\tau^{(u)}_j$ and the non-onsite piece  $U_{j,j+1}$ in Eq.(\ref{S1symp12}) 
and $W_{j,j+1}$ in Eq.(\ref{eq:Type III_W})
).
This $D$ operator effectively creates domain wall on the state with a kink (at the $r_1$) and anti-kink (at the $r_2$) feature, such as in Fig.\ref{fig:domain_walls_antis}.
The total net charge on this type of domain wall (with equal numbers of kink and anti-kinks) is zero, due to no net soliton insertion (i.e. no net winding, so $\int^L_0 \partial_x \phi \, dx =0$).
\cblue{However, by well-separating kinks and anti-kinks,
we can still compute the phase gained at each single kink.\cite{Supplementary Material}}
We consider the induced charge measurement by $S( D|\psi\rangle) $, which is $(S D S^\dagger) S|\psi\rangle =e^{\ti (\Theta_0+\Theta)} D |\psi\rangle $, where $\Theta_0$ is from the initial charge (i.e. $S|\psi\rangle \equiv e^{\ti \Theta_0} |\psi\rangle$) and $\Theta$ is from the charge gained on the kink. The measurement of symmetry $S$ producing a phase $e^{\ti \Theta}$, implies a nontrivial induced charge trapped at the kink of domain walls. We compute the phase at the left kink on a domain wall for all Type I, II, III SPT classes, and summarize them in Table \ref{table2}.

\begin{widetext}
\begin{center}
\begin{table}[H]
\begin{tabular}{|c||c|c|c|}
\hline
\text{SPT class} & $e^{\ti\Theta_L}$ of $D_{N_u}|\psi\rangle$ acted by $Z_{N_v}$ symmetry $S_{v}$ & $e^{\ti \Theta_L}$ of $D_{N_u}|\psi\rangle$ under a soliton  $\int_0^L dx\, \partial_x \phi_u= 2\pi$ & Fractional charge\\[.5mm]  \hline \hline
Type I $p_{1}$ & $S^{(p_1)}_{N_1} D^{(p_1)}_{N_1} S_{N_1}^{(p_1)\dagger}  $ $\to$  $e^{\ti\Theta_L}=e^{ \ti\,\frac{2\pi p_1}{N_1^2}\,}$ &  $S^{(p_1)}_{N_1} (D^{(p_1)}_{N_1})^{N_1} S_{N_1}^{(p_1)\dagger}  $ $\to$  $e^{\ti\Theta_L}=e^{ \ti\,\frac{2\pi p_1}{N_1}\,}$  & No \\[1mm]  \hline
Type II $p_{12}$      & 
$S^{(p_{12})}_{N_2} D^{(p_{12})}_{N_1} S_{N_2}^{(p_{12})\dagger} \to e^{\ti \Theta_L}=e^{ \ti\,\frac{2\pi }{ N_2} \mathbf{\frac{p_{12}}{N_{12}}} }$  & 
$S^{(p_{12})}_{N_2} (D^{(p_{12})}_{N_1})^{N_{12}}  S_{N_2}^{(p_{12})\dagger}$ $\to$ $e^{\ti\Theta_L}=e^{ \ti\,\frac{2\pi p_{12}}{N_{2} }\,}$ &  Yes (Eq.(\ref{eq:induced charge type II}),(\ref{eq: G-W Q}))  \\[1mm] \hline 
Type III $p_{123}$ &  $S^{(p_{123})}_{N_2}D^{(p_{123})}_{N_1} S_{N_2}^{(p_{123})\dagger} $ $\to$  $e^{\ti\Theta_L}=e^{ \ti \,\frac{2\pi p_{123} n_3}{N_{123} } \,}$  &  
$S^{(p_{123})}_{N_2} (D^{(p_{123})}_{N_1})^{N_{123}} S_{N_2}^{(p_{123})\dagger} $ $\to$ $e^{\ti\Theta_L}=1$ & No \\[1mm] \hline
\end{tabular}
\caption{  The phase $e^{i\Theta_L}$ on a domain wall $D_u$ acted by $Z_{N_v}$ symmetry $S_{v}$. This phase is computed at the left kink (the site $r_1$).
The first column shows SPT class labels $p$. The second and the third columns show the computation of phases.
The last column interprets whether the phase indicates a nontrivial induced $Z_{N}$ charge.
Only Type II SPT class with $p_{12}\neq 0$ contains nontrivial induced $Z_{N_2}$ charge with a unit of $\mathbf{{p_{12}}/{N_{12}}}$ trapped at the kink of $Z_{N_1}$-symmetry breaking domain walls.
Here $n_3$ is the exponent inside the $W^{\text{III}}_{}$ matrix, 
$n_3=0,1,\dots,N_3-1$ for each subblock within the total $N_3$ subblocks.\cite{Supplementary Material} 
$N_{12}\equiv\gcd(N_1,N_2)$ and $N_{123}\equiv{\gcd(N_1,N_2,N_3)}$. }
\label{table2}
\end{table}
\end{center}
\end{widetext}
In Table \ref{table2}, although we obtain $e^{\ti\Theta_L}$ for each type, but there are some words of caution for interpreting it. \\
\noindent
{\bf (i)} For Type I class,
with the $Z_{N_1}$-symmetry breaking domain wall, there is no notion of induced $Z_{N_1}$ charge since there is no $Z_{N_1}$-symmetry (already broken) to respect.\\
\noindent
\cblue{
{\bf (ii)} 
$(D^{(p)}_{N})^n$ captures $n$ units of $Z_N$-symmetry-breaking domain wall.
The calculation $S^{(p)}_{N} (D^{(p)}_{N})^n S^{(p)\dagger}_{N}$ renders a $e^{\ti\Theta_L}$ phase for the left kink and 
a $e^{\ti\Theta_R}=e^{-\ti\Theta_L}$ phase for the right anti-kink.
Our formalism is analogous to Ref.\onlinecite{Lu:2013wna}, 
where we choose the domain operator as a segment of symmetry transformation.
For Type II class, if we have operators 
$(D^{(p_{12})}_{N_1})^0$ acting on the interval $[0,x_1)$,
while $(D^{(p_{12})}_{N_1})^1$ acting on the interval $[x_1,x_2)$, $\dots$, 
and $(D^{(p_{12})}_{N_1})^{N_{12}}$ acting on the interval  $[x_{N_{12}-1},x_{N_{12}}=L)$,
then we create the domain wall profile shown in Fig.\ref{fig:domain_walls_antis}. 
It is easy to see that due to charge cancellation on each kink/anti-kink,
the $S^{(p_{12})}_{N_2} (D^{(p_{12})}_{N_1})^{N_{12}}  S_{N_2}^{(p_{12})\dagger}$ measurement on a left kink  
captures the same amount of charge trapped by a nontrivial soliton: $\int_0^L dx\, \partial_x \phi_u= 2\pi$.\cite{Supplementary Material}
}\\
\noindent
{\bf (iii)} For Type II class, we consider $Z_{N_1}$-symmetry breaking domain wall (broken to a unit of $\Delta\phi_1=2\pi/N_{12}$), and find that there is induced $Z_{N_2}$ charge with a unit of $\mathbf{{p_{12}}/{N_{12}}}$, consistent with field theory approach in Eq.(\ref{eq:induced charge type II}),(\ref{eq: G-W Q}).
For a total winding is $\int_0^L dx\, \partial_x \phi_1= 2\pi$, there is also a nontrivial induced $\mathbf{ p_{12} }$ units of $Z_{N_2}$ charge.
Suppose a soliton generate this winding number $1$ domain wall profile, even if $p_{12}=N_{12}$ is identified as the trivial class as $p_{12}=0$, we can observe 
$\mathbf{N_{12} }$ units of $Z_{N_2}$ charge, which is in general still not $N_2$ units of $Z_{N_2}$ charge. 
This phenomena has no analogs in Type I, 
and can be traced back to the discussion in Sec.\ref{sec:Type II field}.\\ 
\noindent
{\bf (iv)} For Type III class, with a $Z_{N_1}$-symmetry breaking domain wall: 
On one hand, the $\Theta_L$ phase written in terms of $Z_{N_2}$ or $Z_{N_3}$ charge unit is non-fractionalized but integer.
On the other hand, we will find in Sec.\ref{sec: degenerate zero A} that the $Z_{N_2}$, $Z_{N_3}$ symmetry transformation surprisingly no longer commute. 
So there is no proper notion of induced $Z_{N_2}$, $Z_{N_3}$ charge at all in the Type III class.

\section{Type III Bosonic Anomaly: Degenerate zero energy modes (projective representation)
\label{sec: Type III zero} }

We apply the tools we develop in Sec.\ref{sec: GC},\ref{sec:Type I} to study the physical measurements for Type III bosonic anomaly.

\subsection{Field theory approach: Degenerate zero energy modes 
trapped at the kink of $Z_N$ symmetry-breaking Domain Walls \label{sec: degenerate zero A} }
We propose the experimental/numerical signature for certain SPT with Type III symmetric class $p_{123} \neq 0$ under the case of (at least) three symmetry group $Z_{N_1} \times Z_{N_2} \times Z_{N_3}$.
Under the presence of a $Z_{N_1}$ symmetry-breaking domain wall (without losing generality, we can also assume it to be any $Z_{N_u}$), we can detect that the
remained unbroken symmetry $Z_{N_2}$, $Z_{N_3}$ carry projective representation. More precisely,
under the $Z_{N_1}$ domain-wall profile,
\be
\int^{L}_{0}\,dx\,\partial_{x} \phi_{1} =\phi_{1}(L) -\phi_{1}(0) = 2\pi \frac{n_1}{N_1},
\ee
we compute the commutator between two unbroken symmetry operators Eq.(\ref{eq:globalS_Type III}): 
\cblue{
\bea \label{eq:commutator_S2_S3}
&&S^{(p_{231})}_{N_2} S^{(p_{312})}_{N_3} =S^{(p_{312})}_{N_3} S^{(p_{231})}_{N_2} e^{\ti \frac{2\pi\; n_1}{N_{123}} p_{123}}\\
&& [\log S^{(p_{231})}_{N_2}, \log S^{(p_{312})}_{N_3}]=\ti \frac{2\pi\; n_1}{N_{123}} p_{123},
\eea
}where we identify the index $(p_{231}+p_{312} ) \to p_{123}$ as the same one.
This non-commutative relation Eq.(\ref{eq:commutator_S2_S3}) indicates that $S^{(p_{231})}_{N_2}$ and $S^{(p_{312})}_{N_3}$ are not in a linear representation, but in a projective representation of $Z_{N_2}$, $Z_{N_3}$ symmetry.
This is analogous to the commutator $[T_x,T_y]$ of magnetic translations $T_x$, $T_y$ along $x,y$ direction on a $\mathbb{T}^2$ torus for a filling fraction $1/k$ fractional quantum hall state
(described by U$(1)_k$ level-$k$ Chern-Simons theory):\cite{Wen:1989zg}
\cblue{
\bea
&& e^{\ti T_x} e^{\ti T_y} =e^{\ti T_y} e^{\ti T_x} e^{\ti\, 2\pi /k}\\
&&[T_x,T_y]= -\ti\, 2\pi /k,
\eea}%
where one studies its ground states on a $\mathbb{T}^2$ torus with a compactified $x$ and $y$ direction gives $k$-fold degeneracy.
\cblue{The k degenerate ground states are $| \psi_m \rangle$ with $m=0, 1, \dots, k-1$, while $| \psi_m \rangle = | \psi_{m+k} \rangle$.
The ground states are chosen to satisfy: $e^{\ti T_x} | \psi_m \rangle =e^{\ti \frac{2\pi m}{k}} | \psi_m \rangle $, $e^{\ti T_y} | \psi_m \rangle = | \psi_{m+1} \rangle $}.
Similarly, for Eq.(\ref{eq:commutator_S2_S3}) we have a $\mathbb{T}^2$ torus compactified in $\phi_2$ and $\phi_3$ directions, such that:\\
(i)  There is a ${N_{123}}$-fold degeneracy for zero energy modes at the domain wall. 
\cblue{We can count the degeneracy by constructing the orthogonal ground states:
consider the eigenstate $| \psi_m \rangle $ of unitary operator $S^{(p_{231})}_{N_2}$, it implies that
$S^{(p_{231})}_{N_2} | \psi_m \rangle =e^{\ti \frac{2\pi\; n_1}{N_{123}} p_{123} m} | \psi_m \rangle$.
$S^{(p_{312})}_{N_3} | \psi_m \rangle =| \psi_{m+1} \rangle $. As long as $\gcd(n_1\,p_{123} ,{N_{123}})=1$,
we have ${N_{123}}$-fold degeneracy of $| \psi_m \rangle $ with $m=0,\dots,{N_{123}}-1$.}\\
(ii) 
Eq.(\ref{eq:commutator_S2_S3}) means the symmetry is realized projectively for the trapped zero energy modes at the domain wall.

We observe these are the signatures of Type III bosonic anomaly. 
This Type III anomaly in principle can be also captured by the perspective
of \emph{decorated $\Z_{N_1}$ domain walls} of Ref.\onlinecite{Chen:201301} with 
projective $\Z_{N_2} \times \Z_{N_3}$-symmetry. 

\subsection{Cocycle approach: Degenerate zero energy modes from $Z_N$ symmetry-preserving monodromy defect (branch cut)
- dimensional reduction from 2D to 1D \label{sec: degenerate zero A}}

\begin{figure}[h!]
\includegraphics[width=0.5\textwidth]{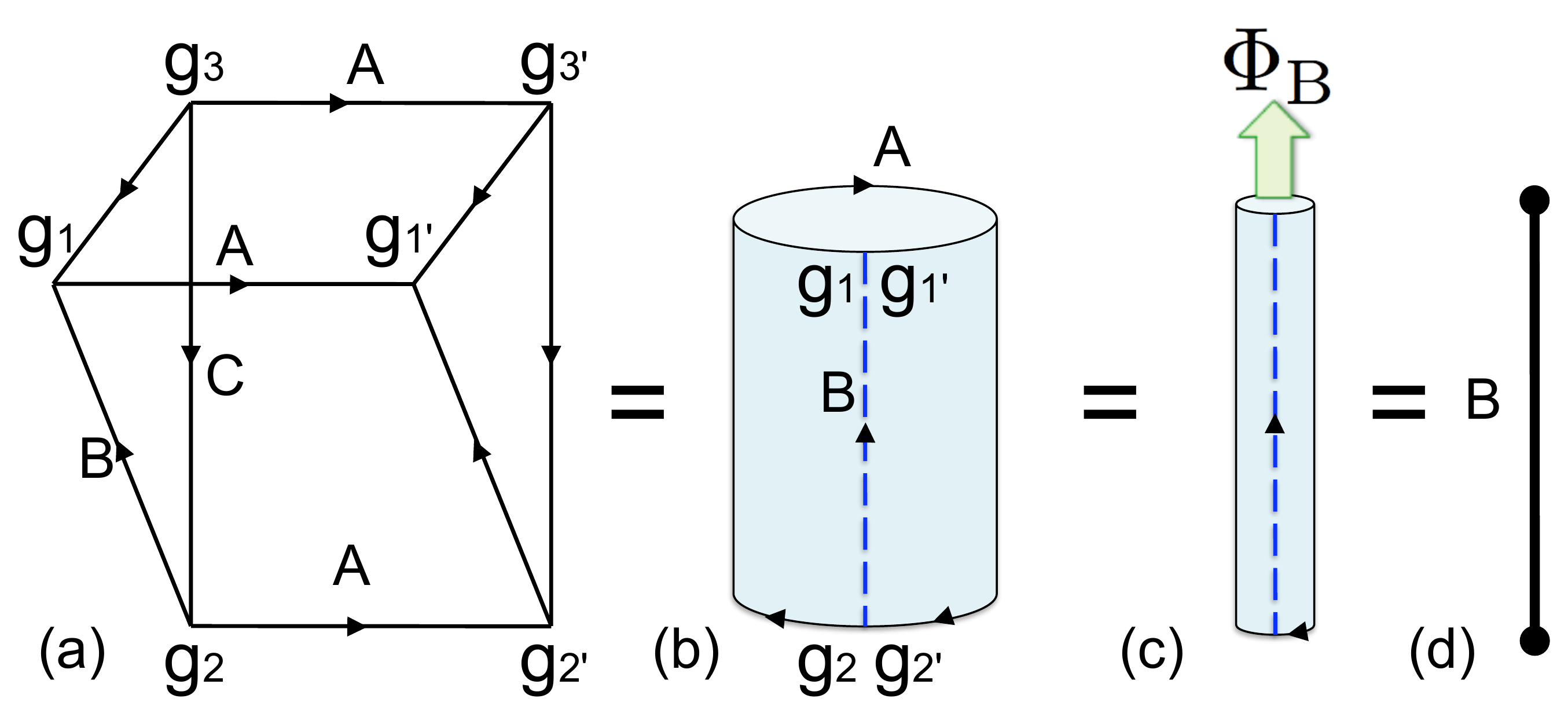} 
\caption{ (a) The induced 2-cocycle from a 2+1D $M^3=M^2 \times  I^1$ topology with a symmetry-preserving $Z_{N_u}$ flux $A$ insertion 
(b) Here $M^2=S^1 \times I^1$ 
is a 2D spatial cylinder, composed by $A$ and $B$, with another extra time dimension $I^1$.
Along the $B$-line we insert a monodromy defect of $Z_{N_1}$, such that $A$ has a nontrivial group element value $A=g_{1'} g_{1}^{-1} =g_{2'} g_{2}^{-1}
=g_{3'} g_{3}^{-1}  \in Z_{N_1}$. The induced 2-cocycle $\beta_A(B,C)$ is a nontrivial element in $\cH^2(Z_{N_v} \times Z_{N_w},\tU(1))$ $=\mathbb{Z}_{N_{vw}}$ (here $u,v,w$ cyclic as $\epsilon^{uvw}=1$), thus which carries a projective representation.
(c) A monodromy defect can viewed as a branch cut induced by a $\Phi_B$ flux insertion (both modifying the Hamiltonians). 
(d) This means that when we do dimensional reduction on the compact ring $S^1$ and view the reduced system as a 1D line segment, there are $N_{123}$ degenerate zero energy modes 
(due to the nontrivial projective representation).}
\label{fig:induced_2-cocycle}
\end{figure}

In Sec.\ref{sec: degenerate zero A}, we had shown the symmetry-breaking domain wall would induce 
degenerate zero energy modes for Type III SPT. In this section,
we will further show that,
a symmetry-preserving $Z_{N_1}$ flux insertion (or a monodromy defect or branch cut modifying the Hamiltonian as in Ref.\onlinecite{Santos:2013uda},\onlinecite{Wen:2013ue})
can also have degenerate zero energy modes. 
This is the case that, see Fig.\ref{fig:induced_2-cocycle}, when we put the system on a 2D cylinder and dimensionally reduce it to a 1D line along the monodromy defect.
In this case there is no domain wall, and the $Z_{N_1}$ symmetry is not broken (but only translational symmetry is broken near the monodromy defect / branch cut).

In the below discussion, we will directly use 3-cocycles $\omega_3$ from cohomology group $\cH^3(G,\tU(1))$ to detect the Type III bosonic anomaly.
For convenience we use the non-homogeneous cocycles  (the lattice gauge theory cocycles), though there is no difficulty to convert it to homogeneous cocycles (SPT cocycles).
The definition of the lattice gauge theory $n$-cocycles are indeed related to SPT $n$-cocycles:\cite{ {Hung:2012dx}, {Chen:2011pg}, {Wen:2013ue},{Hung:2013cda},{Hu:2012wx}}
\bea
&&\omega_n(A_1,A_2, \dots,A_n)=\nu_n(A_1A_2 \dots A_n, A_2 \dots A_n, \dots, A_n, 1) \nonumber\\
&&=\nu_n(\tilde{A}_1, \tilde{A}_2, \dots, \tilde{A}_n, 1)
\eea
here $\tilde{A}_j\equiv A_j A_{j+1} \dots A_n$. For 3-cocycles
\bea
&&\omega_3(A,B,C)=\nu_3(ABC,BC,C,1)\\
&&\Rightarrow \omega_3(g_{01},g_{12},g_{23})=\omega_3(g_{0} g_{1}^{-1}, g_{1} g_{2}^{-1} ,g_{2} g_{3}^{-1}) \nonumber \\
&&=\nu_3(g_{0}g_{3}^{-1} ,g_{1}g_{3}^{-1},g_{2}g_{3}^{-1},1)= \nu_3(g_{0},g_{1},g_{2},g_{3}) \nonumber
\eea
Here $A=g_{01}$, $B=g_{12}$, $C=g_{23}$, with $g_{ab} \equiv g_a g_b^{-1}$. We use the fact that SPT $n$-cocycle $\nu_n$ belongs to the $G$-module, such that 
for $r$ are group elements of $G$, it obeys $r \cdot \nu_n({r}_0, {r}_1, \dots, {r}_{n-1}, 1)=\nu( r {r}_0, r {r}_1, \dots, r {r}_{n-1}, r)$ 
 (here we consider only Abelian group $G=\prod_i Z_{N_i}$).
In our case, we do not have time reversal symmetry, so group action $g$ on the $G$-module is trivial.
%
%

In short, there is no obstacle so that we can simply use the lattice gauge theory 3-cocycle $\omega(A,B,C)$  to study the SPT 3-cocycle $\nu(ABC,BC,C,1)$.
Our goal is to design a geometry of 3-manifold $M^3=M^2 \times  I^1$ with $M^2$ the 2D cylinder with flux insertion (or monodromy defect) and with the $I^1$ time direction
(see Fig.\ref{fig:induced_2-cocycle}(a))
with a sets of 3-cocycles as tetrahedra filling this geometry (Fig.\ref{fig:decomp_2-cocycle}).
All we need to do 
is computing the 2+1D SPT path integral $\textbf{Z}_{\text{SPT}}$ (i.e. partition function) using 3-cocycles $\omega_3$,\cite{{Wen:2013ue}} 
\be
\textbf{Z}_{\text{SPT}}=|G|^{-N_v} \sum_{\{ g_{v}\}} \prod_i (\omega_3{}^{s_i}(\{ g_{v_a} g_{v_b}^{-1} \}))
\ee
Here $|G|$ is the order of the symmetry group, $N_v$ is the number of vertices, $\omega_3$ is 3-cocycle, and ${s_i}$ is the exponent 1 or $-1$(i.e. $\dagger$) depending on the orientation of each tetrahedron($3$-simplex). The summing over group elements $g_{v}$ on the vertex produces a symmetry-perserving ground state.
We consider a specific $M^3$, a $3$-complex of Fig.\ref{fig:induced_2-cocycle}(a), which can be decomposed into tetrahedra (each as a $3$-simplex) shown in Fig.\ref{fig:decomp_2-cocycle}.
There the 3-dimensional spacetime manifold is under triangulation 
(or cellularization) into three tetrahedra. 

We now go back to remark that 
the 3-cocycle condition in Eq.(\ref{eq:cocycle-conditions}) indeed means that the path integral $\textbf{Z}_{\text{SPT}}$ on the 3-sphere $S^3$ (as the surface the 4-ball $B^4$) will be trivial as 1.
The 3-coboundary condition in Eq.(\ref{eq:coboundary-conditions}) means to identify the same topological terms (i.e. 3-cocycle) up to total derivative terms.
There is a specific way (called the {\it branching structure}) to determine the orientation of tetrahedron,
thus to determine the sign of $s$ for 3-cocycles $\omega_3{}^{s}$ by the determinant of volume, $s\equiv\det(\vec{v_{32}},\vec{v_{31}},\vec{v_{30}})$. 
Two examples of the orientation with $s=+1, -1$ are:
\bea
&&{\CocycleTriangle{g_0}{g_1}{g_3}{g_2}{1} } = \tetrahedra{g_0}{g_1}{g_2}{g_3}  \label{eq:simplex1}\\ 
&&=\omega_3(g_0 g_1{}^{-1},g_1 g_2{}^{-1}, g_2 g_3{}^{-1}) \label{eq:w3-1}
\eea
\bea
&&{\CocycleTriangleTWO{g_1}{g_0}{g_3}{g_2}{1} } = \tetrahedraTWO{g_1}{g_0}{g_2}{g_3} \label{eq:simplex2}\\ 
&&=\omega_3{}^{-1}(g_0 g_1{}^{-1},g_1 g_2{}^{-1}, g_2 g_3{}^{-1}). \label{eq:w3-2}
\eea
Here we define the numeric ordering $g_{1'}<g_{2'}<g_{3'}<g_{4'}<g_1<g_2<g_3<g_4$, and our arrows connect from the higher to lower ordering.

Now we can 
compute the induced 2-cocycle (the dimensional reduced 1+1D path integral) with a given inserted flux $A$,
determined from three tetrahedra of 3-cocycles, see Fig.\ref{fig:decomp_2-cocycle} and Eq.(\ref{eq:2-cocycle}).
\begin{widetext}
\onecolumngrid
\begin{figure}[h!]
\includegraphics[width=0.65\textwidth]{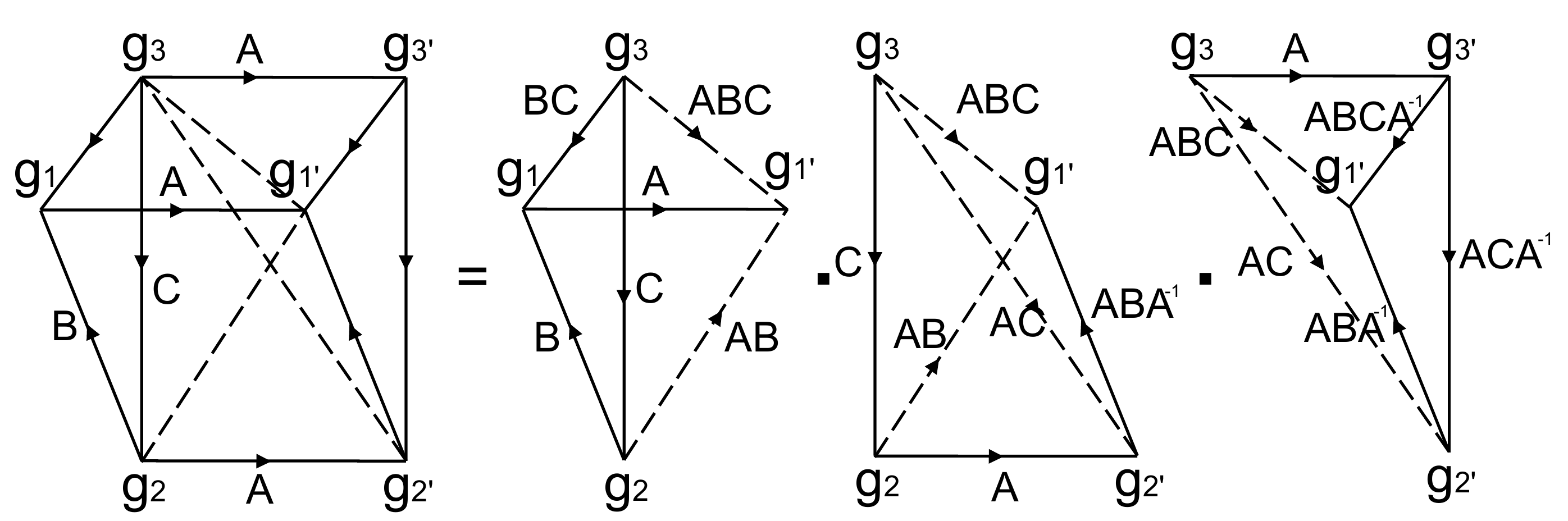} 
\caption{The  
triangulation of a $M^3=M^2 \times  I^1$ topology (here $M^2$ is a spatial cylinder composed by the $A$ and $B$ direction, with a $I^1$ time) into three tetrahedra with branched structures. 
}
\label{fig:decomp_2-cocycle}
\end{figure}
\bea
&&\beta_A(B,C) \equiv {\CocycleTriangle{g_1}{g_2}{g_{1'}}{g_3}{4}  } \centerdot {\CocycleTriangleTWO{g_{2}}{g_{2'}}{g_{1'}}{g_3}{4} } \centerdot {\CocycleTriangle{g_{1'}}{g_{2'}}{g_{3'}}{g_3}{2}  } \label{eq:2-cocycle} \\
&&=\frac{\omega(A,B,C)^{-1} \cdot  \omega(AB A^{-1},A,C)}{  \omega(AB A^{-1},ACA^{-1},A) }
=\frac{\omega(B,A,C)}{\omega(A,B,C) \omega(B,C,A)}
=\frac{\omega(g_1 g_{2}^{-1},  g_{1'} g_1^{-1}, g_2 g_{3}^{-1} )}{\omega(g_{1'} g_1^{-1},g_1 g_{2}^{-1},g_2 g_{3}^{-1}) \omega(g_1 g_{2}^{-1},g_2 g_{3}^{-1},g_{1'} g_1^{-1})} \;\;\;\;\;
\eea
\end{widetext}
\twocolumngrid

We show that among the Type I, II, III 3-cocycles discussed in Sec.\ref{sec: GC},
only when $\omega_3$ is the Type III 3-cocycle $\omega_{\text{III}}$ (of Eq.\ref{type3}), this induced 2-cochain is nontrivial (i.e. a 2-cocycle but not a 2-coboundary). 
In that case,
\be \label{eq:induced 2-cocycle}
\beta_A(B,C)=\exp[\ti \frac{2\pi}{N_{123}}(b_1 a_2 c_3 -a_1 b_2 c_3 -b_1 c_2 a_3)]
\ee 
If we insert $Z_{N_1}$ flux $A=(a_1,0,0)$, then 
we shall compare Eq.(\ref{eq:induced 2-cocycle}) with the nontrivial 2-cocycle $\omega_2(B,C)$ in $\cH^2(Z_{N_2} \times Z_{N_3},\tU(1))$ $=\mathbb{Z}_{N_{23}}$,
\be \label{eq:induced 2-cocycle-2}
\omega_2(B,C)=\exp[\ti \frac{2\pi}{N_{23}}( b_2 c_3 )].
\ee
\cblue{The $\beta_A(B,C)$ is indeed nontrivial 2-cocycle as $\omega_2(B,C)$ 
in the second cohomology group $\cH^2(Z_{N_2} \times Z_{N_3},\tU(1))$. 
Below we like to argue that this Eq.(\ref{eq:induced 2-cocycle-2}) implies the projective representation of the symmetry group $Z_{N_2} \times Z_{N_3}$.
Our argument is based on two facts. 
First, the dimensionally reduced 
SPTs in terms of spacetime partition function Eq.(\ref{eq:induced 2-cocycle-2}) is 
a nontrivial 1+1D SPTs.\cite{{Wang:2014pma}}  
We can physically understand it from the symmetry-twist as a branch-cut modifying the Hamiltonian\cite{Hung:2013cda,{Wang:2014pma}} (see also Sec.\ref{sec:flux}).
Second, from Ref.\onlinecite{Chen:2011pg}'s Sec VI, we know that the 1+1D SPT symmetry transformation $\otimes_x U^x(g)$ along
the 1D's x-site is dictated by 2-cocycle.
The onsite tensor $S(g) \equiv \otimes_x U^x(g)$ acts on a chain 
of 1D SPT renders
\bea
S(g) | \alpha_L, \dots ,\alpha_R\rangle =\frac{\omega_2(\alpha_L^{-1}g^{-1},g)}{\omega_2(\alpha_R^{-1}g^{-1},g)} | g\alpha_L, \dots ,g\alpha_R\rangle,\;\;\;\;\;
\eea
where $\alpha_L$ and $\alpha_R$ are the two ends of the chain, with $g, \alpha_L, \alpha_R, \dots \in G$ all in the symmetry group.
We can derive the effective degree of freedom on the 0D edge $|\alpha_L\rangle$ forms a projective representation of symmetry, we find: 
\bea
&&S(B) S(C) |\alpha_L\rangle \nonumber\\
&&=\frac{ {\omega_2(\alpha_L^{-1}{C}^{-1} B^{-1},B )}  {\omega_2(\alpha_L^{-1}{C}^{-1},B )}  }{ {\omega_2(\alpha_L^{-1}{C}^{-1} B^{-1},BC)}} S(BC) |\alpha_L\rangle \nonumber\\
&&=\omega_2(B,C) S(BC) |\alpha_L\rangle
\eea
In the last line, we implement the 2-cocycle condition of $\omega_2$: $\delta \omega_2(a,b,c)=\frac{\omega_2(b,c) {\omega_2(a,bc)} }{ {\omega_2(ab,c)} {\omega_2(a,b)}}=1$.
The projective representation of symmetry transformation $S(B) S(C) =\omega_2(B,C) S(BC)$ is explicitly derived,
and the projective phase is the 2-cocycle $\omega_2(B,C)$ classified by $\cH^2(G,\tU(1))$.
Interestingly, the symmetry transformations on two ends together will form a linear representation,
namely $S(B) S(C) |\alpha_L, \dots ,\alpha_R\rangle = S(BC) |\alpha_L, \dots ,\alpha_R\rangle$.\cite{Chen:2011pg}
}


The same argument holds when $A$ is $Z_{N_2}$ flux or $Z_{N_3}$ flux.
\cblue{From Sec.\ref{sec: Type III zero}}, the projective representation of symmetry implies  
the nontrivial ground state degeneracy
if we view the system as a dimensionally-reduced 1D line segment as in Fig.\ref{fig:induced_2-cocycle}(d).
From the ${N_{123}}$ factor in Eq.(\ref{eq:induced 2-cocycle}), we conclude there is ${N_{123}}$-fold degenerated zero energy modes. 

\begin{figure}[h!]
\includegraphics[width=0.45\textwidth]{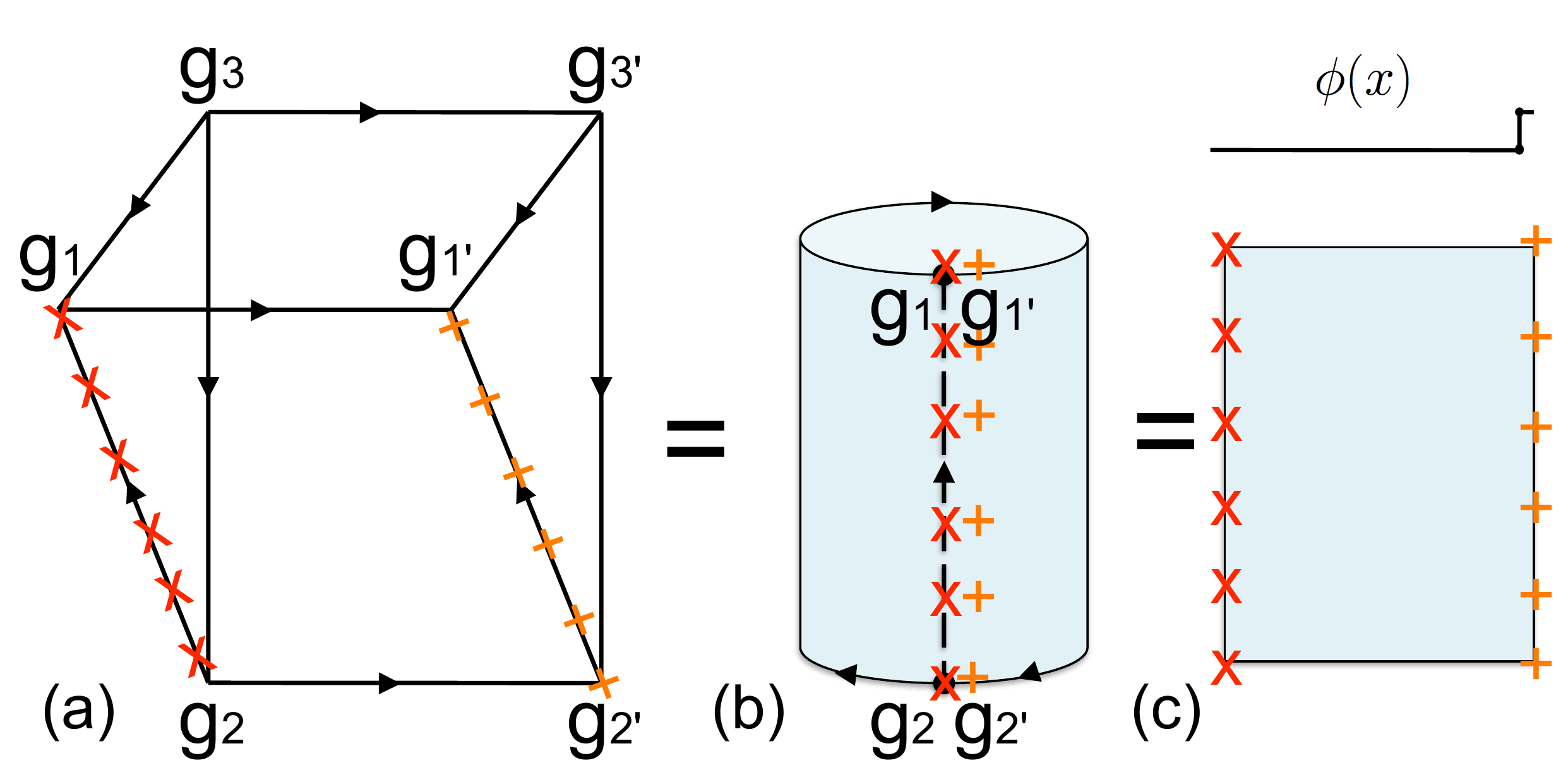} 
\caption{ The $Z_{N_1}$ {\bf symmetry breaking} domain wall along the red $\times$ mark and/or orange + mark, which induces ${N_{123}}$-fold degenerate zero energy modes. 
The situation is very similar to Fig.\ref{fig:induced_2-cocycle} (however, there was $Z_{N_1}$ {\bf symmetry-preserving} flux insertion). 
We show that both cases the induced 2-cochain from calculating path integral $\textbf{Z}_{\text{SPT}}$ renders a nontrivial 2-cocycle of $\cH^2(Z_{N_2} \times Z_{N_3},\tU(1))$ $=\mathbb{Z}_{N_{23}}$,
thus carrying nontrivial projective representation of symmetry.
}
\label{fig:induced_TypeIII_2-cocycle_domain}
\end{figure}

We should make two more remarks: \\
(i) The precise 1+1D path integral is actually summing over $g_{v}$ with a fixed flux $A$ as
$
\textbf{Z}_{\text{SPT}}=|G|^{-N_v} \sum_{\{ g_{v}\}; \text{fixed $A$}}   \beta_A(B,C)
$, but overall our discussion above still holds.\\
(ii) We have used 3-cocycle to construct a symmetry-preserving SPT ground state under $Z_{N_1}$ flux insertion.
We can see that indeed
a $Z_{N_1}$ symmetry-breaking domain wall of Fig.\ref{fig:induced_TypeIII_2-cocycle_domain}
 can be done in almost the same calculation - using 3-cocycles filling a 2+1D spacetime complex(Fig.\ref{fig:induced_TypeIII_2-cocycle_domain}(a)). Although there in Fig.\ref{fig:induced_TypeIII_2-cocycle_domain}(a), we need to fix the group elements $g_1=g_2$ on one side  
 (in the time independent domain wall profile, we need to fix $g_1=g_2=g_3$) and/or fix $g_1'=g_2'$ on the other side.
Remarkably, we conclude that both the {\bf $Z_{N_1}$-symmetry-preserving flux insertion} and {\bf $Z_{N_1}$ symmetry-breaking domain wall} both
provides a ${N_{123}}$-fold degenerate ground states (from the nontrivial projective representation for the $Z_{N_2}$, $Z_{N_3}$ symmetry).
The symmetry-breaking case is consitent with Sec.\ref{sec: degenerate zero A}.


\section{Type I, II, III class observables: Flux insertion and non-dynamically ``gauging'' the non-onsite symmetry \label{sec:flux}}

\begin{figure}[h!]
\includegraphics[width=0.5\textwidth]{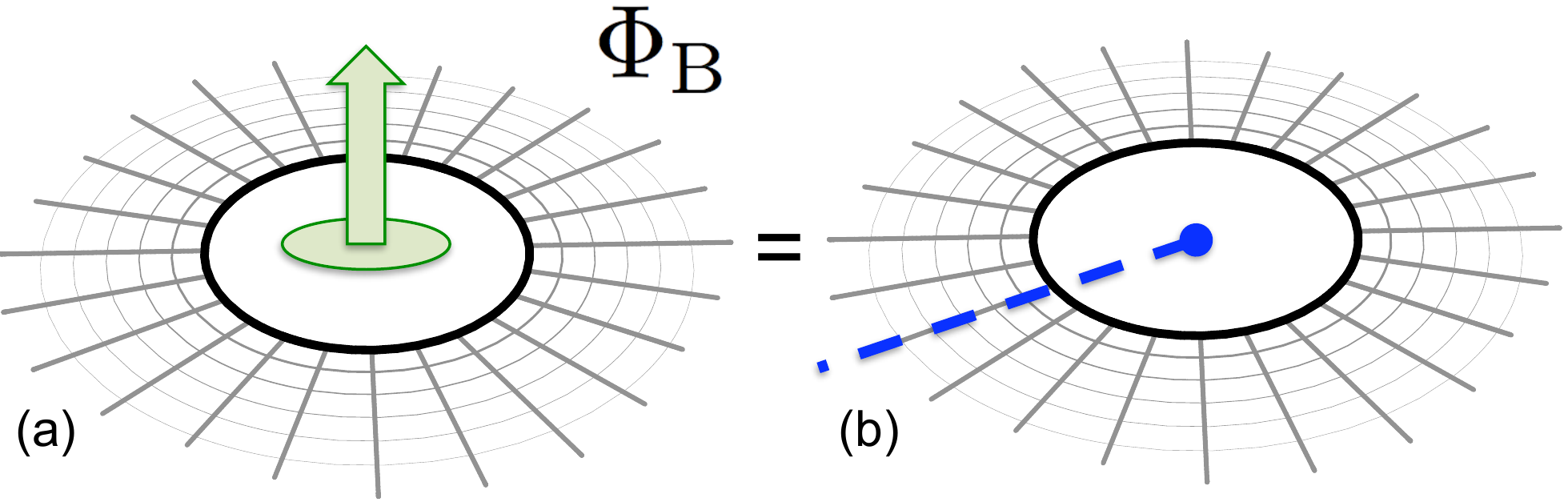} 
\caption{
(a) Thread a gauge 
flux $\Phi_B$ through a 1D ring (the boundary of 2D SPT).
(b) The gauge 
flux is effectively captured by 
a branch cut (the dashed line in the blue color). Twisted boundary condition is applied on the branch cut.
The (a) and (b) are equivalent in the sense that both cases capture the equivalent physical observables, such as the energy spectrum.
}
\label{fig:flux_twist}
\end{figure}

With the Type I, Type II, Type III SPT lattice model built in Sec.\ref{sec:Type I}, 
in principle we can perform numerical simulations to measure their physical observables, such as (i) the energy spectrum, (ii) the entanglement entropy and (iii) the central charge of the edge modes.
Those are the physical observables for the ``untwisted sectors'', and we would like to further achieve more physical observables on the lattice, by
applying the parallel discussion in Ref.\onlinecite{Santos:2013uda}, using $Z_N$ gauge 
flux insertions through the 1D ring. The similar idea can be applied to detect SPTs numerically.\cite{Zaletel}
The gauge flux insertion on the SPT edge modes (lattice Hamiltonian) is like {\it gauging its non-onsite symmetry in a non-dynamical way}. We emphasize that {\it gauging in a non-dynamical way} because the gauge flux is not a local degree of freedom on each site, but a global effect.
The Hamiltonian affected by gauge flux insertions can be realized as the Hamiltonian with twisted boundary conditions, see an analogy made in Fig.\ref{fig:flux_twist}. 
Another way to phrase the flux insertion is that it creates a monodromy defect\cite{Wen:2013ue} (or a branch cut) which modify both the bulk and the edge Hamiltonian.
\cblue{Namely, our flux insertion acts effectively as the \emph{symmetry-twist}\cite{Hung:2013cda,{Wang:2014pma}} modifying the Hamiltonian}.
Here we 
outline the twisted boundary conditions on the Type I, Type II, Type III SPT lattice model of Sec.\ref{sec:Type I}. 


\begin{figure}[h!]
 \includegraphics[width=0.2\textwidth]{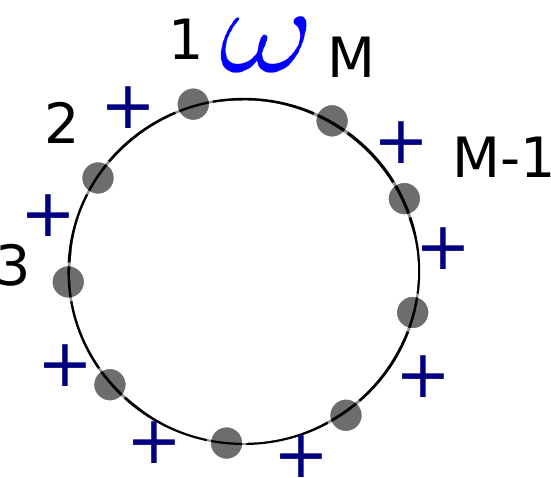}
\caption{
The illustration of an effective 1D lattice model with $M$-sites on a compact ring under a discrete $Z_N$ flux insertion. 
Effectively the gauge 
flux insertion is  captured by a branch cut located between the site-$M$ and the site-$1$.
This results in a $Z_N$ variable $\omega$ insertion as a twist effect modifying the lattice Hamiltonian around the site-$M$ and the site-$1$.
}
\label{fig:flux_twist_ab}
\end{figure}



We firstly review the work done in Ref.\onlinecite{Santos:2013uda} of Type I  SPT class and then extends it to Type II, III class. 
(We skip some tedious calculation 
to Appendix.\ref{App B}.) 
We aim to 
build a lattice model with twisted boundary conditions
to capture the edge modes physics in the presence of a unit of 
${Z}_{N}$ flux insertion. 
Since the gauge flux effectively introduces a branch cut
breaking the translational symmetry of $T$ (as shown in Fig.~\ref{fig:flux_twist}),
the gauged (or twisted) Hamiltonian, say $\tilde{H}^{(p)}_{N}$, is {\it not} invariant respect to translational operator $T$, say $[\tilde{H}^{(p)}_{N}, T]\neq 0$.
The challenge of constructing $\tilde{H}^{(p)}_{N}$ is to firstly find a new (so-called {\it magnetic or twisted}) translation operator $\tilde{T}^{(p)}$ incorporating the gauge flux effect at the branch cut, in Fig.~\ref{fig:flux_twist} (b)
and in Fig.\ref{fig:flux_twist_ab}, say the branch cut is between the site-$M$ and the site-$1$.
We propose two principles to construct the twisted lattice model.
The first general principle is that
a string of $M$ units of {\it twisted translation operator} $\tilde{T}^{(p)}$ renders 
a 
{\it twisted symmetry transformation} $\tilde{S}^{(p)}_{N}$ incorporating a $Z_N$ unit flux,
\be
\label{Type I principle2-1}
\bullet \;
\tilde{S}^{(p)}_{N} \equiv
( \tilde{T}^{(p)} )^{M}
=\tilde{S}^{(p)}_{N}  \cdot (U^{(N,p)}_{M,1} [ {\sigma}^{\dagger}_{M} {\sigma}^{}_{1} ])^{-1} \cdot U^{(N,p)}_{M,1} [ \omega {\sigma}^{\dagger}_{M} {\sigma}^{}_{1}], \;\, 
\ee
with the unitary operator $(\tilde{T}^{(p)})$, i.e. $(\tilde{T}^{(p)})^\dagger\tilde{T}^{(p)}=\openone$.
We clarify that $U^{(N,p)}_{M,1}$ is from Eq.(\ref{eq:Type I symmetry explicit}), 
where $U^{(N,p)}_{M,1} [\dots] \equiv U^{(N,p)}_{M,1} \circ [\dots] $ 
means $U^{(N,p)}_{M,1}$ is a function of 
$\dots$ variables. For example, $ U^{(N,p)}_{M,1} [ \omega {\sigma}^{\dagger}_{M} {\sigma}^{}_{1}]$ means that the variable ${\sigma}^{\dagger}_{M} {\sigma}^{}_{1}$ in Eq.(\ref{eq:Type I symmetry explicit}) is replaced by $\omega {\sigma}^{\dagger}_{M} {\sigma}^{}_{1}$ with an extra $\omega$ insertion.
The second principle is that the twisted Hamiltonian is invariant in respect of 
the twisted translation operator, thus also invariant in respect of twisted symmetry transformation, i.e.
\be \label{Type I principle2-2}
\bullet\;\;\; 
[ \tilde{H}^{(p)}_{N}, \tilde{T}^{(p)} ] =0,\;\;\; 
[\tilde{H}^{(p)}_{N}, \tilde{S}^{(p)}_{N} ]=0.
\ee
%
%
We solve Eq.(\ref{Type I principle2-1}) by finding the twisted lattice translation operator 
\begin{equation}
\tilde{T}^{(p)}
= T\, (U^{(N,p)}_{M,1} ( {\sigma}^{\dagger}_{M} {\sigma}^{}_{1}) )
\tau_{1},
\end{equation}  
for each 
$p \in \mathbb{Z}_N$
classes. 
For the $s$ units of $Z_N$ flux, we have the generalization of $\tilde{T}^{(p)}$ from a unit $Z_N$ flux as, 
\be
\tilde{T}^{(p)}|_{s}
=
T\,
(U^{(N , p)}_{M,1} [ {\sigma}^{\dagger}_{M} {\sigma}^{}_{1}] )^s \,\tau^{s}_{1}.
\ee 

Indeed, there is no difficulty to extend this construction to Type II, III classes.
For Type II SPT classes (with nonzero indices $p_{12}$ and $p_{21}$ of Eq.(\ref{S1symp12}), 
while $p_1=p_2=0$) 
the non-onsite symmetry transformation can be reduced from NNN to NN coupling term 
$U^{(N_1,p_{12})}_{j,j+2} \to U^{(N_1,p_{12})}_{j,j+1}$ , also from   
$U^{(N_2,p_{21})}_{j,j+2} \to U^{(N_2,p_{21})}_{j,j+1}$.
The Type II twisted symmetry transformation has exactly the same form as Eq.(\ref{Type I principle2-1}) except replacing the $U$.
For Type III SPT classes, the Type III twisted symmetry transformation also has the same form as Eq.(\ref{Type I principle2-1})
 except replacing the $U$ to $W$ in Eq.(\ref{eq:Type III_W}).
The second principle in Eq.(\ref{Type I principle2-2}) also follows. 
\\




\noindent
{\bf Twisted Hamiltonian}

The twisted Hamiltonian 
$
\tilde{H}^{(p_1,p_2,p_{12})}_{N_1,N_2}
$
can be readily constructed from
$
{H}^{(p_1,p_2,p_{12})}_{N_1,N_2}
$
of Eq.~(\ref{eq:Type II Hamiltonian lattice}),
with the condition Eq.(\ref{Type I principle2-2}).
(An explicit example for Type I SPT 1D lattice Hamiltonian with a gauge flux insertion has been derived in Ref. \onlinecite{Santos:2013uda}, which we shall not repeat here.)

Notice that 
the twisted non-trivial
Hamiltonian breaks the SPT global symmetry (i.e. 
if
$p
\neq
0
$
mod($N$), then 
$
[\tilde{H}^{(p)}_{N}, S^{(p)}_{N}  ]
\neq
0
$),
which can be regarded as the sign of $Z_N$ anomaly.\cite{Wen:2013oza}
On the other hand, in the trivial state $p=0$, 
Eq.~(\ref{Type I principle2-1}) yields
$
\tilde{S}^{(p=0)}_{N} 
=
S^{(p=0)}_{N}
=
\prod^{M}_{j=1}\tau_{j}
$,
where the twisted trivial Hamiltonian still \textit{commutes}
with the global ${Z}_{N}$ onsite symmetry,
and the twisted boundary effect is nothing but the usual toroidal 
boundary conditions.~\cite{Henkel} (See also a discussion along the
context of SPT and the orbifolds.\cite{Sule:2013qla})\\

The twisted Hamiltonian 
provides distinct low energy spectrum due to the gauge flux insertion (or the symmetry-twist). 
The energy spectrum thus can be
physical observables to distinguish SPTs. Analytically we can use the field theoretic mode expansion for multiplet scalar chiral bosons
$
\Phi_I(x) ={\phi_{0}}_{I}+K^{-1}_{IJ} P_{\phi_J} \frac{2\pi}{L}x+\ti \sum_{n\neq 0} \frac{1}{n} \alpha_{I,n} e^{-in x \frac{2\pi}{L}},
$
with zero modes ${\phi_{0}}_{I}$ and winding modes $P_{\phi_J}$ satisfying the commutator $[{\phi_{0}}_{I},  P_{\phi_J}]=\ti\delta_{IJ}$. The Fourier modes satisfies a generalized Kac-Moody algebra: 
$[\alpha_{I,n} , \alpha_{J,m} ]= n K^{-1}_{IJ}\delta_{n,-m}$.
The low energy Hamiltonian, in terms of various quadratic mode expansions, becomes
\be
H= \frac{(2\pi)^2}{{4\pi} L} [ V_{IJ} K^{-1}_{I l1} K^{-1}_{J l2} P_{\phi_{l1}} P_{\phi_{l2}}+\sum_{n\neq0} V_{IJ} \alpha_{I,n} \alpha_{J,-n}]+\dots
\ee
Following the procedure outlined in Ref.\onlinecite{Santos:2013uda} with gauge flux (compared to the ungauged case in Ref.\onlinecite{Chen:201301}), 
taking into account the twisted boundary conditions, we expect the conformal dimension of gapless edge modes of central charge $c=1$ free bosons
labeled by the primary states $| n_1,m_1,n_2,m_2 \rangle$ (all parameters are integers) with the same compactification radius $R$ for Type I and Type II SPTs
(for simplicity, we assume $N_1 =N_2  \equiv N$):
\bea
&& \label{}
\tilde{\Delta}^{(p_1,p_2,p_{12})}_{N}(n_1,m_1,n_2,m_2;R)\\
&&=
\frac{1}{R^2}\left(n_1+\frac{p_1}{N} +\frac{p_{21}}{N_{}}\right)^{2}
+
\frac{R^2}{4}\left(m_1 + \frac{1}{N}\right)^{2} \nonumber\\
&&+
\frac{1}{R^2}\left(n_2+\frac{p_2}{N}+\frac{p_{12}}{N_{}}\right)^{2}
+
\frac{R^2}{4}\left(m_2 + \frac{1}{N}\right)^{2} \nonumber
\eea
which is directly proportional to the energy of twisted Hamiltonian. ($p_{12}$ or $p_{21}$ can be used interchangeably.)
The conformal dimension $\tilde{\Delta}^{(p_1,p_2,p_{12})}_{N}({\mathcal{P}_u},{\mathcal{P}_{uv}})$ is intrinsically related to the SPT class labels: $p_1,p_2,p_{12}$,
and is a function of momentum ${\mathcal{P}_u} \equiv (n_u+\frac{p_u}{N} +\frac{p_{uv}}{N_{}})(m_u + \frac{1}{N})$
and ${\mathcal{P}_{uv}} \equiv (n_u+\frac{p_u}{N} +\frac{p_{uv}}{N_{}})(m_v + \frac{1}{N})$.
Remarkably, for Type III SPTs, the nature of \emph{non-commutative symmetry} generators will play the key rule, as if the gauged conformal field theory (CFT)
and its correspoinding gauged dynamical bulk theory has \emph{non-Abelian} features,
we will leave this survey for future works. The bottom line is that different classes of SPT's CFT spectra respond to the flux insertion distinctly, thus we can in principle 
distinguish Type I, II and III SPTs.



\section{Conclusion}

Quantum anomalies have 
recently been 
emphasized to be intimately related 
to classifying and characterizing symmetry-protected topological  
states (SPTs) and topologically ordered states.\cite{Wen:2013oza}
While fermionic anomalies are more familiar to the high-energy particle physics communities (such as Adler-Bell-Jackiw anomaly,\cite{{Adler:1969gk},{Bell:1969ts}}
see Supplemental Material\cite{Supplementary Material}), the bosonic anomalies
in our work are less discussed in the literature.
For particle physicists, one may attempt to compute the anomaly 
through (i) a 1-loop Feynman diagram of chiral fermions\cite{{Adler:1969gk},{Bell:1969ts}} 
or (ii) Fujikawa path integral method\cite{Fujikawa:1979ay} by a Jacobi integral 
measure variation under the symmetry transformation. 
However, here, in our work, we instead seek 
another route, a fully bosonic language, to capture bosonic anomalies. 
We ask {\it what are the anomalous signals for these bosonic anomalies}. The result is summarized in Table \ref{table3}.


Since some recent papers also discuss the issues of 
anomalies in the context of SPTs or condensed matter setting\cite{Kapustin:2014lwa,Kapustin:2014tfa,{GYCho},{LiuWen},{Ringel},{Wang:2014ajz},{literature-anomaly},{Kravec:2013pua}}
we shall stress the meaning of \emph{quantum anomaly} more clearly.
We shall also ask: 

\frm{``How does the \emph{bosonic anomaly} of our study relate to the context of the known quantum anomaly in the language of \emph{high energy physics}?''}

To answer this question, we have defined, 

\frm{The {quantum anomaly} is \emph{the obstruction of a symmetry of a theory to be fully-regularized for the full quantum theory as an onsite symmetry on the UV-cutoff lattice 
in the same spacetime dimension}.} 

First, this understanding is consistent with the cases of ABJ anomaly, 
where the symmetry of a classical action \emph{cannot} be a symmetry of any regularization of the full quantum theory.
For example, in chiral U(1)-anomaly at quantum level, the axial U(1)$_A$ symmetry is in conflict with the vector U(1)$_V$ symmetry conservation.\cite{{Adler:1969gk},{Bell:1969ts},{Fujikawa:1979ay}}

Second, one can further ask, ``how can we fully regularize the edge theory with bosonic anomalies on the same spacetime dimension(1+1D) if it has quantum anomalies?''
The answer is that, ``because the (anomalous) symmetry is realized as a \emph{non-onsite} symmetry instead of as a \emph{onsite} symmetry, we can still realize the edge theory 
on the lattice \emph{anomalously}.''
Again, this agrees with our result and the known previous work.\cite{{2011PhRvB..84w5141C},{Chen:2011pg},{Chen:2012hc},{Santos:2013uda},{Wang:2013yta}}
This regularization with \emph{non-onsite} symmetry indeed is 
analogues to the Ginsparg-Wilson fermion approach\cite{Ginsparg:1981bj} dealing with the fermion doubling problem for chiral fermions
using \emph{non-onsite} symmetry.\cite{Wang:2013yta}
The \emph{non-onsite symmetry} is an \emph{anomalous symmetry}; 
thus that is why it is difficult to {\it gauge the non-onsite symmetry locally and dynamically} (see Ref.\onlinecite{Wang:2013yta} for 
a connection between Ginsparg-Wilson fermions 
and SPTs).

Furthermore, 
another way to understand the anomaly is 
that one can regularize the quantum theory with onsite symmetry, if the regularization is done with an \emph{extra dimensional bulk}\cite{Chen:2011pg} 
(thus not in the same spacetime dimension as the boundary).
Again, this realization agrees with the quantum anomaly picture leaking quantum numbers through an extra dimensional bulk, shown in Fig.\ref{fig:flux_cut_analogy}.

Let us now summarize the Type I, II, III bosonic anomalies using the above understanding.
To detect Type II bosonic SPTs, 
we find that the classic model studied by Jackiw-Rebbi\cite{Jackiw:1975fn} or Goldstone-Wilczek\cite{Goldstone:1981kk} offers a similar prototype observable. 
More precisely, the {\bf induced fractional quantum number} is found in 
$p_{12}$ class in $G=Z_{N_1}\times Z_{N_2}$ symmetry. 
For Type II SPTs, the $Z_{N_1}$-symmetry-breaking domain wall will gap the edge and then induce a ${\frac{p_{12}}{N_{12}}}$ fractional unit of $Z_{N_2}$ charge (Fig.\ref{fig:domain_walls_kink}).
The fermionized language shown in Fig.\ref{fig:soliton_current}, can capture the 1-loop effect analogous to ABJ anomaly's 1-loop calculation.\cite{{Adler:1969gk},{Bell:1969ts}}

Type III SPTs' bosonic anomaly provides different 
phenomena. The $N_{123}$-fold {\bf degenerate ground states} are induced  
from either the symmetry-breaking domain wall on the 1D edges
(Fig.\ref{fig:induced_TypeIII_2-cocycle_domain}) or the symmetry-preserving 
monodromy defect connecting edges through the bulk of a cylinder (which
can be viewed as a dimensional-reduced 1D line system in Fig.\ref{fig:induced_2-cocycle}).
We show that the induced projective representation of symmetry under the above two circumstances implies 
the $N_{123}$-fold degenerate zero energy modes.\cite{finite_size}  
%
%

We shall stress that the Type III edge's symmetry transformation provides a new kind of symmetry charge $Q$ coupling as 
$Q \int \epsilon^{u v w} \partial_x \phi^{}_{v}(x) \phi^{}_{w}(x) dx$
in the current term Eq.(\ref{eq:globalS_Type III}), which is rather distinct from the conventional 
symmetry charge $q$ coupling as 
$q \int  \partial_x \phi^{}_{u}(x) dx$.
While the work done in Ref.\onlinecite{{Lu:2012dt},{Ye:2013upa}} cannot accommodate Type III class ($p_{123}\neq 0$) SPTs,
our approach with a new charge vector $Q$ goes beyond previous work; 
thus we expect to obtain the new refined classification for the field theory also for other finite symmetry groups using Eq.(\ref{eq:globalS_Type III})
and its generalizatoin.   

For Type II and Type III SPT classes, we can 
characterize them by dimensional reduction to 
a lower dimensional boundary, and look for its induced quantum number or topological defects(
similar effects happen in Majorana zero modes for free-fermion SPT cases\cite{Santos:2010gq}).
For Type I class $p_{1} \in { \Z_{N_1}}$, however, the 
physical observables we found so far is a bulk probe, instead of having a dimensional-reduction 
to a lower dimensional system trapped with 
nontrivial quantum number. For Type I SPT probe, either the flux insertion goes through the bulk cylinder, 
or the branch cut/monodromy defects connects from the edges to the bulk (Fig.\ref{fig:flux_cut_analogy}).
One can calculate the conformal dimension $\tilde{\Delta}(\mathcal{P})$(both analytically and numerically) as a function of momentum $\mathcal{P}$\cite{DiFrancesco}
in the twisted sector under monodromy defects, and one can show that each SPT class has distinct spectral shift.\cite{Santos:2013uda}

Meanwhile, this type of probe such as flux insertion/monodromy defect which connects from the boundary to the bulk is essentially a signal of edge anomalous physics.
In a sense, we develop an {\it effective 1D lattice Hamiltonian} with non-onsite symmetry 
which signals the existence of higher dimensional bulk, just like
the edge chiral boson theory signals the bulk Chern-Simons theory.
Only through a `` non-dynamically'' gauge-flux insertion, are we able to achieve {\it gauging the non-onsite symmetry effectively with a monodromy defect branch cut}, shown in Fig.\ref{fig:induced_TypeIII_2-cocycle_domain},\ref{fig:flux_cut_analogy}.
This provides yet another way to interpret the edge anomaly - the 1D edge modified twisted Hamiltonian incorporating a branch cut does {\it not} preserve the original symmetry $G$ 
(i.e. $[\tilde{H}^{(p)}_{N}, S^{(p)}_{N}  ]\neq 0$ in Sec.\ref{sec:flux}).
However, one can readily check the full bulk-edge Hamiltonian description $\tilde{H}^{(p)}_{N,\text{cilinder}}$ such as a cylinder with two edges in Fig.\ref{fig:flux_cut_analogy} will preserve the symmetry $G$ 
(i.e. $[\tilde{H}^{(p)}_{N,\text{cilinder}}, S^{(p)}_{N}  ] = 0$).

We 
emphasize that, 
thanks to realizing the symmetry as a \emph{non-onsite symmetry} on the lattice,
all our SPT edge lattice constructions are successfully regularized on discrete space lattice with finite dimensional Hilbert space on the 1D ring. 
All our lattice models are ready for performing 
numerical simulations. 
For future directions, it will be 
interesting to numerically study its physical observables to 
detect the distinct SPT classes, and also to study the charge transport 
with two edges on the cylinder {\it talking to each other}
by quantum number pumping process in Fig.\ref{fig:flux_cut_analogy}.
This may require a full construction of the extra dimensional 2D bulk lattice,
which can 
address
what we mean by {\it quantum anomalies} as 
{\it some lower dimensional theory leaks certain quantum numbers to an extra dimensional bulk 
}.\\

 \section*{Acknowledgements} 
JW is grateful to Frank Wilczek and Yusuke Nishida for inspiring discussions some years ago about Goldstone-Wilczek method,
also to Roman Jackiw pointing out 
the first use of 3-cocycle in physics in Ref.\onlinecite{{Jackiw:1984rd},{Treiman:1986ep}}.
JW thanks Ling-Yan Hung for very helpful feedback on the manuscript. 
This research is supported by NSF Grant No.
DMR-1005541, NSFC 11074140, and NSFC 11274192. 
Research at Perimeter Institute is supported by the Government of Canada through Industry Canada and by the Province of Ontario through the Ministry of Economic Development \& Innovation.

\appendix

\begin{center}

{\bf Supplemental Materials}\\

\end{center}


\section{Chiral Fermionic Adler-Bell-Jackiw Anomalies and Topological Phases}


In contrast to the \emph{bosonic anomalies of discrete symmetries} studied in our main text, 
here we present a \emph{chiral fermionic anomaly} (ABJ anomalies\cite{{Adler:1969gk},{Bell:1969ts}}) of \emph{a continuous} U(1) \emph{symmetry} realized in topological phases 
in condensed matter. 

\begin{figure}[h!] 
{\includegraphics[width=.24\textwidth]{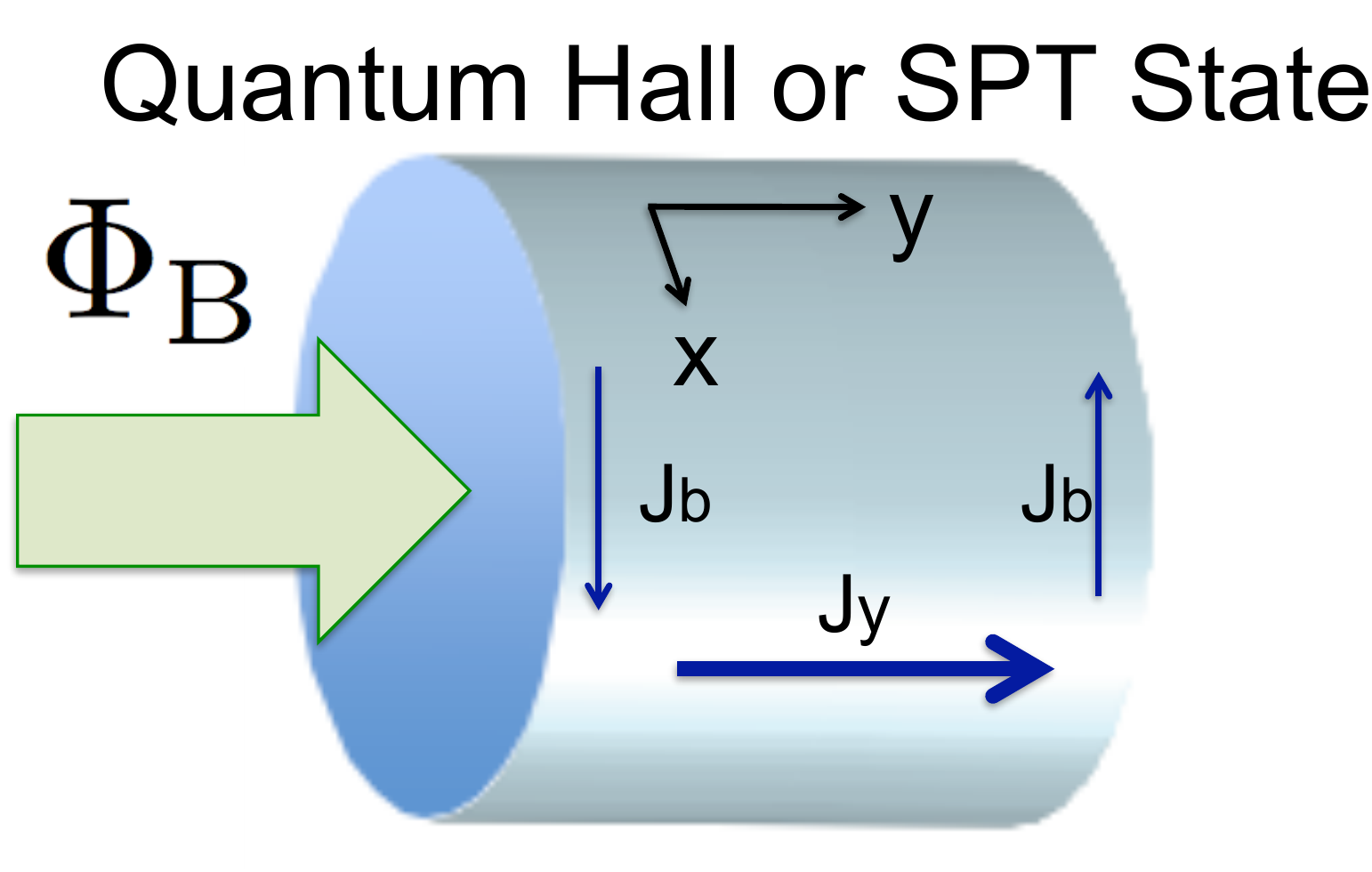}}
\caption{For topological phases, the anomalous current $J_b$ of the boundary theory along $x$ direction leaks to $J_y$ along $y$ direction in the extended bulk system. 
$\Phi_B$-flux insertion $d\Phi_B/dt=-\oint E \cdot d L$ induces the electric $E_x$ field along the $x$ direction.
The effective Hall effect dictates that $J_y=\sigma_{xy}E_x=\sigma_{xy}\varepsilon^{\mu\nu}\,\partial_{\mu} A_{\nu}$, with the effective Hall conductance $\sigma_{xy}$
probed by an external U(1) gauge field $A$. 
}
\label{cylinder}
\end{figure}

Specifically we consider an 1+1D U(1) quantum anomaly realization through 1D edge of U(1) quantum Hall state, such as in Fig.\ref{cylinder}.
We can formulate a Chern-Simons action $S=\int\big(\frac{K}{4\pi} \;a\wedge d a+ \frac{q}{2\pi}A \wedge d a\big)$ with an internal statistical gauge field $a$ and an external U(1) electromagnetic gauge field $A$. 
Its 1+1D boundary is described by a (singlet or multiplet-)chiral boson theory of a field $\Phi$ ($\Phi_L$ on the left edge, $\Phi_R$ on the right edge). 
Here the field strength $F=dA$ is equivalent to the external U(1) flux in the 
flux-insertion thought experiment 
threading through the cylinder
(see a precise derivation in the Appendix of Ref.\onlinecite{Santos:2013uda}). 
Without losing generality, 
let us first focus on 
the boundary with only one edge mode. We derive its equations of motion as 
\bea 
\partial_{\mu}\,j_{\textrm{b} }^{\mu}
&=&
\frac{\sigma_{xy}}{2}\,
\varepsilon^{\mu\nu}\,F_{\mu\nu}
={\sigma_{xy}}\,
\varepsilon^{\mu\nu}\,\partial_{\mu} A_{\nu}
=
J_{y}, \label{eq:J=sy}\\
\partial_{\mu}\,  j_{\textrm{L}}&=&\partial_{\mu} (\frac{q}{2\pi}\epsilon^{\mu\nu} \partial_\nu \Phi_L)=\partial_{\mu} (q\bar{\psi} \gamma^\mu  P_L  \psi)=+J_{y},\;\;\;\;\; \\
\partial_{\mu}\,  j_{\textrm{R}}&=&-\partial_{\mu} (\frac{q}{2\pi}\epsilon^{\mu\nu} \partial_\nu \Phi_R)=\partial_{\mu} (q\bar{\psi}  \gamma^\mu P_R \psi)=-J_{y}.\;\;\;\;\;\;\;
\eea
We show the Hall conductance
from its definition $J_y ={\sigma_{xy}} E_x $ in Eq.(\ref{eq:J=sy}), as 
$\sigma_{xy}=qK^{-1}q/(2\pi)$.

\begin{figure}[h!] 
{\includegraphics[width=.4\textwidth]{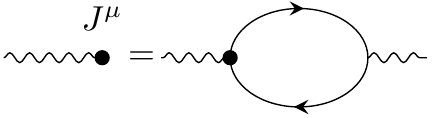}}
\caption{In the fermionic language, the 1+1D chiral fermions (represented by the solid line) and the external U(1) gauge field (represented by the wavy curve) 
contribute to a 1-loop Feynman diagram correction to the axial current $j^\mu_A$. This leads to the non-conservation of $j^\mu_A$ as the anomalous current
$\partial_{\mu}\,  j_{\textrm{A}}^{\mu}= \varepsilon^{\mu\nu} (qK^{-1}q/2\pi)\,F_{\mu\nu}.$ }
\label{ABJ_1loop_1D}
\end{figure}

Here $j_{\textrm{b}}$ stands for the edge current. 
A left-moving current $j_L=j_{\textrm{b}}$ is on one edge, and a right-moving current $j_R=-j_{\textrm{b}}$ is on the other edge, shown in Fig.\ref{cylinder}. 
By bosonization, we convert a compact bosonic phase $\Phi$ to the fermion field $\psi$.
The vector current is $ j_{\textrm{L}}+j_{\textrm{R}} \equiv j_{\textrm{V}}$, and the U(1)$_V$ current is conserved.
The axial current is $ j_{\textrm{L}}-j_{\textrm{R}} \equiv j_{\textrm{A}}$, and we derive the famous ABJ U(1)$_A$ anomalous current in 1+1D 
(or Schwinger's 1+1D quantum electrodynamic [QED] anomaly\cite{Schwinger:1962tp}).
\bea
\partial_{\mu}\,  j_{\textrm{V}}^{\mu}&=&\partial_{\mu}\, ( j_{\textrm{L}}^{\mu}+j_{\textrm{R}}^{\mu} )= 0,\\ 
\partial_{\mu}\,  j_{\textrm{A}}^{\mu}&=&\partial_{\mu}\, ( j_{\textrm{L}}^{\mu}-j_{\textrm{R}}^{\mu} )=\sigma_{xy} \varepsilon^{\mu\nu}\,F_{\mu\nu}. 
\eea
This simple bulk-edge derivation is consistent with field theory 1-loop calculation through Fig.\ref{ABJ_1loop_1D}.
It shows that the combined boundary theory on the left and right edges (living on the edges of a 2+1D U(1) Chern-Simons theory) can
be viewed as an 1+1D anomalous world of Schwinger's 1+1D QED.\cite{Schwinger:1962tp}
This is an example of chiral fermionic anomaly of a continuous U(1) symmetry when $K$ is an odd integer.
(When $K$ is an even integer, it becomes a chiral bosonic anomaly of a continuous U(1) symmetry.)

\color{black}

\section{Matrix Product Operators and Lattice Regularization \label{sec:Appendix A}}

In this Appendix, we provide detailed calculation about the Matrix Product Operators (MPO) formalism. 
Contracting three neighbored sites tensor $T(g_a),T(g_b),T(g_c)$ of $G$-symmetry transformation $S$ (with $g \in G$) in different order will render a relative projective phase. 
Importantly, if this phase is nontrivial 3-cocycle,
then it readily verifies that our lattice construction maps to the nontrivial class of cohomology group.
We also show the details of lattice regularizations in Sec.\ref{sec:Type I}. 

We now formulate the unitary operator $S^{(p)}_N$ as the 
MPO with the form: 
\be \label{eq:MPO_general-A}
S^{(p)}_N=\sum_{ \{j,j'\}} \tr[T^{j_1 j_1'}_{\alpha_1 \alpha_2}  T^{j_2 j_2'}_{\alpha_2 \alpha_3} \dots T^{j_M j_M'}_{\alpha_M \alpha_1}] |   j'_1, \dots, j'_M\rangle \langle j_1, \dots, j_M |.
\ee
This is the operator formalism of matrix product states(MPS).
Here {\it physical indices} $j_1,j_2, \dots, j_M$ and $j'_1,j'_2, \dots, j'_M$ are labeled by input/output physical eigenvalues (here ${Z}_N$ rotor angle), the subindices $1,2,\dots, M$ are the physical site indices.
%
There are also {\it virtual indices} $\alpha_1, \alpha_2, \dots, \alpha_M$ which are traced in the end.
Summing over all the operation from $\{j,j'\}$ indices, we shall reproduce the symmetry transformation operator $S^{(p)}_N$. 

To find out the projective representation $e^{i \theta(g_a,g_b,g_c)}$. 
Below we use the facts of tensors $T(g_a), T(g_b), T(g_c)$ acting on three neighbored sites $a$, $b$, $c$ with group elements ${g_a,g_b,g_c}$. 
There is a generic projective relation
\be \label{eq:projective-a}
T(g_a\cdot g_b)=P_{g_a,g_b}^\dagger T(g_a) T(g_b)P_{g_a,g_b}.
\ee
Here $P_{g_a,g_b}$ is the projection operator.
We contract three neighbored-site tensors in two different orders,
\be \label{eq:projective-3cocycle app}
{(P_{g_a,g_b} }{ \otimes I_3) P_{g_a g_b, g_c}} \simeq e^{\ti \theta(g_a,g_b,g_c)} {( I_1 \otimes P_{g_b,g_c} ) P_{g_a,g_bg_c}}.
\ee
The left-hand-side contracts the sites $a,b$ first then with the site $c$, while the right-hand-side contracts the sites $b,c$ first then with the site $a$.
Here $\simeq$ means the equivalence is up to a projection out of un-parallel states. 
If the projective phase $e^{i \theta(g_a,g_b,g_c)}$ happens to be the nontrivial 3-cocycle in a cohomology group, then
we reach our goal - this verifies that our SPT lattice constructions (thus also the low energy field theory) maps to the nontrivial class of the cohomology group $\mathcal{H}^3(G,\tU(1))$.
This is the emphasis of this Appendix.

\subsection{Type I and II classes  \label{sec:Appendix-Type II}}

We first write down the 
$\tilde{\sigma}^{(1)}_{j}$, $\tilde{\sigma}^{(2)}_{j}$ operators
in the lattice regularization for Type II symmetry transformation in Sec.\ref{sec: lattice Type I II}: 
\begin{widetext}
\bea \label{eq: Type II tilde sigma 1}
&& \tilde{\sigma}^{(u)}_{j}  =  {\begin{pmatrix} 
\left( \begin{smallmatrix}
1 & 0 & 0 & 0 \\ 0 & {\omega}_{uv} & 0 & 0 \\
0 & 0 & . & 0 \\
0 & 0 & 0 & {\omega}_{uv}^{\gcd(N_u,N_v)-1}
\end{smallmatrix}
\right)  & 0 & 0 & 0 \\
0 & \left( \begin{smallmatrix}
1 & 0 & 0 & 0 \\ 0 & {\omega}_{uv} & 0 & 0 \\
0 & 0 & . & 0 \\
0 & 0 & 0 & {\omega}_{uv}^{\gcd(N_u,N_v)-1}
\end{smallmatrix}
\right)  & 0 & 0\\  
0 & 0 & \ddots  & 0\\  
0 & 0 & 0 &  \left( \begin{smallmatrix}
1 & 0 & 0 & 0 \\ 0 & {\omega}_{uv} & 0 & 0 \\
0 & 0 & . & 0 \\
0 & 0 & 0 & {\omega}_{uv}^{\gcd(N_u,N_v)-1}
\end{smallmatrix}
\right)
\end{pmatrix}}_j =  \langle \phi_{u,j} | e^{\ti \tilde{\phi}_{u,j}}   | \phi_{u,j} \rangle,  \;\;\;\;\;\; 
\eea
\end{widetext}

The  $\tilde{\sigma}^{(u)}_{j}$ matrix has $N_u \times N_u$ components. It is block diagonalizable
with $\frac{N_u}{N_{12}} $ subblocks, and each subblock with ${N_{12}} \times {N_{12}}$ components. 
We now verify that our symmetry transformations Eq.(\ref{S1symp12})
(thus the lattice Hamiltonian Eq.(\ref{eq:Type II Hamiltonian lattice})) corresponds to non-trivial 3-cocycles in the third cohomology group in $\cH^3({Z}_{N_1} \times {Z}_{N_2},\tU(1))=\mathbb{Z}_{N_1} \times \mathbb{Z}_{N_2} \times \mathbb{Z}_{\gcd(N_1,N_2)}$.
In this subsection we will focus on $p_1$, $p_2$ Type I and $p_{12}$ Type II class.
The test below will verify that we indeed construct lattice model of the nontrivial SPT of the $p_{12}$ class with $p_{u} \in \mathbb{Z}_{N_u}$, 
$p_{12} \in \mathbb{Z}_{\gcd(N_1,N_2)}$.

The tensor $T(g)$ and the unitary operator ${S}^{(p_1,p_{12})}_{N_1} \cdot {S}^{(p_2,p_{21})}_{N_2}$ 
as the matrix product operators (MPO) already appeared in the main text, we should not repeat it here.
To find out the projective representation $e^{i \theta(g_a,g_b,g_c)}$ of three tensors $T(g_a), T(g_b), T(g_c)$ acting on three neighbored sites, we follow the fact in Eq.(\ref{eq:projective}),
and derive the Type I, II projection operator:
\begin{widetext}
\bea \label{eq: P Type II}
&&P^{(p)}_{N_1,N_2} \equiv P^{(p)}_{N_1,N_2,(m^{(1)}_a,m^{(1)}_b),(m^{(2)}_a,m^{(2)}_b)}
=
\int d{\phi'}^{(1)}_{in} d {\phi'}^{(2)}_{in}   ( |  {\phi'}^{(1)}_{in}+\frac{2\pi {m^{(1)}_b} }{N_1} \rangle   |  {\phi'}^{(1)}_{in} \rangle \langle  {\phi'}^{(1)}_{in}  | )
(|  {\phi'}^{(2)}_{in}+\frac{2\pi {m^{(2)}_b} }{N_2} \rangle   |  {\phi'}^{(2)}_{in} \rangle \langle  {\phi'}^{(2)}_{in}  |) \nonumber \\
&& \cdot e^{\ti p_1{\phi'}^{(1)}_{in} \big( [{m^{(1)}_a+m^{(1)}_b}]_{N1}-  (m^{(1)}_a+ {m^{(1)}_b})  \big)/N_1}  \cdot
 e^{\ti p_2{\phi'}^{(1)}_{in} \big( [{m^{(2)}_a+m^{(2)}_b}]_{N2}-  (m^{(2)}_a+ {m^{(2)}_b})  \big)/N_2}  \nonumber \\
&& \cdot e^{\ti p_{21} (\tilde{\phi'}^{(1)}_{in})_r\Big( [m^{(2)}_a+{m^{(2)}_b}]_{N_2} - (m^{(2)}_a+{m^{(2)}_b}) \Big) /N_2} 
\cdot e^{\ti p_{12} (\tilde{\phi'}^{(2)}_{in})_r\Big( [m^{(1)}_a+{m^{(1)}_b}]_{N_1} - (m^{(1)}_a+{m^{(1)}_b}) \Big) /N_1}, 
\eea
\end{widetext}
where $[m_a+m_b]_N$ with subindex $N$ means taking the value module $N$.
$P_{g_1,g_2}$ inputs one state $\langle  {\phi'}^{(1)}_{in} | \langle  {\phi'}^{(2)}_{in}  |$ and outputs two states 
$( |  {\phi'}^{(1)}_{in}+\frac{2\pi {m^{(1)}_b} }{N_1} \rangle   |  {\phi'}^{(1)}_{in} \rangle)(|  {\phi'}^{(2)}_{in}+\frac{2\pi {m^{(2)}_b} }{N_2} \rangle   |  {\phi'}^{(2)}_{in} \rangle)$.
To derive the projective phase $e^{i\theta(g_a, g_b, g_c)}$, we start by contracting $T(g_b)$ and $T(g_c)$ firstly, and then the combined tensor contracts with $T(g_a)$ gives:
\begin{widetext}
\bea
&&( I_1 \otimes P_{g_b,g_c} ) P_{g_a,g_b g_c}
=\int d{\phi''}^{(1)}_{in} d {\phi''}^{(2)}_{in}   
( |  {\phi''}^{(1)}_{in}+\frac{2\pi ({m^{(1)}_b} + {m^{(1)}_c}) }{N_1} \rangle_a   |  {\phi''}^{(1)}_{in}+\frac{2\pi {m^{(1)}_c} }{N_1} \rangle_b   |  {\phi''}^{(1)}_{in} \rangle_{c} \langle  {\phi''}^{(1)}_{in}  |_{abc}  ) \nonumber\\ 
&&\cdot ( |  {\phi''}^{(2)}_{in}+\frac{2\pi ({m^{(2)}_b} + {m^{(2)}_c}) }{N_2} \rangle_a   |  {\phi''}^{(2)}_{in}+\frac{2\pi {m^{(2)}_c} }{N_2} \rangle_b   |  {\phi''}^{(2)}_{in} \rangle_{c} \langle  {\phi''}^{(2)}_{in}  |_{abc} ) \nonumber\\
&&\cdot 
e^{\ti p_1{\phi''}^{(1)}_{in} \big( ({m^{(1)}_a+m^{(1)}_b}+m^{(1)}_c)_{N1}-  m^{(1)}_a- {m^{(1)}_b} - m^{(1)}_c  \big)/N_1}  \cdot
 e^{\ti p_2{\phi''}^{(2)}_{in} \big( ({m^{(2)}_a+m^{(2)}_b}+m^{(2)}_c)_{N2} -  m^{(2)}_a - {m^{(2)}_b} -{m^{(2)}_c}   \big)/N_2} \nonumber \\
&& \cdot e^{\ti p_{21} (\tilde{\phi''}^{(1)}_{in})_r\big( (m^{(2)}_a+{m^{(2)}_b}+{m^{(2)}_c})_{N_2} - m^{(2)}_a - {m^{(2)}_b} -{m^{(2)}_c} \big) /N_2} 
\cdot e^{\ti p_{12} (\tilde{\phi''}^{(2)}_{in})_r\big( (m^{(1)}_a+{m^{(1)}_b}+{m^{(1)}_c})_{N_1} -  m^{(1)}_a - {m^{(1)}_b} -{m^{(1)}_c} \big) /N_1}
\eea
\end{widetext}
which inputs one state $\langle \phi_{in} |$ and outputs three states $ |\phi_{in}+\frac{2\pi}{N}(m_b+m_c) \rangle$, $|\phi_{in}+\frac{2\pi}{N}m_c \rangle$ and  $| \phi_{in} \rangle$.
Similarly we can derive $(P_{g_a,g_b} \otimes I_3) P_{g_a g_b, g_c}$ by contracting $T(g_a)$ and $T(g_b)$ firstly, and then the combined tensor contracts with $T(g_c)$.
By computing Eq.(\ref{eq:projective-3cocycle app}), 
with only $p_{1}$ index (i.e. setting $p_2=$$p_{12}=0$),
we can derive Type I 3-cocycle:
\bea \label{eq:proj Type I}
&& e^{\ti \theta(g_a,g_b,g_c)} =
e^{\ti p_{1} ({ \frac{2\pi {m^{(1)}_c} }{N_1}  }) \frac{(m^{(1)}_a+{m^{(1)}_b})_{N_2} - (m^{(1)}_a+{m^{(1)}_b})}{N_1}} \nonumber\\
&& =\omega_{\text{I}}^{(i)}(m_c,m_a,m_b). 
\eea
By 
computing Eq.(\ref{eq:projective-3cocycle app}) 
with only $p_{21}$ index (i.e. setting $p_1=$$p_2=$$p_{12}=0$),
we can recover Type II 3-cocycle, 
\bea \label{eq:proj Type II}
&&e^{\ti \theta(g_a,g_b, g_c)}=e^{\ti p_{21} ({ \frac{2\pi {m^{(1)}_c} }{N_1}  })\big( [m^{(2)}_a+{m^{(2)}_b}]_{N_2} - (m^{(2)}_a+{m^{(2)}_b}) \big) /N_2} \nonumber \\
&&=\omega_{\text{II}}^{(ij)}(m_c,m_a,m_b), 
\eea
up to the index redefinition $p_{21} \to -p_{12}$. 
We thus derive that the projective representation $e^{i \theta(g_a,g_b, g_c)}$ from MPS tensors corresponds to 
the group cohomology approach.\cite{Chen:2011pg} 
From here we learn that the inserted $p_{12}$ and $p_{21}$ are indeed the same indices because 
$e^{\ti p_{21} ({ \frac{2\pi {m^{(1)}_c} }{N_1}  })\big( (m^{(2)}_a+{m^{(2)}_b})_{N_2} - (m^{(2)}_a+{m^{(2)}_b}) \big) /N_2}$ and $e^{\ti p_{12} ({ \frac{2\pi {m^{(2)}_c} }{N_2} } ) \big( (m^{(1)}_a+{m^{(1)}_b})_{N_1} - (m^{(1)}_a+{m^{(1)}_b}) \big) /N_1}$  are equivalent 3-cocycles up to 3-coboundaries,\cite{deWildPropitius:1996gt}
meanwhile $p_{12}=p_{12} \text{ mod } \gcd(N_1,N_2)$. 
This demonstrates that our lattice construction fulfills all $\mathbb{Z}_{\gcd(N_1,N_2)}$ Type II classes of SPT with $Z_{N_1} \times Z_{N_2}$-symmetry, and also 
Type I $\mathbb{Z}_{N_1},\mathbb{Z}_{N_2}$ classes as we desired.

\subsection{Type III class  \label{sec:Appendix-Type III}}

We first motivate our construction of matrix product operators  
by observing that Type III 3-cocycle in Eq.(\ref{type3}) inputs, for example, $a_1 \in Z_{N_1}$, $b_2 \in Z_{N_2}$, $c_3 \in Z_{N_3}$ and outputs a U(1) phase.
This implies that the ${Z}_{N_1}$ symmetry transformation will affect the mixed ${Z}_{N_2},{Z}_{N_3}$ rotor models, while similarly ${Z}_{N_2},{Z}_{N_3}$ global symmetry will cause the same effect.
This observation guides us to write down the tensor $T(g)$ and the symmetry transformation $S^{(p)}_N=S^{(p_{123})}_{N_1,N_2,N_3}$ defined in Sec.\ref{sec:MPO-Type I}.
We propose the tensor $T(g)$ and $S^{(p_{123})}_{N_1,N_2,N_3}$ already in the main text, which we shall not repeat. 
Let us first understand how to regularize the symmetry operator on the lattice.
\begin{widetext}

\begin{center}
\bf{Lattice Regularization}
\end{center}

We derive the non-onsite symmetry transformation $W^{\text{III}}_{j,j+1}$, acting on the site $j$ and $j+1$ as:
\bea \label{eq:App_Type III_W}
\bullet \;\;W^{\text{III}}_{j,j+1}&&=\prod_{u,v,w \in \{1,2,3\}} \exp \Big( { \ti {\frac{N_1N_2N_3}{2\pi\gcd({N_1,N_2,N_3})}} \epsilon^{u v w} \frac{p_{123}}{N_{u}}  \big(   \phi^{j+1,(v)}_{in} \phi^{j,(w)}_{in} \big) } \Big)\\
&&={\prod_{(v,w)=(2,3),(3,1),(1,2)}} e^{ \ti   {p_{123}} \big(  ( \phi^{j+1,(v)}_{in}-\phi^{j,(v)}_{in}) \phi^{j,(w)}_{in}  -  (\phi^{j+1,(w)}_{in}-\phi^{j,(w)}_{in} ) \phi^{j,(v)}_{in} \big)  {\frac{N_v N_w}{2\pi\gcd({N_1,N_2,N_3})}} }\\
&&=\prod_{(v,w)=(2,3),(3,1),(1,2)}  \Big( (\sigma_{j}^{(v)\dagger} \sigma_{j+1}^{(v)} )^{\phi^{j,(w)}_{in}} ((\sigma_{j}^{(w)} \sigma_{j+1}^{(w)\dagger})^{\phi^{j,(v)}_{in}} ) \Big)^{ p_{123} {\frac{N_v N_w}{2\pi\gcd({N_1,N_2,N_3})}} }\\
&&={ \prod_{u,v,w \in \{1,2,3\}}  \Big( \sigma_{j}^{(v)\dagger} \sigma_{j+1}^{(v)}  \Big)^{ \epsilon^{u v w}   p_{123} {  \frac{ {\log(\sigma_{j}^{(w)})} N_v N_w}{2\pi \ti \gcd({N_1,N_2,N_3})}} }}\\
&&\equiv W^{\text{III}}_{j,j+1; N_1} \cdot W^{\text{III}}_{j,j+1; N_2} \cdot W^{\text{III}}_{j,j+1; N_3} 
\eea
where we separate $Z_{N_1}$,$Z_{N_2}$,$Z_{N_3}$ non-onsite symmetry transformation to $W^{\text{III}}_{j,j+1; N_1}$,$W^{\text{III}}_{j,j+1; N_2}$,$W^{\text{III}}_{j,j+1; N_3}$ respectively.
More explicitly, we have $Z_{N_1}$ non-onsite symmetry transformation:
\bea \label{eq:Type III_W_N1}
W^{\text{III}}_{j,j+1; N_1}&&=e^{ \ti {p_{123}} \big(  ( \phi^{j+1,(2)}_{in}-\phi^{j,(2)}_{in}) \phi^{j,(3)}_{in}  -  (\phi^{j+1,(3)}_{in}-\phi^{j,(3)}_{in} ) \phi^{j,(2)}_{in} \big) {\frac{N_2N_3}{2\pi\gcd({N_1,N_2,N_3})}} }\\
&&={\Big( (\sigma_{2,j}^\dagger \sigma_{2,j+1} )^{\log(\sigma_{3,j})}((\sigma_{3,j} \sigma_{3,j+1}^\dagger)^{\log(\sigma_{2,j})}) \Big)^{ p_{123} {\frac{N_2N_3}{2\pi \ti \gcd({N_1,N_2,N_3})}} }},
\eea
and $W^{\text{III}}_{j,j+1; N_2}$,$W^{\text{III}}_{j,j+1; N_3}$ have the analogous forms.
We first attempt to regularize this $W^{\text{III}}_{j,j+1}$ operator by defining
\bea \label{eq:Type III phi}
{\phi^{j,(u)}_{in}}\equiv \ti^{-1}\log(\sigma_{u,j})
=\ti^{-1} {\begin{pmatrix} 
\log[1] & 0 & 0 & 0 \\
0 & \log[\omega_u] & 0 & 0\\  
0 & 0 & \ddots  & 0\\  
0 & 0 & 0 & \log[\omega_u^{N_u-1}]
\end{pmatrix}}_j = {\begin{pmatrix} 
0 & 0 & 0 & 0 \\
0 & \frac{2\pi}{N_u} & 0 & 0\\  
0 & 0 & \ddots  & 0\\  
0 & 0 & 0 & \frac{2\pi{(N_u-1)}}{N_u}
\end{pmatrix}}_j,
\eea
here $u \in \{1,2,3\}$.
The challenge of the lattice regularization is to understand what exactly does this operator 
${ (\sigma_{v,j}^\dagger \sigma_{v,j+1} )^{ p_{123} {\frac{{\log(\sigma_{w,j})} N_vN_w}{2\pi \ti \gcd({N_1,N_2,N_3})}} }}$ in Eq.(\ref{eq:Type III_W})
mean on the lattice.
Without losing generality, let us take
${\Big( (\sigma_{v,j}^\dagger \sigma_{2,j+1} )^{\log(\sigma_{3,j})}\Big)^{ p_{123} {\frac{N_2 N_3}{2\pi \ti \gcd({N_1,N_2,N_3})}} }}$ in $W^{\text{III}}_{j,j+1; N_1}$ of Eq.(\ref{eq:Type III_W_N1}) as an example.
The answer to this question is that we should view how this operator acts on the combined $Z_{N_2} \times Z_{N_3}$ states: $| \phi^{(2)} \rangle \otimes | \phi^{(3)} \rangle$.
The $W^{\text{III}}_{j,j+1; N_1}$ operator is a $((N_2)^2 \times (N_3)^2 ) \times ((N_2)^2 \times (N_3)^2 )$-component matrix acting on the $(N_2)^2 \times (N_3)^2$-dimensional Hilbert space spanned by 
the all $| \phi^{(2)}_j \rangle \otimes | \phi^{(2)}_{j+1} \rangle \otimes | \phi^{(3)}_{j}  \rangle \otimes | \phi^{(3)}_{j+1}  \rangle$  states at the site $j$ and $j+1$.
The key is regularizing this operator $W^{\text{III}}_{j,j+1;N_1}$ explicitly, using Eq.(\ref{eq:Type III phi}) as 
\bea 
&&(\sigma_{j}^{(2)\dagger} \sigma_{j+1}^{(2)} )^{{  {\frac{  p_{123} \log(\sigma_{3,j})  N_2 N_3}{2\pi \ti \gcd({N_1,N_2,N_3})}} }}=
{\begin{pmatrix} 
{(\sigma_{j}^{(2)\dagger} \sigma_{j+1}^{(2)} )}^{\log[1]} & 0 & 0 & 0 \\
0 & {(\sigma_{j}^{(2)\dagger} \sigma_{j+1}^{(2)} )}^{\log[\omega_3]} & 0 & 0\\  
0 & 0 & \ddots  & 0\\  
0 & 0 & 0 & {(\sigma_{j}^{(2)\dagger} \sigma_{j+1}^{(2)} )}^{\log[\omega_3^{N_3-1}]}
\end{pmatrix}}_j^{{  {\frac{p_{123}  N_2 N_3}{2\pi \ti \gcd({N_1,N_2,N_3})}} }} \\
&&={\begin{pmatrix} 
(\sigma_{j}^{(2)\dagger} \sigma_{j+1}^{(2)} )^{0} & 0 & 0 & 0 \\
0 & (\sigma_{j}^{(2)\dagger} \sigma_{j+1}^{(2)} )^{ {{  {\frac{p_{123}  N_2 }{\gcd({N_1,N_2,N_3})}} }}  } & 0 & 0\\  
0 & 0 & \ddots  & 0\\  
0 & 0 & 0 & (\sigma_{j}^{(2)\dagger} \sigma_{j+1}^{(2)} )^{  {{  {\frac{p_{123}  N_2 (N_3-1) }{\gcd({N_1,N_2,N_3})}} }} }
\end{pmatrix}}_j  \label{eq: Type III_sigma_2}
\eea
We emphasize that each sub-block involving $(\sigma_{2,j}^\dagger \sigma_{2,j+1} )$ is a $(N_2)^2  \times (N_2)^2$-component matrix. 
(Here $\sigma_{2,j+1}$ is a $N_2  \times N_2$-component matrix.)
There are totally $N_3 \times N_3$ sub-blocks.
We recall that $\sigma_{2}$ are operators defined in this manner in Eq.(\ref{eq: sigma operator}), i.e. $\sigma_{2} \sim e^{i \phi^{(2)}}$, with $\phi^{(2)}$ a $Z_{N_2}$ variable.
Thus, the operator in each sub-block has the form
\bea
(W^{\text{III}}_{j,j+1; N_1}) 
&= &\Big( (\sigma_{2,j}^\dagger \sigma_{2,j+1} )^{n_3 N_2 }((\sigma_{3,j} \sigma_{3,j+1}^\dagger)^{n_2 N_3 }) \Big)^{ \frac{p_{123}}{\gcd({N_1,N_2,N_3})} }
\eea
The notation $n_u$(above $u=2$ or $3$) denotes an integer which corresponds to the $Z_{N_u}$ values for $| \phi^{(u)} = n_u (2\pi/N_u) \rangle$ state in different sub-blocks. 
First, we notice that $p_{123}$ is identified by $p_{123} =p_{123} \text{ mod } {\gcd({N_1,N_2,N_3})}$. 
In addition, when $p_{123}$ is a multiple of ${\gcd({N_1,N_2,N_3})}$, we have $(W^{\text{III}}_{j,j+1; N_1})=1$ 
(here $1$ really means $\openone_{N_2\times N_2,j} \otimes \openone_{N_2\times N_2,j+1} \otimes \openone_{N_3\times N_3,j} \otimes \openone_{N_3\times N_3,j+1}$, the identity operator of $Z_{N_2}$, $Z_{N_3}$ states on sites $j,j+1$).
When $p_{123}$ is not a multiple of ${\gcd({N_1,N_2,N_3})}$,
our lattice construction represents a nontrivial non-onsite symmetry transformation ($W^{\text{III}}_{j,j+1} \neq 1$), thus produces a nontrivial SPT labeled by $p_{123} \in \mathbb{Z}_{\gcd({N_1,N_2,N_3})}$.
One may expect to full-regularize Eq.(\ref{eq: Type III_sigma_2}), we need to solve a constraint $(W^{\text{III}}_{j,j+1; N_1})^{N_1}$ 
analogous to Eq.(\ref{eq:constraint Type I}),(\ref{eq: Type II constraint}).
But we do not have to: 
the exponent in Eq.(\ref{eq: Type III_sigma_2}) 
is already an integer, e.g. ${{  {\frac{p_{123}  N_2 n_3 }{\gcd({N_1,N_2,N_3})}} }}$ is necessarily an integer.
We note that, as we expected, when $p_{123}=\gcd({N_1,N_2,N_3})$, we have $(W^{\text{III}}_{j,j+1; N_1})^{N_1}=1$; when 
$p_{123} \neq \gcd({N_1,N_2,N_3})$, we have $(W^{\text{III}}_{j,j+1; N_1})^{N_1} \neq 1$.
Therefore, we have shown Eq.(\ref{eq: Type III_sigma_2}) as the fully-regularized $Z_{N_2}$ operator acting on the $Z_{N_2} \times Z_{N_3}$ states. 

It is straightforward to apply the above $W^{\text{III}}_{j,j+1; N_1}$ discussion to $S^{(p_{123})}_{N_1,N_2,N_3}$, $W^{\text{III}}_{j,j+1}$. 
We should just regard $S^{(p_{123})}_{N_1,N_2,N_3}$, $W^{\text{III}}_{j,j+1}$
as operators acting on the Hilbert space with 
$Z_{N_1}\times Z_{N_2} \times Z_{N_3}$ states.
We can show that all terms in 
$W^{\text{III}}_{j,j+1; N_1} \cdot W^{\text{III}}_{j,j+1; N_2} \cdot W^{\text{III}}_{j,j+1; N_3}$ can be regularized in the same way.




\begin{center}
\bf{Matrix Product Operators and Cocycles}
\end{center}

Below we calculate in details on Type III analog of Eq.(\ref{eq:projective-3cocycle app}) to derive the nontrivial projective phase in MPO formalism, equivalent to the Type III 3-cocycles Eq.(\ref{type3}).
We use the fact Eq.(\ref{eq:projective-a}) 
to derive the projection tensor $P_{g_a,g_b}$,
{
\bea \label{eq: P Type III}
&&P^{(p)}_{N_1,N_2,N_3}\equiv P^{(p)}_{N_1,N_2,N_3,(m^{(1)}_a,m^{(1)}_b),(m^{(2)}_a,m^{(2)}_b),(m^{(3)}_a,m^{(3)}_b)}\\
&&= 
\prod_{u,v,w \in \{1,2,3\}}  \int d{\phi'}^{(u)}_{in}  ( |  {\phi'}^{(u)}_{in}+\frac{2\pi {m^{(u)}_b} }{N_1} \rangle   |  {\phi'}^{(u)}_{in} \rangle \langle  {\phi'}^{(u)}_{in}  | )
   \cdot
 e^{ \ti 2\pi p_{123} \epsilon^{u v w  } {\phi'}^{(u)}_{in} \big( \frac{m^{(v)}_a}{N_v} \frac{m^{(w)}_b}{N_w} \big)\frac{N_1N_2N_3}{2\pi\gcd({N_1,N_2,N_3})} } \nonumber
\eea
}
\end{widetext}
Similar to Eq.(\ref{eq: P Type II}), $P_{g_1,g_2}$ inputs one state $\langle  {\phi'}^{(1)}_{in} | \langle  {\phi'}^{(2)}_{in}  | \langle  {\phi'}^{(3)}_{in}  |$ and outputs two states 
$( |  {\phi'}^{(1)}_{in}+\frac{2\pi {m^{(1)}_b} }{N_1} \rangle   |  {\phi'}^{(1)}_{in} \rangle)(|  {\phi'}^{(2)}_{in}+\frac{2\pi {m^{(2)}_b} }{N_2} \rangle   |  {\phi'}^{(2)}_{in} \rangle)(|  {\phi'}^{(3)}_{in}+\frac{2\pi {m^{(3)}_b} }{N_3} \rangle   |  {\phi'}^{(2)}_{in} \rangle)$. 
For $( I_1 \otimes P_{g_b,g_c} ) P_{g_a, g_b g_c}$, we start by contracting $T(g_b)$ and $T(g_c)$ firstly, and then the combined tensor contracts with $T(g_a)$ gives:
\begin{widetext}
\bea \label{eq:Type III_P_{g_a, g_b g_c}} 
( I_1 \otimes P_{g_b,g_c} ) P_{g_a, g_b g_c}
&&= \prod_{u,v,w \in \{1,2,3\}} \int d{\phi''}^{(u)}_{in}    ( |  {\phi''}^{(u)}_{in}+\frac{2\pi {m^{(u)}_b} }{N_u} +\frac{2\pi {m^{(u)}_c} }{N_u} \rangle_{a}   |  {\phi''}^{(u)}_{in}+\frac{2\pi {m^{(u)}_c} }{N_u} \rangle_{b}
 |  {\phi''}^{(u)}_{in} \rangle_{c} \langle  {\phi''}^{(u)}_{in}  |_{abc} ) \nonumber\\ %
 &&\cdot e^{ \ti 2\pi p_{123} \epsilon^{uvw} {\phi''}^{(u)}_{in} \big( \frac{m^{(v)}_b}{N_v} \frac{m^{(w)}_c}{N_w} \big)\frac{N_1N_2N_3}{2\pi\gcd({N_1,N_2,N_3})} } 
\eea 
\end{widetext}
In Eq.(\ref{eq:Type III_P_{g_a, g_b g_c}}), we have dropped an extra factor 
$e^{ \ti 2\pi p_{123} \epsilon^{uvw} {\phi''}^{(u)}_{in} \big( \frac{m^{(v)}_a}{N_v} \frac{[m^{(w)}_b+m^{(w)}_c]_{N_w}}{N_w} \big)\frac{N_1N_2N_3}{2\pi\gcd({N_1,N_2,N_3})} }=1$, 
because we are dealing with $Z_{N}$ variables so the module relation renders the factor to be always trivial as $1$.

On the other hand, to derive $(P_{a,b} \otimes I_3) P_{a b{,}c}$, we start by contracting $T(g_a)$ and $T(g_b)$ firstly, and then the combined tensor contracts with $T(g_c)$:
\begin{widetext}
\bea \label{eq:Type III_P_{g_a g_b, g_c}}
(P_{a,b} \otimes I_3) P_{a b{,}c} 
&&=\prod_{u,v,w \in \{1,2,3\}}  \int  d{\phi''}^{(u)}_{in}    ( |  {\phi''}^{(u)}_{in}+\frac{2\pi {m^{(u)}_b} }{N_u} +\frac{2\pi {m^{(u)}_c} }{N_u} \rangle_{a}   |  {\phi''}^{(u)}_{in}+\frac{2\pi {m^{(u)}_c} }{N_u} \rangle_{b}
 |  {\phi''}^{(u)}_{in} \rangle_{c} \langle  {\phi''}^{(u)}_{in}  |_{abc} ) \nonumber \\
 &&\cdot e^{ \ti 2\pi p_{123} \epsilon^{u  v w} ( \frac{2\pi {m^{(u)}_c} }{N_{u}} )\big( \frac{m^{(v)}_a}{N_v} \frac{m^{(w)}_b}{N_w} \big)\frac{N_1N_2N_3}{2\pi\gcd({N_1,N_2,N_3})} } 
\cdot e^{ \ti 2\pi p_{123} \epsilon^{ u v w } {\phi''}^{(u)}_{in} \big( \frac{m^{(v)}_a}{N_v} \frac{m^{(w)}_b}{N_w} \big)\frac{N_1N_2N_3}{2\pi\gcd({N_1,N_2,N_3})} } 
\eea
\end{widetext}
Compare to Eq.(\ref{eq:projective-3cocycle app}), we can derive $e^{\ti \theta(g_a, g_b, g_c)}$ in Eq.(\ref{eq:proj Type III}).

Adjust $p_{123}$ index (i.e. setting Eq.(\ref{eq:Type III T_2})'s $p_{123} \to p_{123}/2$, $p_{213}=p_{312}=0$ 
), and compute Eq.(\ref{eq:projective-3cocycle app}) with only $p_{123}$ index,
we can recover the projective phase revealing Type III 3-cocycle: 
\bea \label{eq:proj Type III}
&&e^{\ti \theta(g_1,g_2, g_3)}=e^{ \ti 2\pi p_{123} \epsilon^{u  v w}\big( \frac{{m^{(u)}_c} }{N_{u}} \frac{m^{(v)}_a}{N_v} \frac{m^{(w)}_b}{N_w} \big)\frac{N_1N_2N_3}{\gcd({N_1,N_2,N_3})} } \nonumber \\
&& \simeq \omega_{\text{III}}^{(uvw)}(m_c,m_a,m_b).
\eea

\twocolumngrid

\section{Induced Fractionalized Charges and Domain Wall Operators}

Here we fill in more details on computing induced fractionalized charges (Type II bosonic anomaly) via lattice domain wall operators, outlined in Sec.III.C.
The symmetry operator is $S=\prod_{j}\,\tau_{j}\,\prod_{j}\,U_{j,j+1}$ acting on all sites on a 1D compact ring.
We define a chain of domain wall operator from the site $j=r_1$ to the site $j=r_2$ as
$D(r_1,r_2) \equiv \prod^{r_2}_{j=r_1}\,\tau_{j}\,\prod^{r_2}_{j=r_1}\,U_{j,j+1}$ which creates a kink at the site $r_1$ and an anti-kink at the site $r_2$.
In the main text, we prescribe a method to capture the fractionalized charge at the kink/anti-kink based on: 
\begin{widetext}
\bea
&& S\,D(r_1,r_2)^m\,S^{\dagger}
= \Big[ U(\omega^{-1} \sigma^{\dagger}_{r_1-1}  \sigma_{r_1})\,
U^{\dagger}(\sigma^{\dagger}_{r_1-1} \sigma_{r_1}) \Big]^m  
 \cdot
\Big[ U(\omega \sigma^{\dagger}_{r_2} \sigma_{r_2+1})\,
U^{\dagger}(\sigma^{\dagger}_{r_2} \sigma_{r_2+1}) \Big]^m 
\cdot D(r_1,r_2)^m 
\eea
\end{widetext}

Above we 
express a generic onsite symmetry operator $\tau_{j}$ capturing 
$\tau_{j}^{(u)}$ for $\prod_u Z_{N_u}$-symmetry. 
We also express a generic non-onsite symmetry operator in terms of $U_{j,j+1}$.
An explicit calculation for Type I's $U_{j,j+1}$ shows:
\bea
\Big[
U(\omega \sigma^{\dagger}_{r} \sigma_{r+1})\,
U^{\dagger}(\sigma^{\dagger}_{r} \sigma_{r+1})
\Big]^m
&=&
e^ 
{
-\ti\,\frac{2\pi p m}{N^2}\,
\sum^{N-1}_{a=1}\,
\left(
\sigma^{\dagger}_{r} \sigma_{r+1}
\right)^a
} \nonumber \\
&=&
e^ 
{
\ti\,\frac{2\pi p m}{N^2}\,
},\\
\Big[
U(\omega^{-1} \sigma^{\dagger}_{r-1} \sigma_{r})\,
U^{\dagger}(\sigma^{\dagger}_{r-1} \sigma_{r})
\Big]^m
&=&
e^ 
{
\ti\,\frac{2\pi p m}{N^2}\,
\sum^{N-1}_{a=1}\,
\left(
\sigma^{\dagger}_{r-1} \sigma_{r}
\right)^a
} \nonumber \\
&=&
e^ 
{
-\ti\,\frac{2\pi p m}{N^2}\,
}.
\eea
We can define $\Big[ U(\omega^{-1} \sigma^{\dagger}_{r_1-1}  \sigma_{r_1})\,
U^{\dagger}(\sigma^{\dagger}_{r_1-1} \sigma_{r_1}) \Big]^m \equiv e^{\ti \Theta_L}$ as the
fractionalized charge phase measurement on the left kink at $r_1$, since this operator contribute the phase gained exactly at the kink $r_1$. 
And we can define
$\Big[ U(\omega \sigma^{\dagger}_{r_2} \sigma_{r_2+1})\,
U^{\dagger}(\sigma^{\dagger}_{r_2} \sigma_{r_2+1}) \Big]^m \equiv e^{\ti \Theta_R}$
as the fractionalized charge phase measurement on the right kink at $r_2$, 
since this operator contribute the phase gained exactly at the anti-kink $r_2$. 
Below we explicit express a generic non-onsite symmetry operator $U_{j,j+1}$ in terms of non-onsite symmetry operators of 
Type I's $U^{(N_u,p_u)}_{j,j+1}$,
Type II's $U^{(N_u,p_{uv})}_{j,j+1}$,
Type III's $W^{\text{III}}_{j,j+1}$, with $u,v\in \{1,2,3 \}$.
The phases gained at the kink can be computed via the quantities $S\,D(r_1,r_2)^m\,S^{\dagger}$ below:

\noindent
$\bullet$ Type I: $S^{(p_1)}_{N_1} (D^{(p_1)}_{N_1})^m S_{N_1}^{(p_1)\dagger}  $ with
$e^{\ti\Theta_L}=e^{-\ti\Theta_R}=
e^{ \ti\,\frac{2\pi p_1}{N_1^2} m\,}$

\noindent
$\bullet$ Type II: $S^{(p_{12})}_{N_2} (D^{(p_{12})}_{N_1})^m S_{N_2}^{(p_{12})\dagger}$ with
$e^{\ti\Theta_L}=e^{-\ti\Theta_R}=e^{ \ti\,\frac{2\pi p_{12}}{N_{2} N_{12}} m\,} $. \\
 
\noindent
$\bullet$ Type III: $S^{(p_{123})}_{N_2} (D^{(p_{123})}_{N_1})^{m} S_{N_2}^{(p_{123})\dagger}$ with $e^{\ti\Theta_L}=e^{-\ti\Theta_R}=e^{ \ti \,\frac{2\pi p_{123} n_3}{N_{123} } m \,}$.
Here $n_3=0,1,\dots,N_3-1$ is the exponent for each subblock of total $N_3$ subblocks inside the $W^{\text{III}}$ matrix
Eq.(\ref{eq: Type III_sigma_2}).

The systematic interpretation of fractionalzied charge is organized in TABLE III in the main text.

\color{black}

\section{Twisted Sectors: Twisted Hamiltonian and Twisted Non-Onsite Symmetry Transformation  \label{App B}}

\noindent
{\bf Type II }

We can adopt the discussion in Sec.\ref{sec:flux} on the twisted translation operator $\tilde{T}^{(p)}$ and the twisted symmetry transformation $S^{(p)}_{N}$ to Type II symmetry class.
What we will focus on is the 
indices $p_{12}$ and $p_{21}$ of Eq.(\ref{S1symp12}). 
We will set $p_1=p_2=0$ for the sake of simplicity.
With this assumption, 
we can adjust the non-onsite symmetry transformation 
$U^{(N_1,p_{12})}_{j,j+2} \to U^{(N_1,p_{12})}_{j,j+1}$ (from NNN to NN), also from 
$U^{(N_2,p_{21})}_{j,j+2} \to U^{(N_2,p_{21})}_{j,j+1}$. 
Here we explicitly indicates that $U^{(N_1,p_{12})}_{j,j+1}$, $U^{(N_2,p_{21})}_{j,j+1}$ are polynomial functions of 
$(\tilde{\sigma}^{(2)\dagger}_{j}\tilde{\sigma}^{(2)}_{j+1})$, $(\tilde{\sigma}^{(1)\dagger}_{j}\tilde{\sigma}^{(1)}_{j+1})$ respectively,
with $\tilde{\sigma}^{(1)}$, $\tilde{\sigma}^{(2)}$ carefully being defined in Eq.(\ref{eq: Type II tilde sigma 1}). 
The two principles addressed in Sec.\ref{sec:flux} for Type I still valid. The first principle becomes defining the twisted symmetry transformation:
\begin{widetext}
\bea
\label{Type II principle2-1}
\bullet\;\;\;  
\tilde{S}^{(p_{12})}_{N_1} 
&\equiv&( \tilde{T}^{(p_{12})}_{N_1} )^{M}
=
{S}^{(p_{12})}_{N_1} \cdot 
\big(U^{(N_1,p_{12})}_{M,1} [ \tilde{\sigma}^{(2)\dagger}_{M}\tilde{\sigma}^{(2)}_{1}]\big)^{-1} \cdot  
U^{(N_1,p_{12})}_{M,1} [\omega_{12} \tilde{\sigma}^{(2)\dagger}_{M}\tilde{\sigma}^{(2)}_{1}] ,\\
\bullet\;\;\;  
\tilde{S}^{(p_{21})}_{N_2} 
&\equiv&( \tilde{T}^{(p_{21})}_{N_2} )^{M}
=  
 {S}^{(p_{21})}_{N_2} \cdot \big( U^{(N_2,p_{21})}_{M,1} [ \tilde{\sigma}^{(1)\dagger}_{M}\tilde{\sigma}^{(1)}_{1}] \big)^{-1} \cdot
 U^{(N_2,p_{21})}_{M,1} [\omega_{21} \tilde{\sigma}^{(1)\dagger}_{M}\tilde{\sigma}^{(1)}_{1}]. \label{Type II principle2-1:2}
\eea
\end{widetext}
with some unitary twisted translation operator $\tilde{T}^{(p_{12})}_{N_1}$, $\tilde{T}^{(p_{21})}_{N_2}$,
where the $\tilde{S}^{(p_{12})}_{N_1}$ incorporating a $Z_{N_1}$ flux at the branch cut, while the $\tilde{S}^{(p_{21})}_{N_2}$ incorporating a $Z_{N_2}$ flux at the branch cut.
Here we insert ${\omega}_{12} \equiv {\omega}_{21}  \equiv e^{\ti \frac{2 \pi}{\gcd(N_1,N_2)}}$ into the non-onsite symmetry transformation $U^{}_{M,1}$ at the $M$-th and the $1$-st sites to capture the branch cut physics as Fig.\ref{fig:flux_twist_ab}. 
The twisted lattice translation operators solved from Eq.(\ref{Type II principle2-1}),(\ref{Type II principle2-1:2}) are 
\bea
&&\tilde{T}^{(p_{12})}_{N_1} 
=
T\, \cdot
U^{(N_1,p_{12})}_{M,1} [ \tilde{\sigma}^{(2)\dagger}_{M}\tilde{\sigma}^{(2)}_{1}]  \cdot
\tau_{1}^{(1)}, \;\;\;\\
&&\tilde{T}^{(p_{21})}_{N_2} 
=
T\, \cdot
U^{(N_2,p_{21})}_{M,1} [ \tilde{\sigma}^{(1)\dagger}_{M}\tilde{\sigma}^{(1)}_{1}]  \cdot
\tau_{1}^{(2)}.
\eea

The second principle is that the twisted Hamiltonian is invariant respect to twisted translation operators $\tilde{T}$, thus also invariant respect to $\tilde{S}$, i.e. 
\bea \label{Type II principle2-2}
\bullet\;\;\; 
&&[ \tilde{H}^{(p)}_{N}, \tilde{T}^{(p_{12})}_{N_1} ] =[ \tilde{H}^{(p)}_{N}, \tilde{S}^{(p_{12})}_{N_1} ] \nonumber\\
&&=[ \tilde{H}^{(p)}_{N}, \tilde{T}^{(p_{21})}_{N_2} ] =[ \tilde{H}^{(p)}_{N}, \tilde{S}^{(p_{21})}_{N_2} ]=0. 
\eea


The twisted Hamiltonian 
$
\tilde{H}^{(p_1,p_2,p_{12})}_{N_1,N_2}
$
for Type I, II can be readily constructed from
$
{H}^{(p_1,p_2,p_{12})}_{N_1,N_2}
$
of Eq.~(\ref{eq:Type II Hamiltonian lattice}),
with the condition in Eq.(\ref{Type I principle2-2}), Eq.(\ref{Type II principle2-2}).\\


\noindent
{\bf Type III}

We follow the same principles 
to explore the Type III twisted sectors with a flux insertion (or branch cut).
We will focus on Type III class with $p_{123} \neq 0$, and other Type I, II class indices are zeros.
The first principle suggests that a string of M units of 
 {\it twisted translation operator} $\tilde{T}^{(p_{123})}_{N_1,N_2,N_3}$ modifies Eq.(\ref{eq:Type III_S})'s
 $S^{(p_{123})}_{N_1,N_2,N_3}$
  to 
a 
{\it twisted symmetry transformation} 
$\tilde{S}^{(p_{123})}_{N_1,N_2,N_3} \equiv
(\tilde{T}^{(p_{123})}_{N_1} )^{M} \cdot  (\tilde{T}^{(p_{123})}_{N_2} )^{M} \cdot (\tilde{T}^{(p_{123})}_{N_3} )^{M} 
$ 
incorporating a $Z_{N_1},Z_{N_2},Z_{N_3}$ unit flux respectively by, 
\begin{widetext}
\bea \label{Type III principle2-1}
\bullet \tilde{S}^{(p_{123})}_{N_1,N_2,N_3} 
 ={S}^{(p_{123})}_{N_1,N_2,N_3} \cdot 
\big( W^{\text{III}}_{M,1} [\sigma_{M}^{(v)\dagger} \sigma_{1}^{(v)} ]\big)^{-1} \cdot W^{\text{III}}_{M,1} [\omega_{123} \,\sigma_{M}^{(v)\dagger} \sigma_{1}^{(v)} ]   
\eea
where the non-onsite symmetry transformation part $W^{\text{III}}_{j,j+1} \equiv W^{\text{III}}_{j,j+1}[\sigma_{v,j}^\dagger \sigma_{v,j+1} ]$ is defined in Eq.(\ref{eq:Type III_W}) as a polynomial 
of $\sigma_{v,j}^\dagger \sigma_{v,j+1} $, and its $\omega_{123}$ insertion 
\be
W^{\text{III}}_{j,j+1} [\omega_{123} \,\sigma_{j}^{(v)\dagger} \sigma_{j+1}^{(v)} ] \equiv
{ \prod_{u,v,w \in \{1,2,3\}} \epsilon^{u v w}   \Big( \omega_{123}\;\sigma_{j}^{(v)\dagger} \sigma_{j+1}^{(v)} \Big)^{ p_{123} {\frac{{\log(\sigma_{j}^{(w)})}  N_vN_v}{2\pi\gcd({N_1,N_2,N_3})}} }}
\ee
\end{widetext}
captures the $Z_{N_u}$ unit flux effect by the branch cut. 
(In Appendix.\ref{sec:Appendix-Type III}, we show that Eq.(\ref{Type III principle2-1}) is regularized on the lattice.)
Adopted the notation in Eq.(\ref{eq:Type III_W}), the twisted lattice translation operator solved from Eq.(\ref{Type III principle2-1}) is 
\be
\tilde{T}^{(p_{123})}_{N_u} 
=
T\, \cdot
W^{\text{III}{}}_{M,1} ( \tilde{\sigma}^{(v)\dagger}_{M}\tilde{\sigma}^{(v)}_{1})  \cdot \tau_{1}^{(u)},
\ee
here ${u,v,w \in \{1,2,3\}}$.

The second principle is that the twisted Hamiltonian is invariant respect to twisted translation operators, thus also invariant respect to twisted symmetry transformations,
\be \label{Type III principle2-2}
\bullet\;\;\; 
[ \tilde{H}^{(p)}_{N}, \tilde{T}^{(p_{123})}_{N_u}  ] =0, \;\; [ \tilde{H}^{(p)}_{N}, \tilde{S}^{(p_{123})}_{N_1,N_2,N_3} ]=0. 
\ee
Based on Eq.(\ref{Type III principle2-2}), it is straightforward to construct a Type III twisted Hamiltonian
incorporating the symmetry twist (equivalently a gauge flux) at the branch cut.



\end{document}